\documentclass{jfm}

\include{commands}
\usepackage{graphicx}
\usepackage{natbib}
\usepackage{color}
\ifCUPmtlplainloaded \else
  \checkfont{eurm10}
  \iffontfound
    \IfFileExists{upmath.sty}
      {\typeout{^^JFound AMS Euler Roman fonts on the system,
                   using the 'upmath' package.^^J}%
       \usepackage{upmath}}
      {\typeout{^^JFound AMS Euler Roman fonts on the system, but you
                   dont seem to have the}%
       \typeout{'upmath' package installed. JFM.cls can take advantage
                 of these fonts,^^Jif you use 'upmath' package.^^J}%
      }
  \else
  \fi
\fi

\ifCUPmtlplainloaded \else
  \checkfont{msam10}
  \iffontfound
    \IfFileExists{amssymb.sty}
      {\typeout{^^JFound AMS Symbol fonts on the system, using the
                'amssymb' package.^^J}%
       \usepackage{amssymb}%
       \let\le=\leqslant  \let\leq=\leqslant
       \let\ge=\geqslant  \let\geq=\geqslant
      }{}
  \fi
\fi

\usepackage{ amsmath} 
\usepackage{float}
\usepackage{rotating}

\ifCUPmtlplainloaded \else
  \IfFileExists{amsbsy.sty}
    {\typeout{^^JFound the 'amsbsy' package on the system, using it.^^J}%
     \usepackage{amsbsy}}
    {}
\fi

\newsavebox{\astrutbox}
\sbox{\astrutbox}{\rule[-5pt]{0pt}{20pt}}

\title[Particle suspensions in shear thinning and shear thickening carrier fluids]{Interface-resolved simulations of particle suspensions in Newtonian, shear thinning and shear thickening carrier fluids}

\author[D. Alghalibi, I. Lashgari, L. Brandt and S. Hormozi]%
{D\ls H\ls I\ls Y\ls A\ns A\ls L\ls G\ls H\ls A\ls L\ls I\ls B\ls I $^{1,}$ $^2$, I\ls M\ls A\ls N\ns L\ls A\ls S\ls H\ls G\ls A\ls R\ls I$^1$,
L\ls U\ls C\ls A\ns B\ls R\ls A\ls N\ls D\ls T$^1$
\and S\ls A\ls R\ls A\ls  H\ns H\ls O\ls R\ls M\ls O\ls Z\ls I$^3$\thanks{Email address for correspondence: hormozi@ohio.edu} }

\affiliation{$^1$ Linn$\acute{\textrm{e}}$ Flow Centre and SeRC (Swedish e-Science Research Centre), \\
KTH Mechanics, S-100 44 Stockholm, Sweden \\[\affilskip] 
$^2$ College of Engineering, Kufa  University, Al Najaf, Iraq \\
$^3$ Department of Mechanical Engineering, Ohio University, Athens, OH 45701-2979, USA}

\date{?; revised ?; accepted ?}
\begin{document}

\maketitle
\begin{abstract}
{ We present a numerical study of noncolloidal spherical and rigid particles suspended in Newtonian, shear thinning and shear thickening fluids employing an Immersed Boundary Method. We consider  a linear Couette configuration to explore a wide range of solid volume fractions ($0.1\le \Phi \le 0.4$) and particle Reynolds Numbers ($0.1\le Re_p \le 10$). We report the distribution of solid and fluid phase velocity and solid volume fraction
and show that close to the boundaries inertial effects result in a significant slip velocity between the solid and fluid phase. 
The local solid volume fraction profiles indicate particle layering close to the walls, which increases with  the nominal $\Phi$. This feature is associated with the confinement effects.
We calculate the probability density function of local strain rates and 
compare their mean value with the values estimated  from the homogenization theory  of \cite{Chateau08}, indicating a reasonable agreement in the Stokesian regimes. 
Both the mean value and standard deviation of the local strain rates increase
 primarily with the solid volume fraction and secondarily with the $Re_p$. 
 The wide spectrum of the local shear rate and its dependency on $\Phi$ and $Re_p$ points to the deficiencies of  the mean value of the  local shear rates in estimating the rheology of these noncolloidal complex suspensions.  
 Finally, we show that in the presence of inertia,  the effective viscosity of these noncolloidal  suspensions deviates from that of Stokesian suspensions. We discuss how inertia affects the microstructure and  provide a scaling argument 
to give a closure for the suspension shear stress for  both Newtonian and power-law suspending fluids.  The stress closure is valid for moderate  particle Reynolds numbers, $ O(Re_p)\sim 10$.}\end{abstract}


\section{Introduction}
\label{sec:intro}
The dynamics of suspensions of solid particles in a fluid medium are inherently complex, yet such complex fluids are ubiquitous in nature, e.g.\ lavas, slurries and debris, and are important for a wide variety of industrial processes, such
as paints, pastes, concrete casting, drilling muds, waste disposal, food processing, crude oil flows with rocks and medicine. This wealth clarifies why the behavior of these suspensions has been extensively studied both experimentally,
theoretically, and more recently via numerical simulations. These studies have uncovered numerous complex features of suspension flows, which are, however, not yet fully understood \cite[see][]{Stickel05}. The complexities and challenges 
are attributed to the large variety of interactions among particles (hydrodynamic, contact, interparticle forces), the physical properties of the particles (shape, size, deformability, volume fraction) and the properties of the suspending 
fluid (Newtonian or non-Newtonian). In this work we numerically study suspensions of neutrally buoyant rigid spheres in both Newtonian and inelastic non-Newtonian fluids.

The rheology of neutrally-buoyant non-Brownian particles suspended in a Newtonian fluid has been largely investigated, see  \cite{Batchelor70,Brady88,Larson99,Stickel05}. This is determined by two dimensionless numbers:  the solid volume
fraction $\Phi$ and the particle Reynolds number $Re_p=\rho_f \dot\gamma a^2/\mu$ (where $a$ is the particle radius,   $\rho_f$ the fluid density, $\mu$ the fluid viscosity and $\dot\gamma$ the flow shear rate).
Many studies focused on the limiting case of Stokesian suspensions  when inertia is negligible (i.e., $Re_p\rightarrow0$) and  the effective viscosity of the suspension  depends only on $\Phi$. 
Theoretical works mainly address the limiting cases of $\Phi\rightarrow0$ and $\Phi\rightarrow \Phi_{max}$, where $\Phi_{max}$ is the maximum packing fraction. When a suspension is dilute, its effective viscosity follows the linear behaviour derived 
by \cite{Einstein906,Einstein911} $\mu_{eff}=\mu(1+2.5\Phi)$ (particles interactions are neglected) or the quadratic formulation of \cite{Batchelor77} $\mu_{eff}=\mu(1+2.5\Phi+6.95\Phi^2)$ (with mutual particle interactions included).
The recent theoretical works of Wyart and co-workers \citep[see e.g.,][]{DeGiuli15}  make use of perturbations around the jamming state to show that the effective viscosity diverges as $\Phi\rightarrow \Phi_{max}$. 

At moderate to large values of  $\Phi$, however, the rheology becomes more complex due to multi-body and short-range interactions. The suspension shear viscosity  increases with $\Phi$ before diverging at  $\Phi_{max}$. In addition, normal stresses 
appear when the suspension is subject to shear. A number of studies have been performed  to measure the suspension effective  shear viscosity and the normal  stresses, see the experiments of
\cite{Boyer11,Bonnoit10,Couturier11,Dbouk13,Deboeuf09,Krieger59,Ovarlez06,Singh03,Zarraga00} and numerical simulations of \cite{DboukLemaire13,Sierou02,Yurkovetsky08,Yeo10}.

In the presence of weak inertia ($Re_p \ne 0$) the rheological measurements start to differ from those in the Stokesian regime. For suspensions with $0.02 \le Re_p \le 10$ and $ .1 \le \Phi \le 0.3$, the recent numerical studies of 
\cite{Kulkarni08a,Picano13,Yeo13} show an increase in the suspension stresses as $Re_p$ increases, although discrepancies  exist in the reported values of stresses.  Due to the improvement of computational methods
\cite[see e.g.][]{Kulkarni08a,Yeo11,Yeo13,Lashgari14,Picano15,Fornari16}, interface resolved simulations of solid particles in Newtonian fluids can now reveal details of the suspension microstructure, and shed light on the role
of inertia in the overall dynamics  \cite[][]{Prosperetti15}. More precisely, the study of \cite{Picano13}  shows that  the  inertia affects the suspension microstructure, resulting in an enhancement of effective shear viscosity. 
As $Re_p$ increases, the pair distribution function becomes more anisotropic, almost zero on the rear of the particles, effectively as additional excluded volumes. This increases the effective solid  volume fraction, and consequently, the effective shear viscosity.
Taking into account this excluded volume effect (which depends on both $\Phi$ and $Re_p$), \cite{Picano13} scaled the effective shear viscosity in presence of small  inertia to that of Stokesian suspensions.  
While there are hardly any studies  addressing  the rheology of suspensions for $Re_p \geq10$ and $ .1 \le \Phi \le 0.45$ \cite{Lashgari14,Lashgari16,Linares17,Bagnold54},  there is a considerable  body of research addressing the rheology of dry granular materials
($\Phi \geq 0.45$), where viscous effects are negligible and friction, collision and particle phase momentum govern the flow dynamics \cite{Andreotti13,DeGiuli15,Trulsson12,Amarsid17}.  

The behaviour of suspensions is even more complex when the carrier fluid is non-Newtonian, such as generalized Newtonian or viscoelastic fluids. Only a few studies have been devoted to noncolloidal particles suspended in non-Newtonian fluids,
attempting to address the bulk rheology from a continuum-level closure perspective. These studies mainly focus on non-inertial suspensions with few exceptions e.g., \cite{Hormozi17}. 

On the theoretical front, the homogenization approach is adopted by  \cite{Chateau08} to derive  constitutive laws  for suspensions of non-colloidal particles in yield stress fluids.   The authors consider  a Herschel-Bulkley suspending fluid 
and show that the bulk rheology of suspensions also follow the Herschel-Bulkley model, with an identical power law index, but with yield stress and consistency that increase with the solid volume fraction. Generally, adding large particles to
a fluid enhances both  the effective viscosity of the bulk and the local shear rate of the fluid phase. While the latter has no influence on the viscosity of a Newtonian suspending fluid, it strongly influences the local  apparent viscosity  in
the case of a non-Newtonian suspending fluid. In the homogenization theory of \cite{Chateau08} the mean value of  the local shear rate is estimated via an energy argument and used to derive the suspension constitutive laws. A number of 
experimental works have been carried out \cite[see][]{Chateau08,Coussot09,Mahaut08,Ovarlez12,Ovarlez06,Vu10,Dagois-Bohy15}  showing a general agreement with the homogenization theory. 

It has also been shown that adding large particles to a shear thinning  fluid not only enhances the effective shear viscosity, but also promotes the onset of  the non-Newtonian behavior  to a smaller shear rate \citep{Poslinski88, Liard14}. These features are also observed for  shear thickening suspending fluids in both Continuous Shear  Thickening (CST) and Discontinuous Shear  Thickening (DST) 
scenarios \citep{Cwalina14, Liard14, Madraki17}. The advancement in the  onset of shear thinning, CST or DST, is explained by  the increase of the local shear rate in the suspending fluid due to the presence of larger particles \citep{Chateau08, DeGiuli15}.
Recent studies show that the homogenization theory underestimates the occurrence of DST \citep{Madraki17} and overestimates the values of the yield stress, which depends on the shear history, and consequently, on the suspension microstructure, see \cite{Ovarlez15}.
While the former can be attributed to the fact that the mean value of the local shear rate is not sufficient to predict DST, instead controlled by the extreme values of the 
local shear rate distribution, the latter implies that constitutive laws derived from homogenization theory need to be refined by taking into account the  microstructure and shear history.

 Here, we present a numerical study of rigid-particle suspensions in pseudoplastic (via the Carreau model)  and dilatant fluids (via the power law model)  employing an Immersed Boundary Method.  The method was originally proposed 
by  \cite{Breugem12} to enable us to resolve the dynamics of finite-size neutrally buoyant particles in flows. The simulations are performed in a planar Couette configuration where the rheological parameters of the suspending fluids and the
 particle properties are kept constant while the volume fraction of the solid phase varies in the range of $10\%\leq \Phi \leq40\%$ and the bulk shear rate varies to provide particle Reynolds numbers  in the range of $0.1\leq Re_p \leq10$. 
 The well resolved simulations  benefit the study via  providing access to the local values of particle and fluid phase velocities, shear rate and particle volume fraction. We explore the confinement effects, microstructure and rheology of 
 these complex suspensions when the particle Reynolds number is non-zero. We compare our results with the recently proposed constitutive laws based on the homogenization theory \cite[see][]{Chateau08} and refine these laws when inertia is present. 
 
The paper is organised as follows. The governing equations and numerical method are discussed in $\S{2}$. We present our results in  $\S{3}$, first considering the local distribution of flow profiles and the Stokesian rheology of non colloidal
suspensions with generalized Newtonian suspending fluids. We then discuss the role of inertia and propose new  rheological laws including inertial effects.  A summary of the main conclusions
and some final remarks are presented in $\S{4}$.

\section{Governing equations and numerical method}
\subsection{Governing equations}

We study the motion of finite-size rigid particles suspended in an inelastic  non-Newtonian carrier fluid.
The genaralized incompressible Navier-Stokes equation with shear-dependent viscosity and the continuity equation govern the motion of the fluid phase,
 \begin{align}\label{eq:GNS} 
\rho (\frac{\partial \textbf{u}}{\partial t} + \textbf{u} \cdot \nabla \textbf{u}) = -\nabla P + \nabla \cdot [\hat{\mu}(\textbf{u}) (\nabla \textbf{u} + \nabla \textbf{u}^T ) ] + \rho \textbf{f}, \\ \nonumber
\nabla \cdot \textbf{u} = 0, 
\end{align}
{where $\textbf{u}=(u,v,w)$ is the velocity vector containing the spanwise, streamwise and wall-normal components corresponding to the $(x,y,z)$ coordinate directions respectively (see figure~\ref{fig:box}). The pressure is denoted 
by $P$ while the density of both the fluid and particles is indicated by $\rho$ as we consider neutrally buoyant particles. The fluid viscosity $\hat{\mu}$ varies as a function of the local shear rate $\dot{\gamma} (\textbf{u})$
following the rheological Carreau-law or Power-law models defined below. Finally the body force \textbf{f} indicates the forcing from the dispersed phase on the carrier fluid.     
\begin{figure}
\centering{
\includegraphics[width=0.9\linewidth]{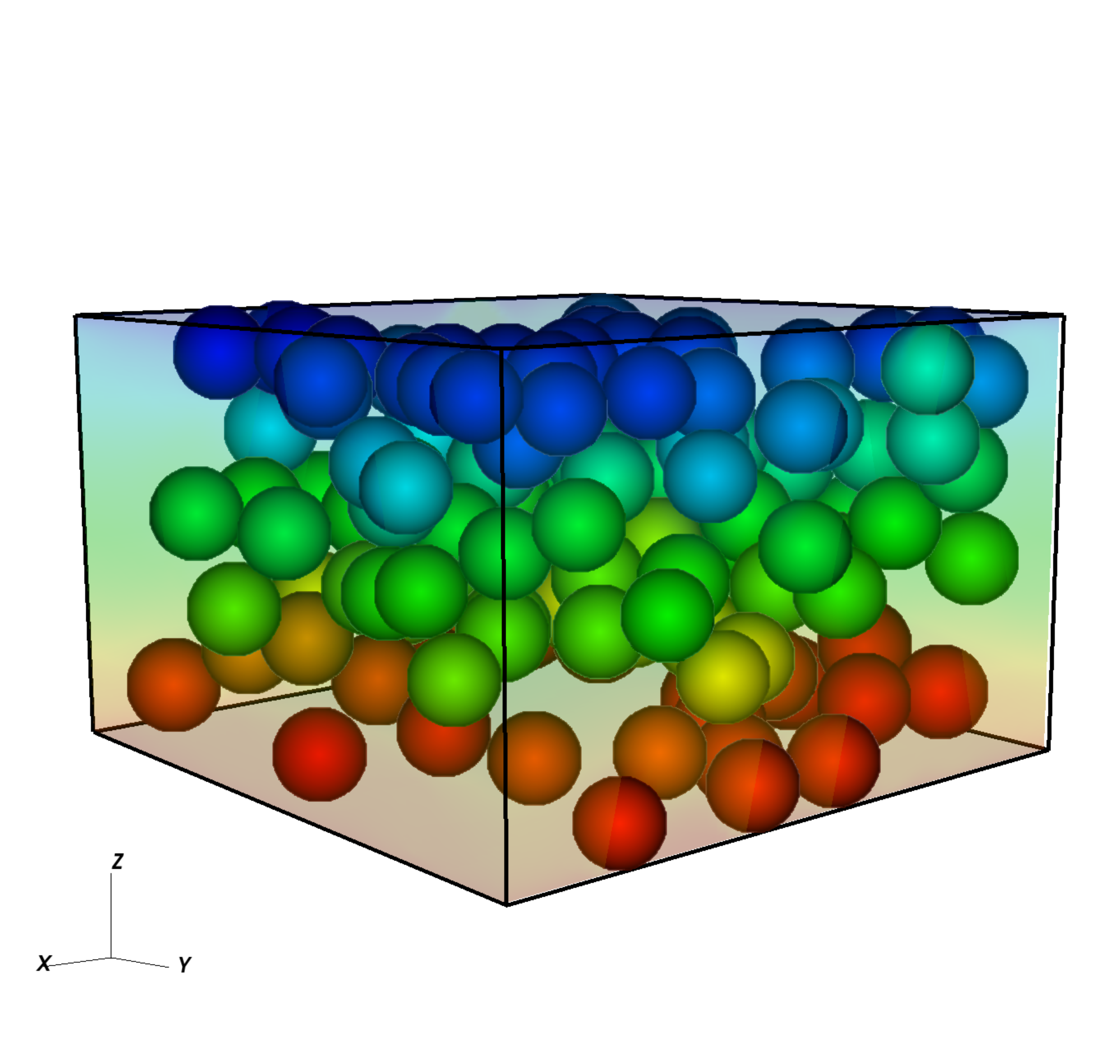}
\put(-100,55){{\large $3.2h $}}
\put(-340,155){{\large $2h$}}
\put(-280,60){{\large $3.2h $}}
\put(-350,70){{\large $wall-normal$}}
\put(-370,35){{\large $ spanwise$}}
\put(-300,35){{\large $ streamwise$}}
\caption{\label{fig:box} 
Instantaneous snapshot of the particle  arrangement for a laminar shear thinning flow, $\dot{\gamma}=1$, $Re_p=0.2004$ and $\Phi=0.21$. {The wall normal, streamwise and spanwise coordinates and particle diameters are shown at their
actual size. The particle diameter is equal to $2h/5$ with $h$ the half-channel width.}}}
\end{figure}
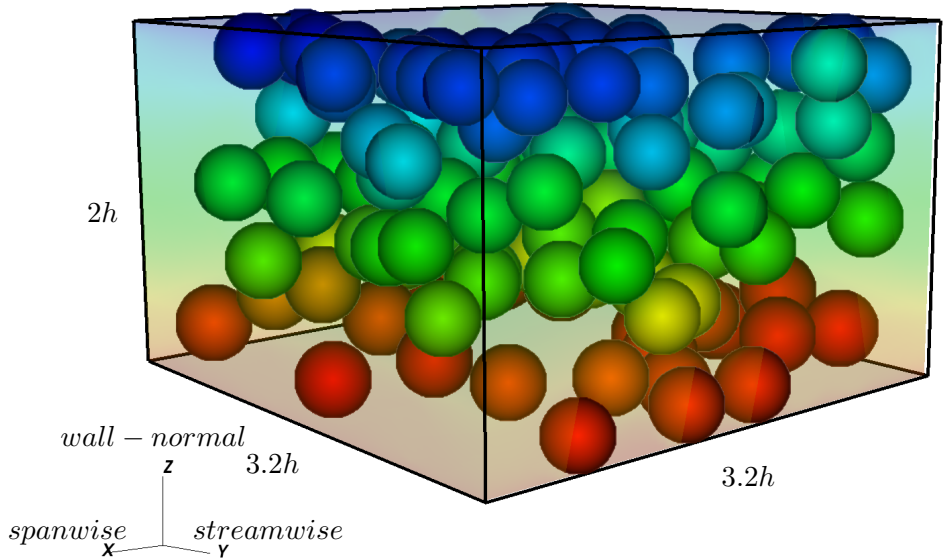
The motion of the rigid spherical particles is described by the Newton-Euler equations,
\begin{align}
m_p \frac{ d \textbf{U}_c^{p}}{dt} = \textbf{F}_p , \\ \nonumber
I_p \frac{ d \pmb{\Omega}_c^{p}}{dt} = \textbf{T}_p , 
\label{eq:PM}  
\end{align}

{where $\textbf{U}_c^p$ and $\pmb{\Omega}_c^p$ are the translational and angular velocity of the particle $p$, while $m_p$ and $I_p$ are the mass and moment-of-inertia, $2m_p a^2/5$, of a sphere with radius $a$}. $\textbf{F}_p$ and
$\textbf{T}_p$ are the net force and momentum resulting from hydrodynamic and particle-particle interactions,

\begin{align}
F_p = \oint_{\partial {V}_p}  [ -P \textbf{I} + \hat{\mu}(\textbf{u}) (\nabla \textbf{u} + \nabla \textbf{u}^T ) ] \cdot  \textbf{n} dS+ \textbf{F}_c, \\ \nonumber
T_p = \oint_{\partial {V}_p}  \textbf{r} \times \big{\{} [ -P \textbf{I} + \hat{\mu}(\textbf{u}) (\nabla \textbf{u} + \nabla \textbf{u}^T ) ] \cdot  \textbf{n}  \big{\}} dS + \textbf{T}_c. 
\label{eq:RPM}  
\end{align}
In {these equations} $\partial {V}_p$ represents the surface of the particles with outwards normal vector $\textbf{n}$ and $\textbf{I}$ the identity tensor. The radial distance from the center to the surface of each particle is 
indicated by $\textbf{r}$. The force and torque, $\textbf{F}_c$ and $\textbf{T}_c$, act on the particle as a result of particle-particle or particle-wall contacts. The no-slip and no-penetration boundary conditions on the surface
of the particles are imposed by forcing the fluid velocity at each point on the surface of the particle, $\textbf{X}$, to be equal to particle velocity at that point, $\textbf{u}(\textbf{X}) = \textbf{U}^p(\textbf{X}) = \textbf{U}_c^p + \pmb{\Omega}_c^p \times \textbf{r} $.
This condition is not imposed directly in the Immersed Boundary Method used in the current study, but instead included via the body force $\textbf{f}$ on the right-hand side of equation (\ref{eq:GNS}).

\subsubsection{Viscosity models }

Several models have been developed to capture the inelastic behaviour of some non-Newtonian fluids such as polymeric solutions. In the current work, we employ the Carreau law  to describe the behavior of shear thinning (pseudoplastic) fluids. 
This model describes the fluid viscosity  well-enough for most engineering calculations \cite[][]{Bird87}. The model assumes an isotropic viscosity proportional to some power of the shear rate $\dot{\gamma}$ \cite[][]{Morrison01},
\begin{equation}
\hat{\mu} = \frac{{\mu}_\infty}{{\mu}_0} + [ 1 - \frac{{\mu}_\infty}{{\mu}_0}][1+(\lambda \dot{\gamma})^2]^{(n-1)/2} .
\label{eq:CMV}  
\end{equation}
In the expression above $\hat{\mu}$ is the non-dimensional viscosity $\hat{\mu}=\mu/\mu_0$, where $\mu_0$ is the zero shear rate viscosity. In this work the non-dimensional viscosity takes the value $\hat{\mu}=1$ at zero shear rate and 
$\hat{\mu}= \frac{{\mu}_\infty}{{\mu}_0} =0.001$ in the limit of infinite shear rate as shown in figure \ref{fig:visc}(a). The second invariant of the strain-rate tensor $\dot{\gamma}$ is determined by the dyadic product of the strain tensor
$ \dot{\gamma}=\sqrt{2s_{ij} : s_{ij}}$, where $\textbf{s} = \frac{1}{2}(\nabla \textbf{u} + \nabla \textbf{u}^T )$ \cite[see][]{Bird87}. The power-index $n$ indicates the non-Newtonian fluid behaviour. For $n<1$ the fluid is shear thinning 
where the fluid viscosity decreases monotonically with the shear rate. The constant $\lambda$ is a dimensionless time, scaled by the flow time scale, and represents the degree of shear thinning. In the present study the power-index and
the time constant are fixed to $n = 0.3$ and $\lambda=10$. We 
report in figure \ref{fig:visc}(a), using markers, the range of shear rate and the corresponding viscosity of the carrier fluid used for the different simulations of particle-laden flows discussed later. 

For shear thickening (dilatant) fluid we employ the Power-law model,    
 \begin{equation}
\hat{\mu} = \hat{m} \dot{\gamma}^{n-1}, 
\label{eq:PLV} 
\end{equation}
which reproduces a monotonic increase of the viscosity with the local shear rate when $n > 1$. The constant $\hat{m}$ is called the consistency index and indicates the slope of the viscosity profile. For a shear thickening fluid we 
use $n=1.5$ and $\hat{m}=1$. This corresponds to $\hat{\mu}=1$ at the lowest shear rate employed in the study, see figure \ref{fig:visc}(b). For a more detailed description of the parameters appearing in the Carreau and power-low models
we refer the readers to the book by \cite{Morrison01}.

\begin{figure}
\centering{
\hspace{-0.3cm}
\includegraphics[width=0.5\linewidth]{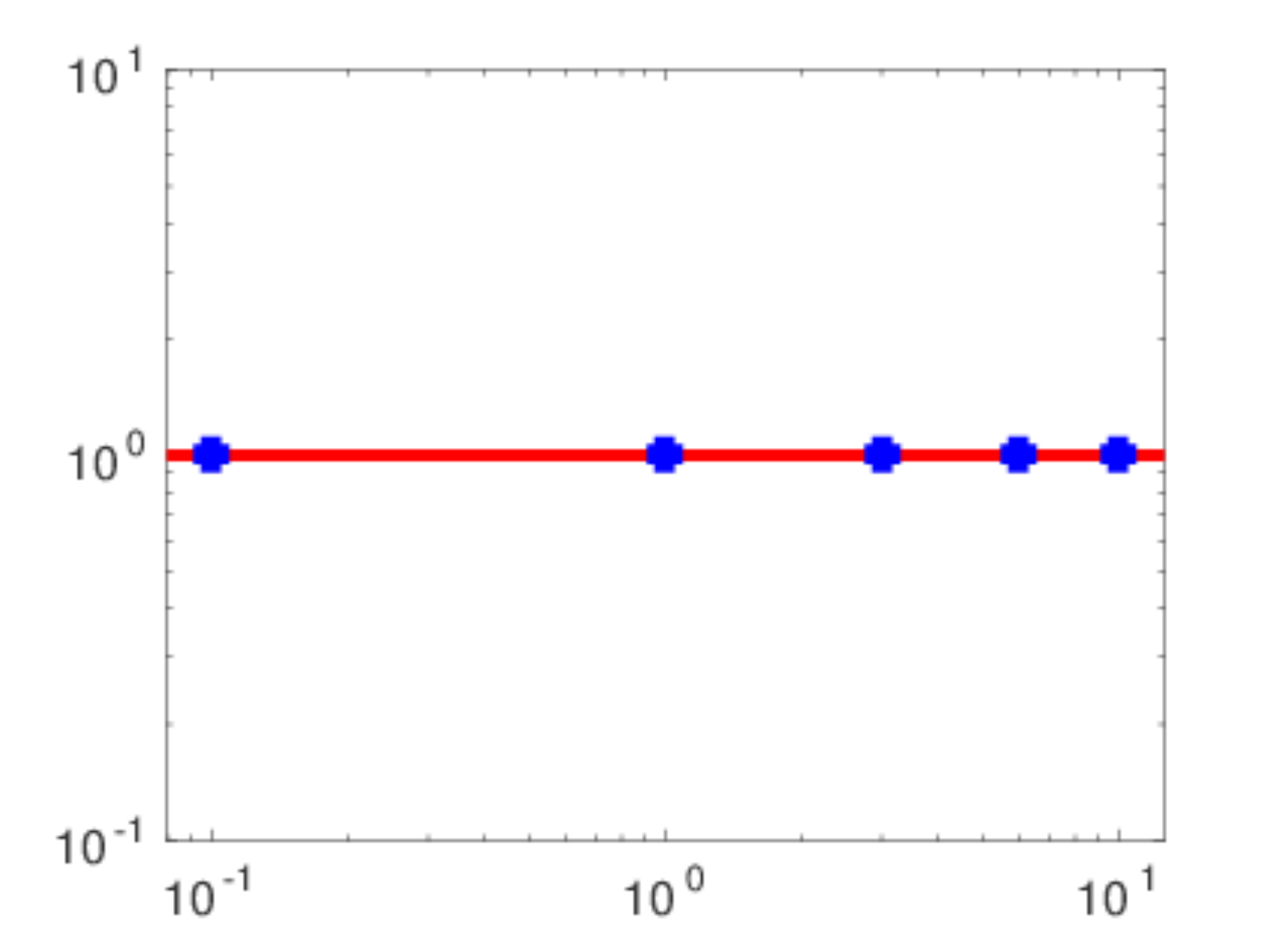}
\put(-170,140){{\large $(a)$}}
\put(-190,70){{$\hat{\mu}$}}
\put(-95,-5){{$\dot{\gamma}$}}
\includegraphics[width=0.5\linewidth]{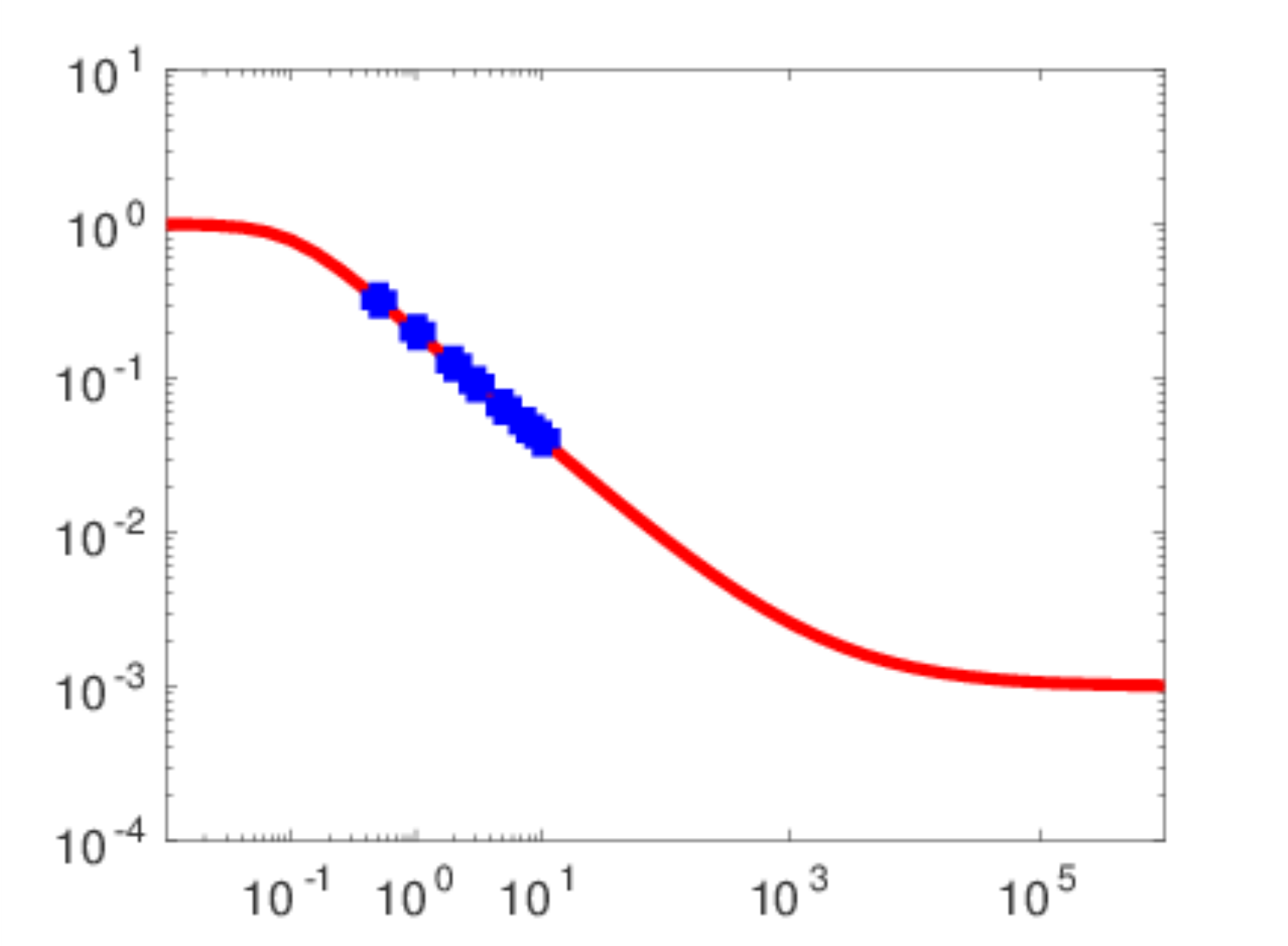}
\put(-170,140){{\large $(b)$}}
\put(-190,70){{$\hat{\mu}$}}
\put(-95,-5){{$\dot{\gamma}$}}
\\
\includegraphics[width=0.5\linewidth]{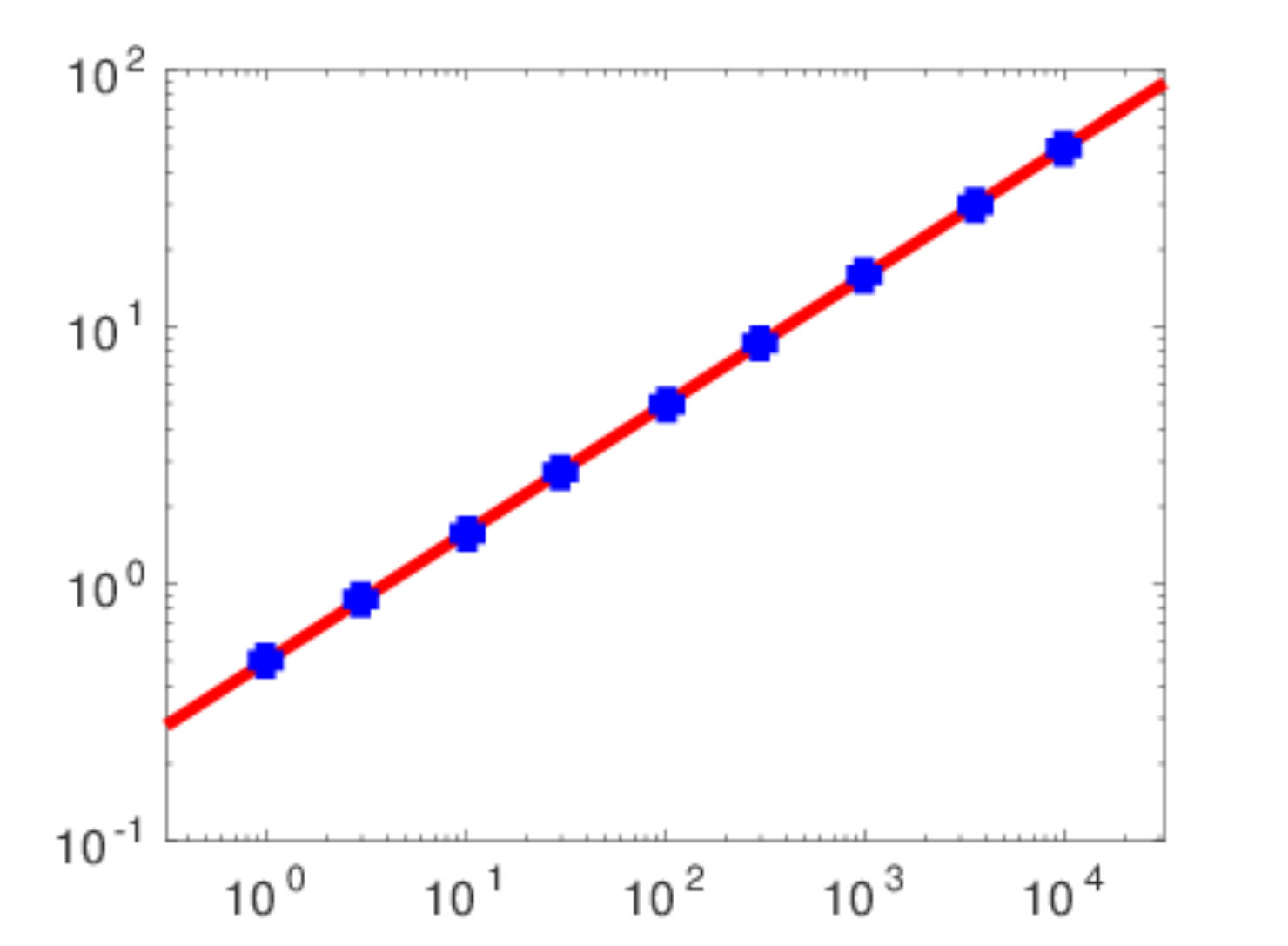}
\put(-170,140){{\large $(c)$}}
\put(-190,70){{$\hat{\mu}$}}
\put(-95,-5){{$\dot{\gamma}$}}
\caption{\label{fig:visc} 
 The non-dimensional fluid viscosity $\hat{\mu}$ versus shear rate $\dot{\gamma}$ for (a) Newtonian, (b) Carreau-law model (shear thinning) with n = 0.3 , $\lambda = 10$  and (c) Power-law model (shear thickening) for n = 1.5. The markers shows
 the shear rates and viscosities of the carrier fluid for the different cases considered here.}}
\end{figure}
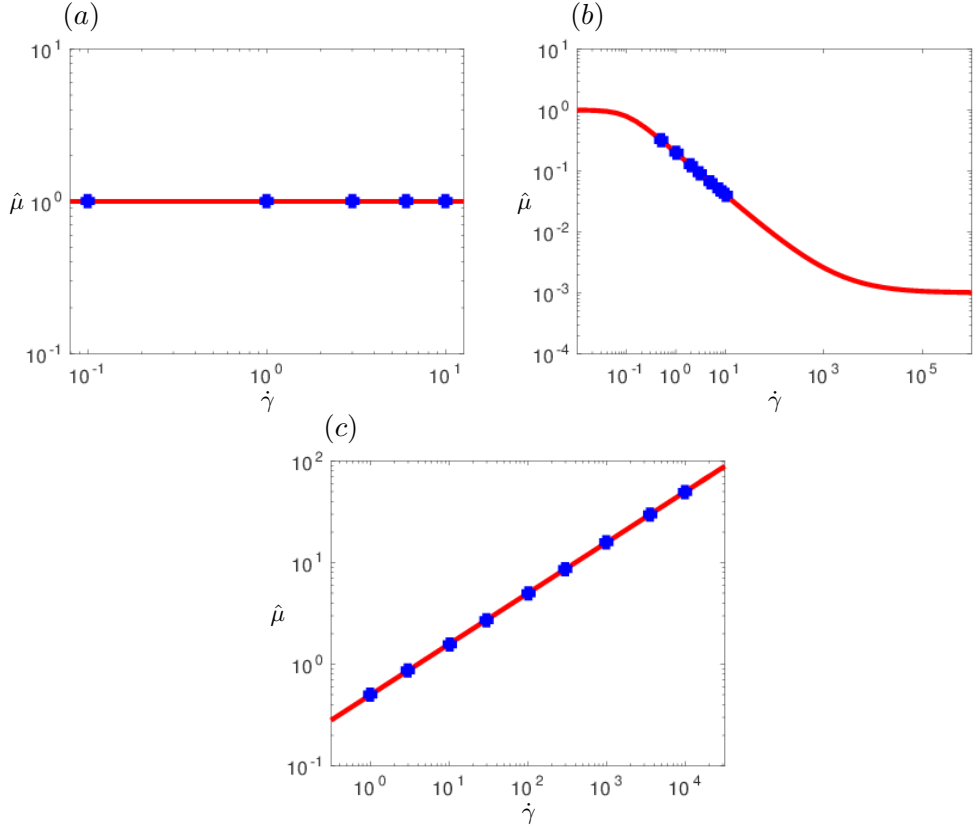

\subsection{Numerical method}

The particle motion in the flow is simulated by means of an efficient Immersed Boundary Method (IBM) coupled with a flow solver for the generalised Navier-Stokes equations. We follow the formulation  by \cite{Breugem12} developed to increase the numerical accuracy and stability for simulations of neutrally buoyant particles.  The governing equations for the fluid phase is discretized using a second order central difference scheme, 
while the Crank-Nicholson scheme is used for the time integration of the viscous term while the nonlinear term is treated explicitly using the three-step Runge-Kutta scheme. A fixed and staggered Eulerian grid is used for the fluid phase whereas a Lagrangian grid 
is attached to the surface of each particle. These two grid points communicate via the IBM forcing to satisfy the no-slip and no-penetration boundary conditions on the surface of the particles. 
 
The interactions between the particles and/or wall are taken into account using the lubrication correction and soft collision model described in detail in  \cite{Costa15}. When the gap distance between the particle-particle or particle-wall
becomes smaller than a certain threshold, a mesh dependent lubrication correction based on the asymptotic solution by \cite{Brenner61} is employed to reproduce correctly the interaction between the particles. At smaller gaps the lubrication
correction is kept constant to account for the surface roughness. Finally, a soft-sphere collision model is activated based on the relative velocity and the overlap between the two particles (particle-wall), where both the normal and tangential
components of the contact force are taken into account. The accuracy of the IBM code is examined extensively among others in the work by \cite{Breugem12,Lambert13,Costa15,Picano15}.
As regards the implementation of the viscosity model in the solver we have tested the code for the unladen channel flow of shear-dependent viscosity fluid against the analytical solution, see \cite{Nouar07}.

\subsection{Flow configuration and numerical setup} 

\begin{table}
\begin{center}
\def~{\hphantom{0}}
  \begin{tabular}{cc|c|cc|c|cccccccc}
      $n$    & $\Phi ($\%$)$        & $N_p$          & $\dot{\gamma}$  & $Re_p$   & $Re_{p,local}(\Phi ($\%$))$    & $\lambda$ & $\mu_0$ & $\mu_\infty$   \\[3pt]\hline
       $1$    & $11, 21, 31.5, 40$   & $67, 128, 193, 245$  & $0.1$   & $0.1$    &                 $-$                & $-$        & $0.05$  & $-$    \\  
      $1$    &  "  &  " & $1$     & $1$      &                 $-$                & $-$        & $0.05$  & $-$      \\  
      $1$    &   " & "  & $3$     & $3$      &                 $-$                & $-$        & $0.05$  & $-$      \\  
      $1$    &   " &  " & $6$     & $6$      &                 $-$                & $-$        & $0.05$  & $-$      \\  
      $1$    &   " &  " & $10$    & $10$     &                 $-$                & $-$        & $0.05$  & $-$    \\   
       \hline
       $0.3$  & $11, 21, 31.5, 40$   & $67, 128, 193, 245$  & $0.5$   & $0.0624$ & $ 0.0725, 0.0847, 0.0991, 0.1054$  & $10$       & $1.25$  & $0.00125$\\
      $0.3$  &    &   & $1$     & $0.2004$ & $ 0.2338, 0.2737, 0.3263, 0.3425$  & $10$       & $1.25$  & $0.00125$\\
      $0.3$  &    &   & $2$     & $0.6473$ & $ 0.7596, 0.8884, 1.0707, 1.1131$  & $10$       & $1.25$  & $0.00125$\\
      $0.3$  &    &   & $3$     & $1.2856$ & $ 1.5138, 1.7964, 2.1423, 2.2120$  & $10$       & $1.25$  & $0.00125$\\
      $0.3$  &    &   & $5$     & $3.0488$ & $ 3.6250, 4.3081, 5.0981, 5.2837$  & $10$       & $1.25$  & $0.00125$\\ 
      $0.3$  &    &   & $7.5$   & $6.0435$ & $ 7.3081, 8.6134, 10.299, 10.506$  & $10$       & $1.25$  & $0.00125$\\  
      $0.3$  &    &   & $10$    & $9.8113$ & $ 11.866, 14.192, 16.832, 16.949$  & $10$       & $1.25$  & $0.00125$\\   
      \hline        
      $1.5$  & $11, 21, 31.5, 40$   & $67, 128, 193, 245$  & $1$     & $0.1000$ & $ 0.0902, 0.0806, 0.0709, 0.0650$  & $-$        & $0.5$   & $-$     \\
      $1.5$  &    &   & $3$     & $0.1732$ & $ 0.1564, 0.1399, 0.1236, 0.1126$  & $-$        & $0.5$   & $-$     \\
      $1.5$  &    &   & $10$    & $0.3162$ & $ 0.2863, 0.2557, 0.2259, 0.2063$  & $-$        & $0.5$   & $-$    \\
      $1.5$  &    &   & $30$    & $0.5477$ & $ 0.4959, 0.4443, 0.3926, 0.3565$  & $-$        & $0.5$   & $-$    \\
      $1.5$  &    &   & $100$   & $1.0000$ & $ 0.9054, 0.8085, 0.7161, 0.6523$  & $-$        & $0.5$   & $-$   \\
      $1.5$  &    &  & $300$   & $1.7321$ & $ 1.5660, 1.3942, 1.2351, 1.1235$  & $-$        & $0.5$   & $-$   \\
      $1.5$  &    &  & $1000$  & $3.1623$ & $ 2.8479, 2.5318, 2.2468, 2.0391$  & $-$        & $0.5$   & $-$  \\
      $1.5$  &    &   & $3600$  & $6.0000$ & $ 5.3618, 4.7675, 4.1731, 3.8068$  & $-$        & $0.5$   & $-$  \\
      $1.5$  &    &  & $10000$ & $10.000$ & $ 8.8478, 7.8553, 6.9061, 6.2513$  & $-$        & $0.5$   & $-$ \\
  \end{tabular}
  \caption{Summary of the simulations performed. $Re_{p,local}(\Phi ($\%$))$ is the local particle Reynolds number  for different values of $\Phi$ and $\bar{\dot{\gamma}}_{local}$.}
  \label{tab:allcases}
  \end{center}
\end{table}

 In this study, we perform interface-resolved simulations of  suspensions of neutrally buoyant spheres in shear thinning and shear thickening fluids. The flow is driven
 by the motion of the upper and lower walls in a plane-Couette configuration. Periodic boundary conditions are imposed in the streamwise and spanwise directions. Similar to  \cite{Picano13}, we use a box size of $2h \times 3.2h \times  3.2h$ with $h$
 the  half-channel width; the number of uniform Eulerian grid points is $ 80\times 128 \times 128$ in the wall-normal, streamwise and spanwise directions. The particles have all the same radius, $a=h/5$, which corresponds to 8 Eulerian grid 
 points per particle radius, whereas 746 Lagrangian grid points are used on the surface of each particle to resolve the fluid-particle interactions. The fluid is sheared in the y-z plane by imposing a constant streamwise velocity of opposite sign 
 $V_w = \dot{\gamma} h$ at the two horizontal walls. 
 
In this work we fix the rheological parameters and vary the wall velocity (the shear rate $\dot{\gamma}$) and the particle volume fraction $\Phi$. We explore a wide range of shear rates,
$0.1\leq \dot\gamma \leq 10$, corresponding to the particle Reynolds numbers  $ 0.1 \leq Re_p= {\rho \dot\gamma a^2}/{\mu} \leq 10 $ for the Newtonian fluid, $0.5 \leq \dot\gamma \leq 10$ corresponding to $0.0624 \leq Re_p \leq 9.8113 $ 
for the shear thinning fluid and $ 1 \leq \dot\gamma \leq 10000$ corresponding to $0.1 \leq Re_p \leq 10 $ for the shear thickening fluid, see figure \ref{fig:visc}. 
Note that $\mu$ in the definition of the particle Reynolds number is, for each case, the non-Newtonian fluid viscosity in the absence of particles. The  range of particle Reynolds numbers, similar for both shear thinning and shear thickening
fluids, is obtained by adjusting, in each case, the zero shear rate viscosity. Four different particle volume fractions, $ \Phi = 0.11, 0.21, 0.315$ and $0.4 $, are examined; this corresponds to $ N_p = 67, 128, 193$ and $245$ particles in the simulation
domain.  The particles are initialised randomly in the channel with velocities equal to the local velocity of the laminar Couette profile. The parameters of the different simulations are summarised in table~\ref{tab:allcases}. Results are collected after the flow reaches a statistically steady state. We ensure the convergence by repeating the analysis using half the number of  samples and comparing the statistics with those from the entire number of samples.

\section{Results}
\label{sec:Results}
In the present study, we investigate the flow of rigid particles in a simple shear where the carrier fluid is Newtonian, shear thinning or shear thickening. We focus on the bulk properties of the suspension as well as its local behavior.
We present the distribution of particle and fluid phase velocity, particle volume fraction and local shear rate. Then we study the rheology of these suspensions and compare our results with predictions from the homogenisation theory presented 
by \cite{Chateau08}, valid for Stokesian suspensions. We therefore focus on how inertia affects the suspension behaviour.

\subsection{Flow profiles }
\label{sec:Flow profiles}
The wall-normal profiles of the local particle volume fraction  $\Phi(y)$ are computed for all the simulations listed in Table \ref{tab:allcases} by averaging the local solid volume fraction over time and in the spanwise direction.
A typical example of the results is shown in Fig.  \ref{fig:phi_wallnorm_thin}. Here the  wall-normal distribution of $\Phi(y)$ is displayed across half of the gap for  four  particle volume fractions, $\Phi =$ [0.11, 0.21, 0.315,  0.4], 
and two particle Reynolds numbers $Re_p =$ [0.1, 6 ] for Newtonian, shear thinning and shear thickening suspending fluids. Three features are evident here. 
First, particles tend to form layers due to the wall confinement; the number of
particles at the wall is larger than in the bulk. Second, the particle layering increases as the bulk solid volume fraction increases. Third, the distribution of $\Phi(y)$ changes slightly with the type of suspending fluids over the range of the particle
Reynolds number studied here (i.e., $0<Re_p \leq 10$), suggesting that the 
local particle volume fraction is mainly controlled by geometry and confinement.

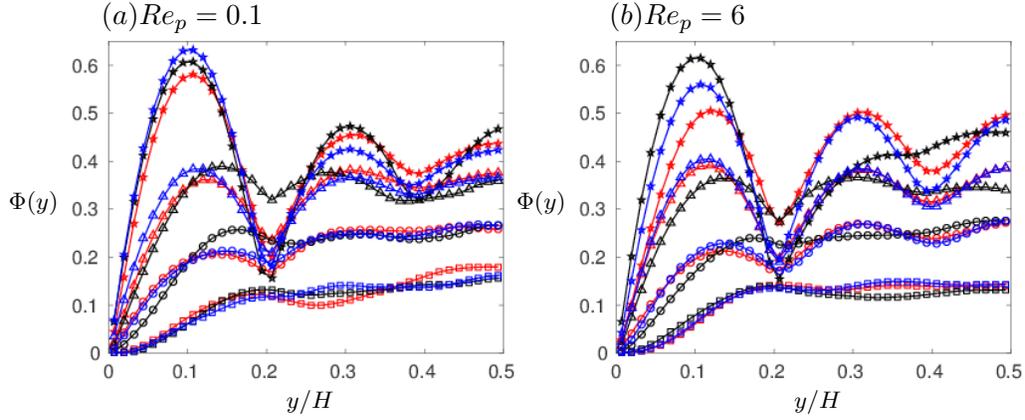
\begin{figure}
\centering{
\hspace{-0.3cm}
\includegraphics[width=0.5\linewidth]{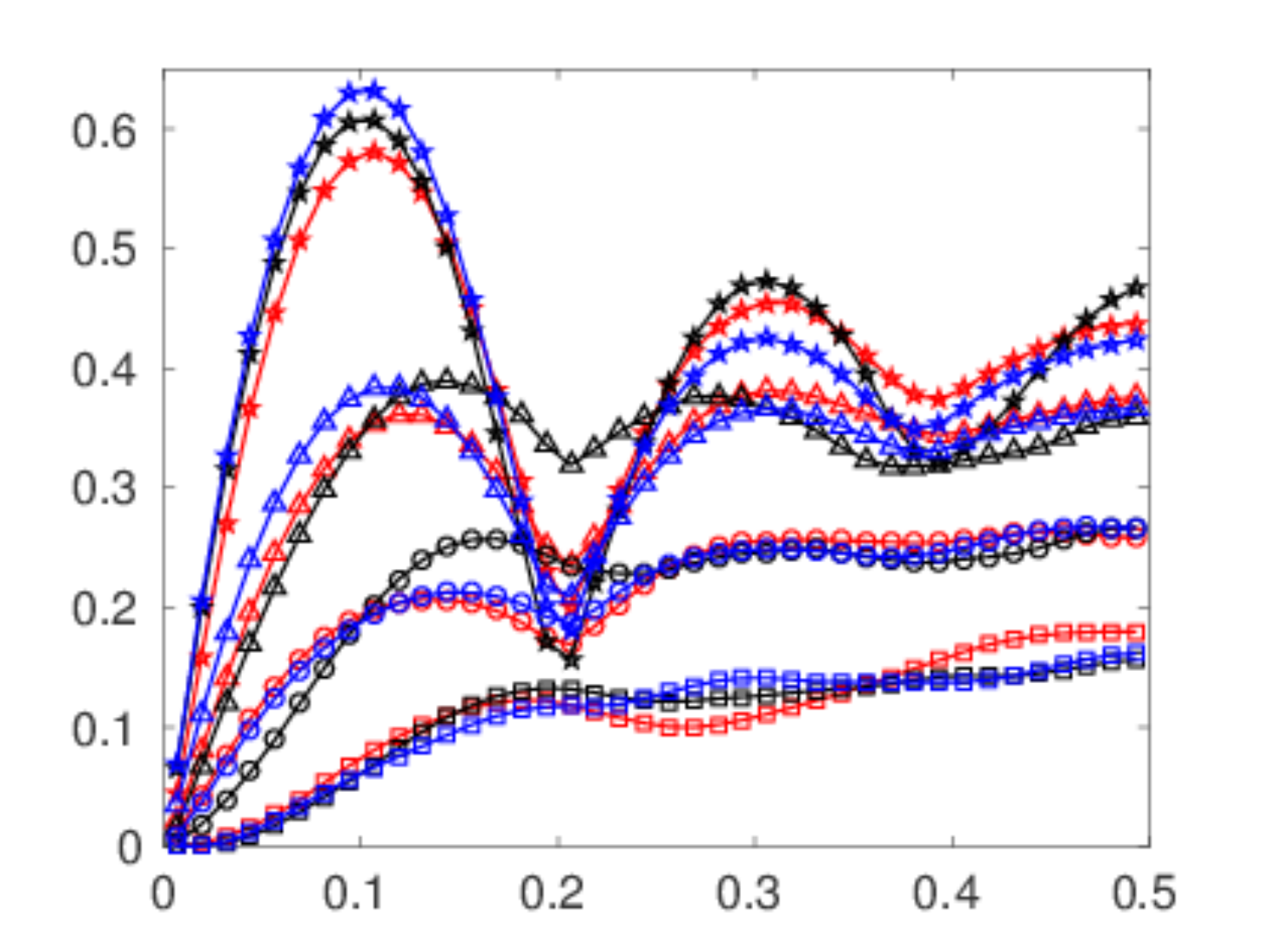}
\put(-170,140){{\large $(a) Re_p=0.1 $}}
\put(-205,70){{$\Phi(y)$}}
\put(-100,-5){{$y/H$}}
\includegraphics[width=0.5\linewidth]{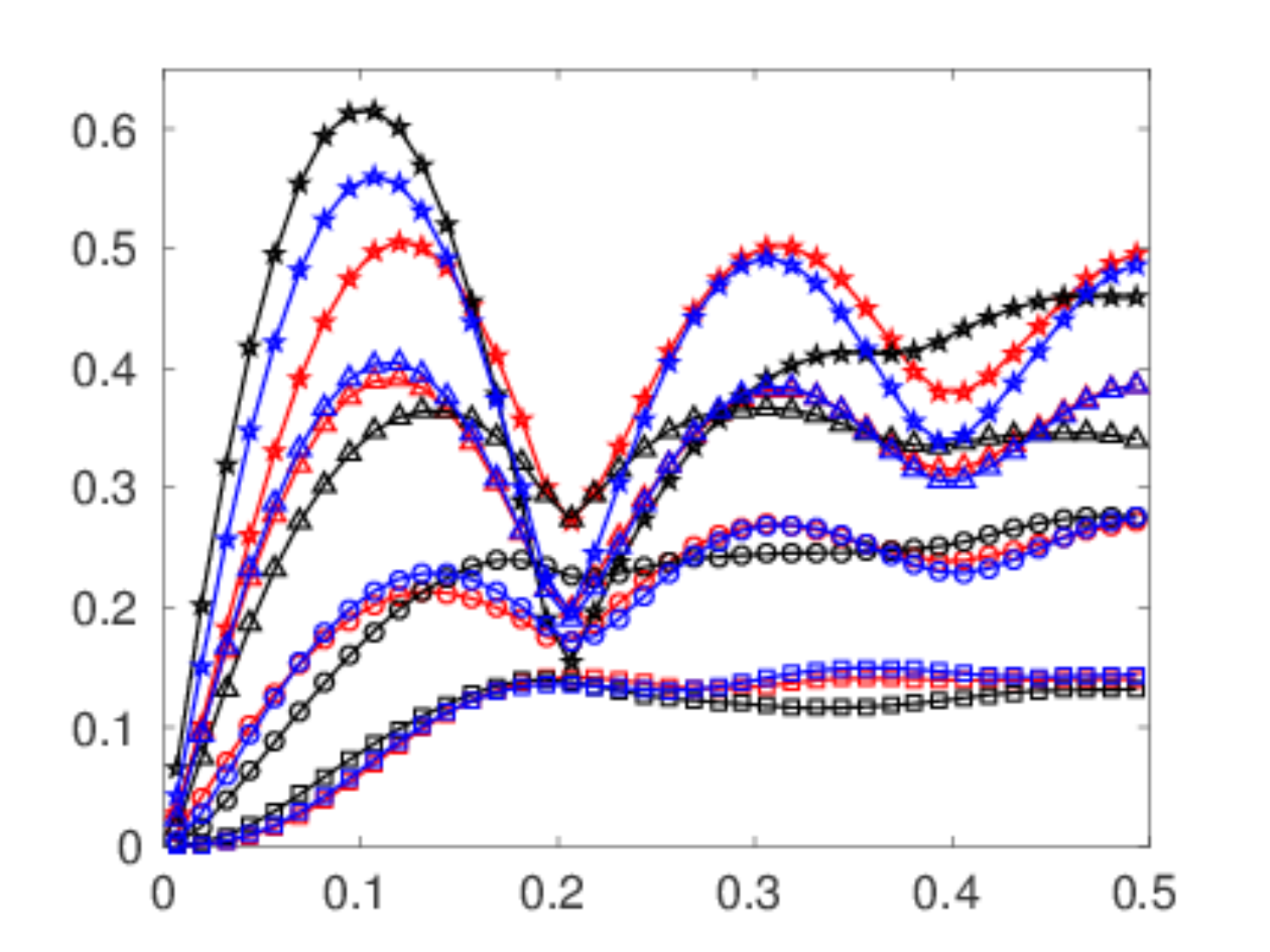}
\put(-170,140){{\large $(b) Re_p=6$}}
\put(-205,70){{$\Phi(y)$}}
\put(-100,-5){{$y/H$}}
\\
\caption{\label{fig:phi_wallnorm_thin} 
 Wall-normal profiles of the local particle volume fraction for a) $Re_p=0.1$; b) $Re_p=6$. The following colors are adopted for suspensions with different types of suspending fluids: the Newtonian suspending fluids: red color; the shear 
 thinning suspending fluid: black color and the  shear thickening suspending fluid: blue color. The following symbols are adopted for different solid volume fractions:
 $\Phi=0.11$: $\square$; $\Phi=0.21$: $\bigcirc$; $\Phi=0.315$: $\bigtriangleup$ and $\Phi=0.40$: $\star$.}}
\end{figure}   

We report the normalized mean fluid $V_f / V_w $ and particle velocity $V_p / V_w $ in Fig. \ref{fig:Velocity_wallnor_thin} for the same values of the particle Reynolds number and bulk solid volume fraction as Fig. \ref{fig:phi_wallnorm_thin}, using same symbol and colour scheme throughout the manuscript. 
The statistics of the fluid phase velocity have been computed neglecting the points occupied by the solid phase in each field (phase ensemble average). Generally, independent of the type of suspending fluid, 
the normalized mean fluid velocity decreases in the intermediate region between the wall and the centerline, $  0.02 \lesssim y/H \lesssim 0.4 $ as the particle volume 
fraction $\Phi$ and shear rate or $Re_p$ increase.  
The larger differences between the different cases are found close to the wall. The deviation of  the normalized mean fluid velocity profile from linearity is more pronounced for the shear thinning suspending fluid and less evident for
the shear thickening fluid.  This is due to the fact that the local viscosity seen by the particles in the case of generalized Newtonian fluids depends on the local shear rate.  Taking this into account results in  larger 
and smaller local particle Reynolds number for the shear thinning and shear thickening fluids, respectively (see section \ref{sec:Stokesian Rheology: homogenisation approach } for more details).  The values of the local particle Reynolds number (see (\ref{eq:Re_plocal})) are reported in Table  \ref{tab:allcases}.

\begin{figure}
\centering{
\hspace{-0.3cm}
\includegraphics[width=0.5\linewidth]{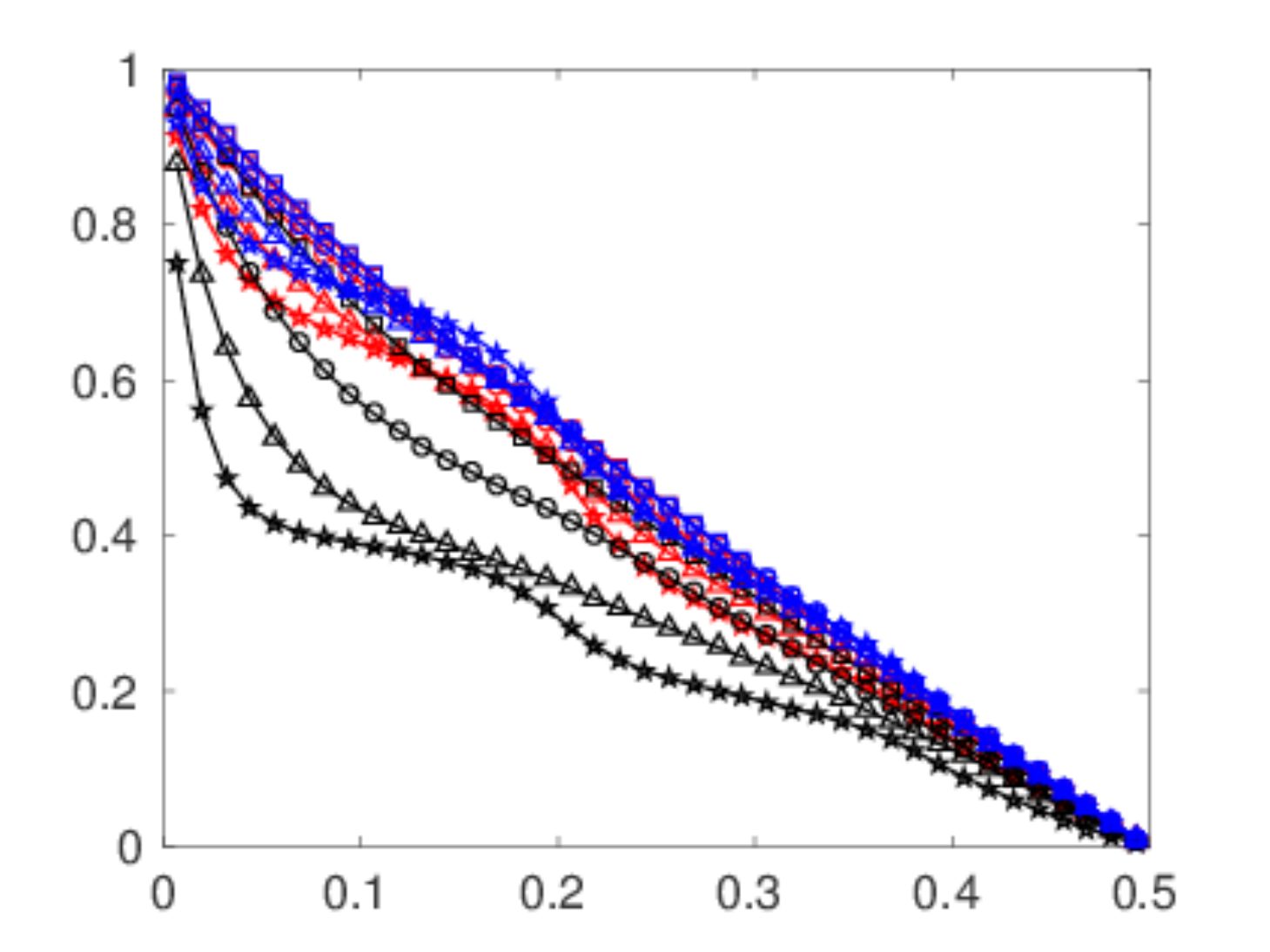}
\put(-170,140){{\large $(a) Re_p=0.1$}}
\put(-195,60){{\rotatebox{90}{$V_f / V_w $}}}
\put(-100,-5){{$y/H$}}
\includegraphics[width=0.5\linewidth]{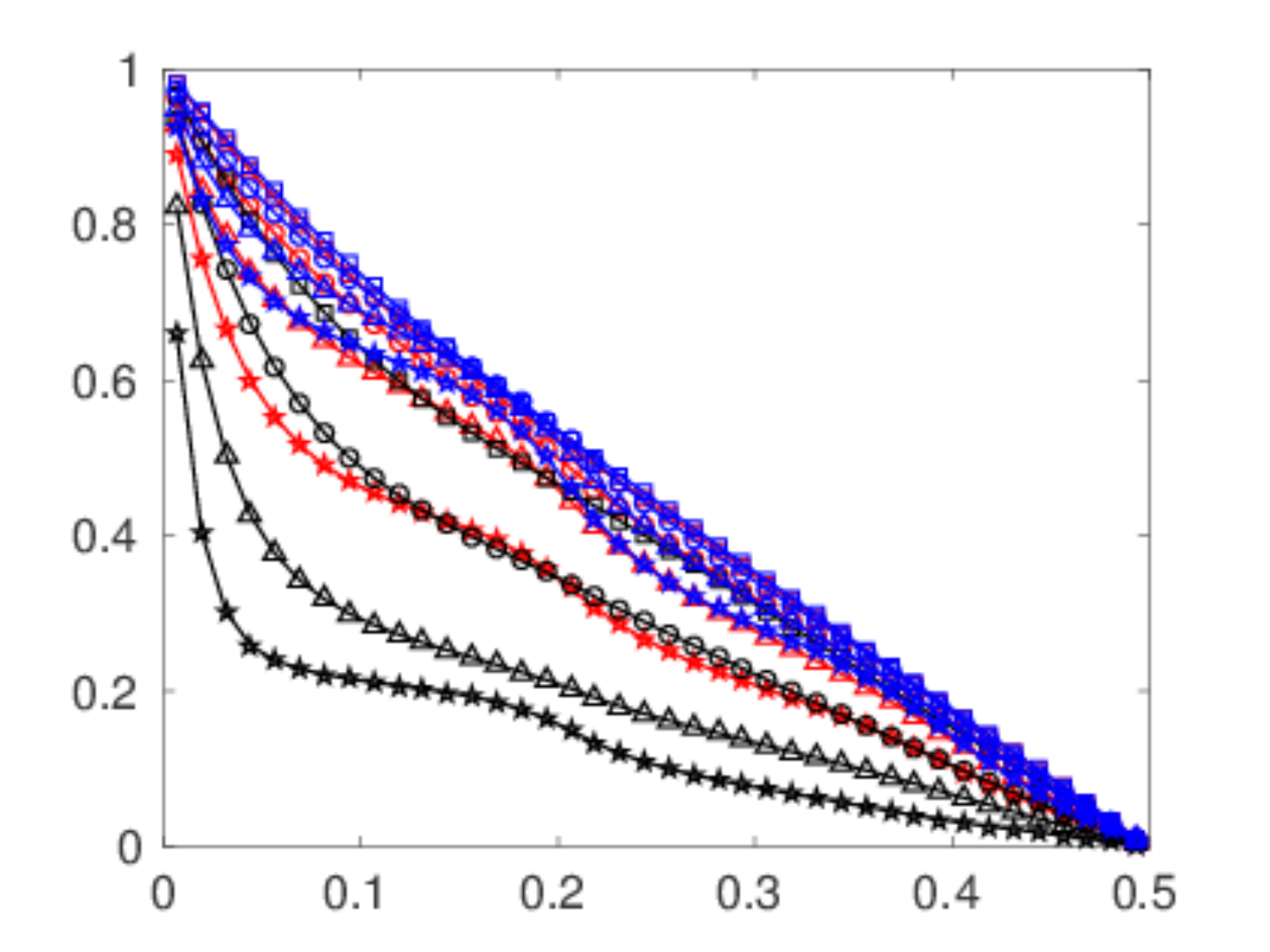}
\put(-170,140){{\large $(b)  Re_p=6$}}
\put(-195,60){{\rotatebox{90}{$V_f / V_w $}}}
\put(-100,-5){{$y/H$}}
\\
\includegraphics[width=0.5\linewidth]{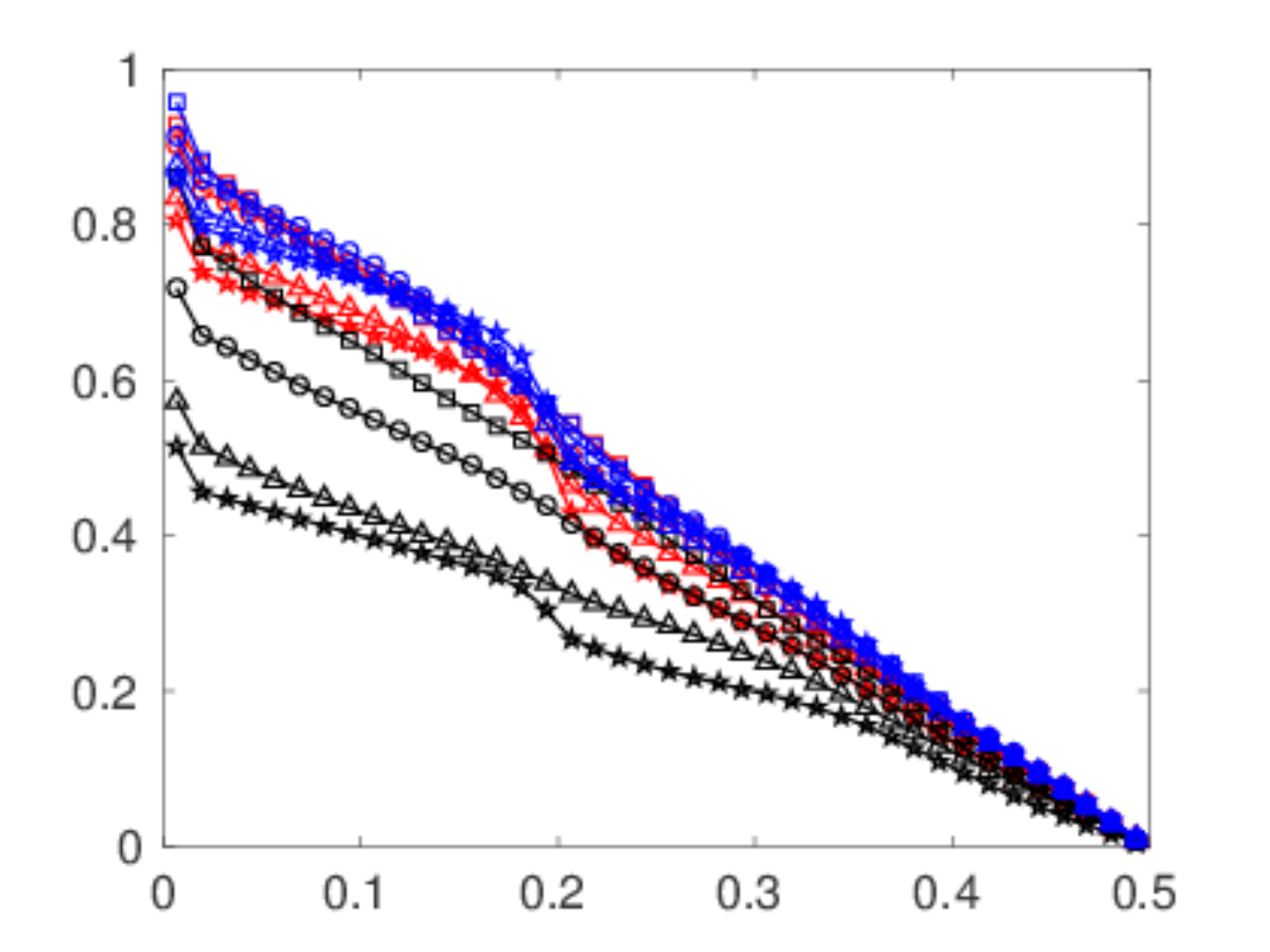}
\put(-170,140){{\large $(c) Re_p=0.1 $}}
\put(-195,60){{\rotatebox{90}{$V_p / V_w $}}}
\put(-100,-5){{$y/H$}}
\includegraphics[width=0.5\linewidth]{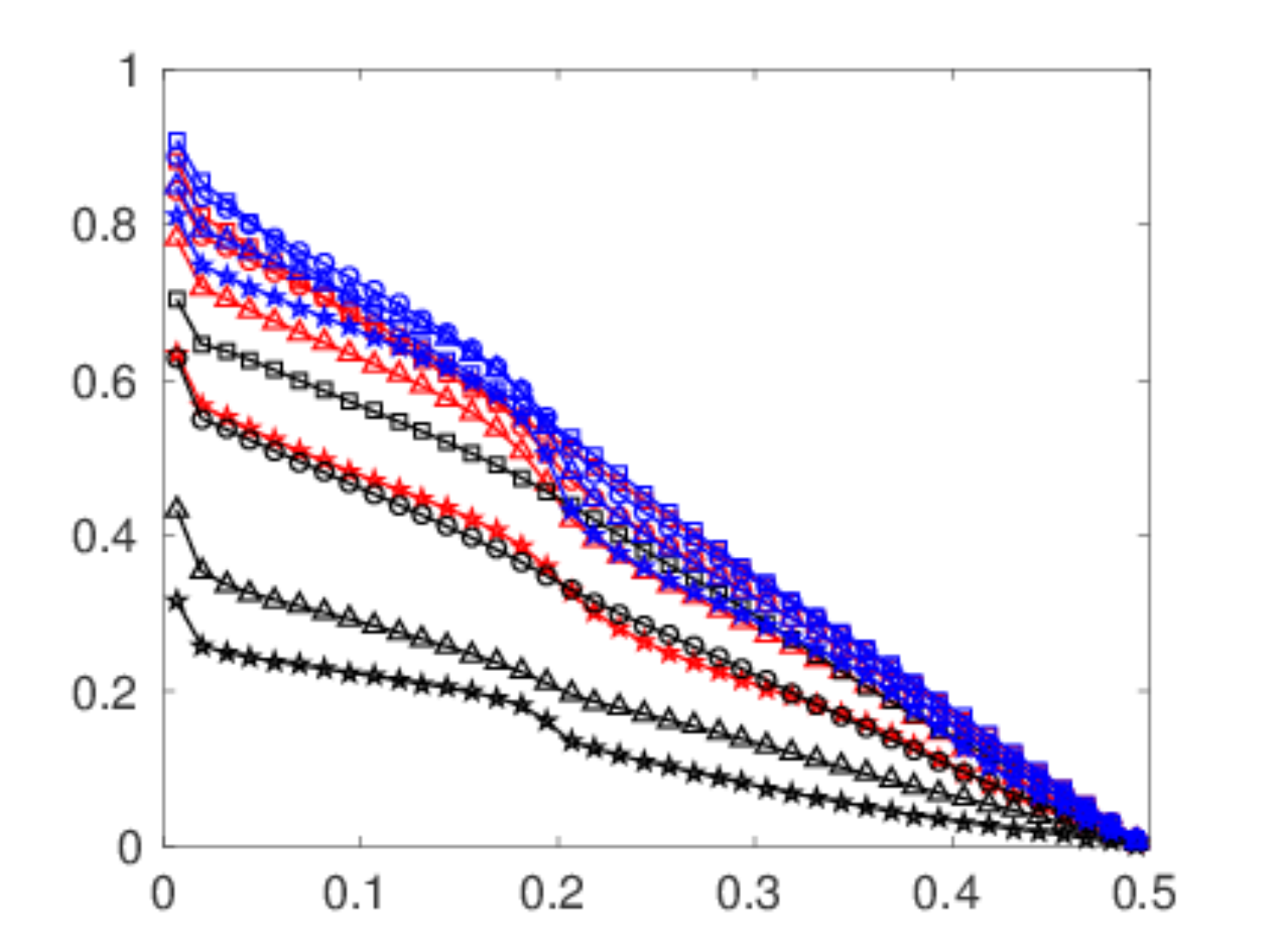}
\put(-170,140){{\large $(d)  Re_p=6 $}}
\put(-195,60){{\rotatebox{90}{$V_p / V_w $}}}
\put(-100,-5){{$y/H$}}
\\
\caption{\label{fig:Velocity_wallnor_thin} 
 Wall-normal profiles of the normalized mean fluid streamwise velocity, $V_f / V_w $: a) $Re_p=0.1$; b) $Re_p=6$.  The normalized mean particle streamwise velocity, $V_p / V_w $: c) $Re_p=0.1$; d) $Re_p=6$. 
 The following colors are adopted for suspensions with different type of suspending fluids: the Newtonian suspending fluids: red color; the shear thinning suspending fluid: black color and the  shear thickening suspending fluid: blue color.  
 The following symbols are adopted for different solid volume fractions: $\Phi=0.11$: $\square$; $\Phi=0.21$: $\bigcirc$; $\Phi=0.315$: $\bigtriangleup$ and $\Phi=0.40$: $\star$.}}\end{figure}
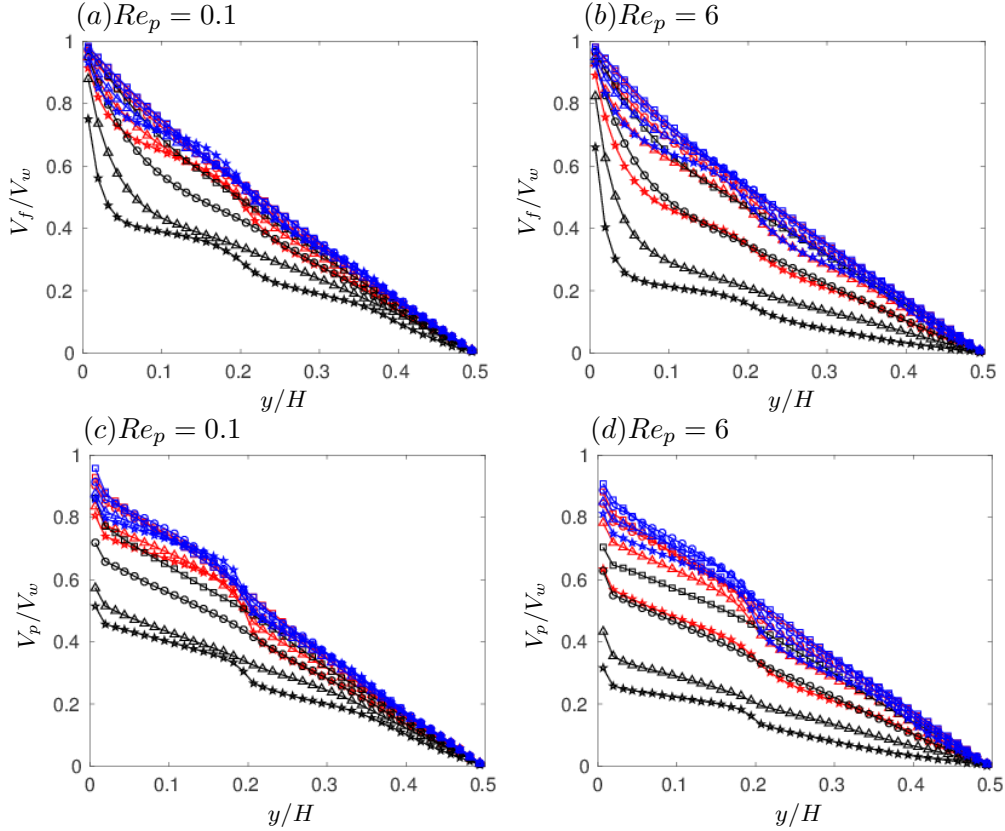

The statistics pertaining the solid phase, depicted in figure Fig.~\ref{fig:Velocity_wallnor_thin}c  and   Fig.~\ref{fig:Velocity_wallnor_thin}d, are calculated using quantities related to each individual particle, and taking  phase-ensemble average
over time and space. As for the carrier fluid, the normalized mean particle velocity decreases when increasing the volume fraction $\Phi$ and inertial effects, $Re_p$. Again this is more significant for the case of shear
thinning suspending fluid due to smaller apparent viscosity at the particle scale (or equivalently the larger local particle Reynolds number).  
Moreover, comparing  Fig. \ref{fig:Velocity_wallnor_thin}a and  Fig. \ref{fig:Velocity_wallnor_thin}c (and similarly  Fig. \ref{fig:Velocity_wallnor_thin}b and  Fig. \ref{fig:Velocity_wallnor_thin}d), we note that the slip velocity between
the solid phase and fluid phase increases close to the wall.  In particular, particles move faster close to the walls, $y/H \lesssim  0.2$, something explained by the different boundary conditions (particles can roll and slide on the wall). 
This slip is more evident for the case of a shear thinning suspending fluid, suggesting that as we increase the local particle Reynolds number, any modelling would need to take into account a boundary layer close to the wall governed by a two-phase equation of motion
instead of mixture equations \citep{Costa16,Dontsov14}.  Indeed, continuum models are more prone to failure when the slip velocity between fluid and particles is not negligible, i.e.\ the particle Stokes number is large.

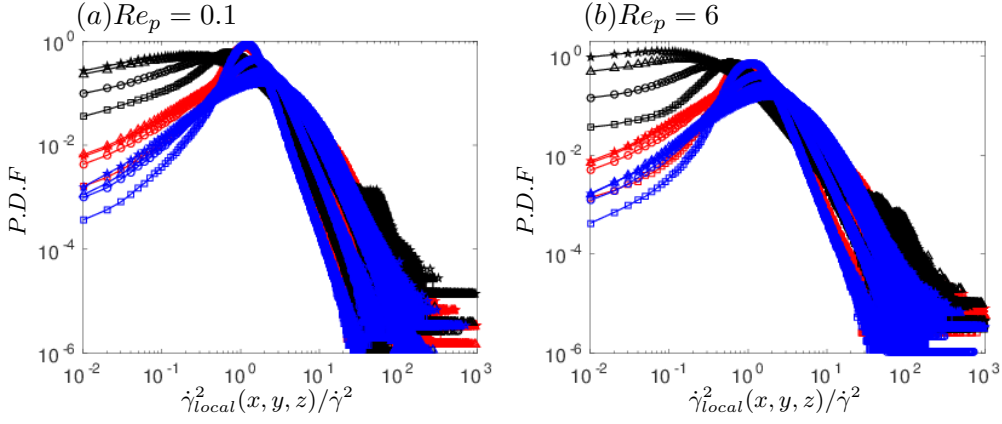
\begin{figure}
\centering{
\hspace{-0.3cm}
\includegraphics[width=0.5\linewidth]{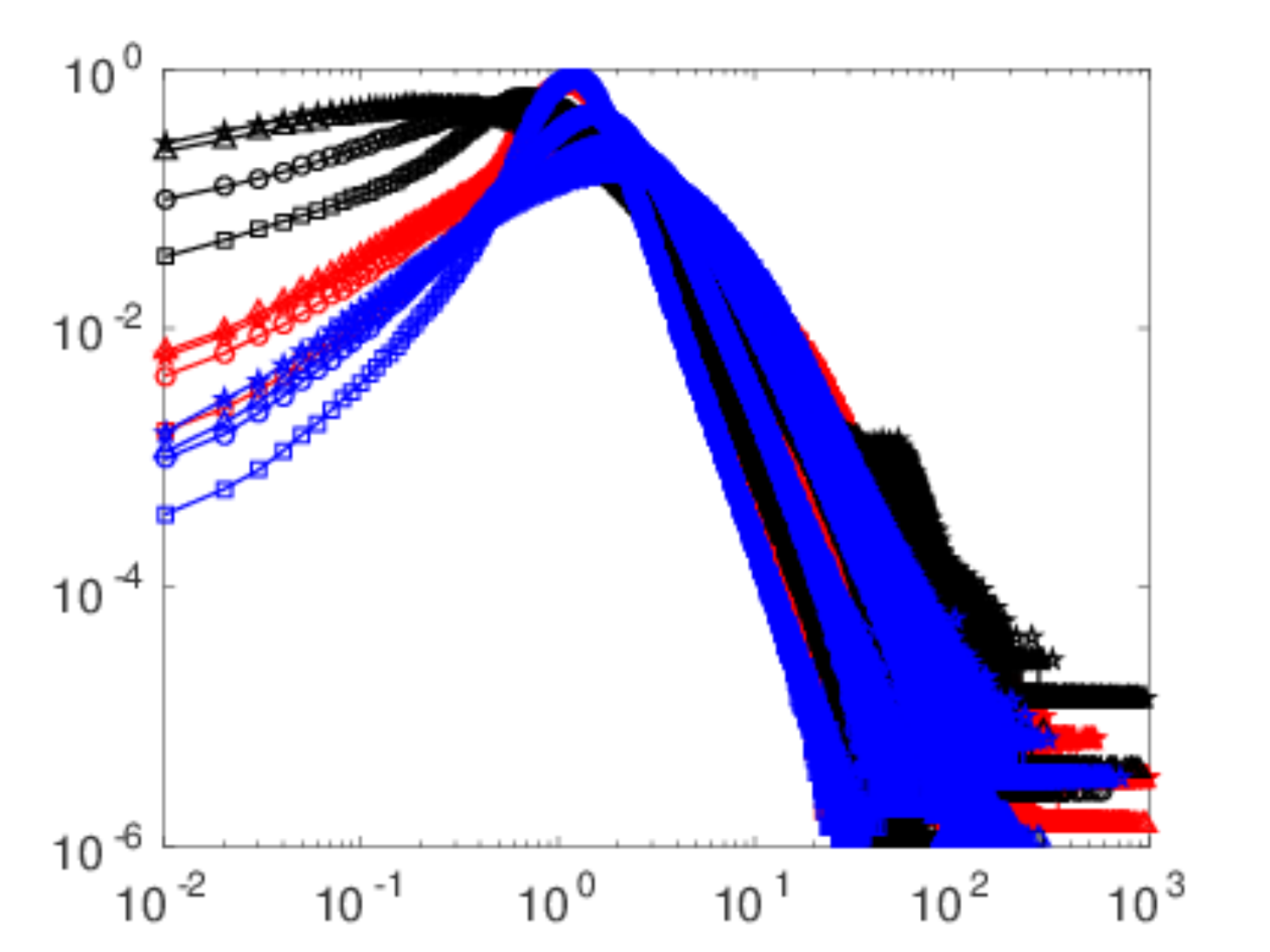}
\put(-170,140){{\large $(a) Re_p= 0.1$}}
\put(-195,60){{\rotatebox{90}{$P.D.F$}}}
\put(-130,-5){{$ \dot{\gamma}^2_{local}(x,y,z) /\dot{\gamma}^2$}}
\includegraphics[width=0.5\linewidth]{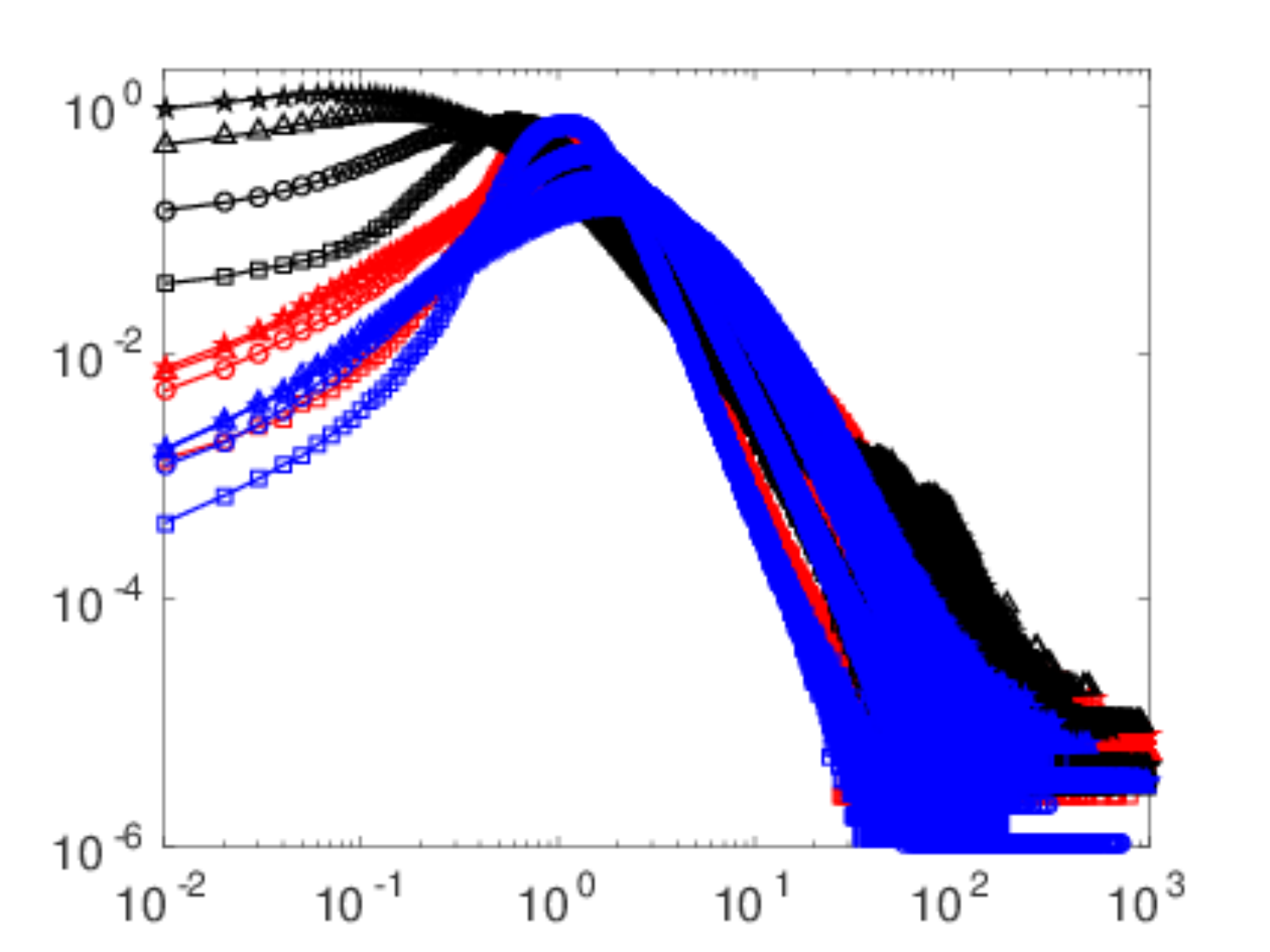}
\put(-170,140){{\large $(b) Re_p= 6$}}
\put(-195,60){{\rotatebox{90}{$P.D.F$}}}
\put(-130,-5){{$ \dot{\gamma}^2_{local}(x,y,z) /\dot{\gamma}^2$}}
\\
\caption{\label{fig:Local_shear_Newt_1} 
Profiles of the Probability Distribution Function (PDF) of the normalized local shear rate, {{$ \dot{\gamma}^2_{local}(x,y,z) /\dot{\gamma}^2$}}, for : a) $Re_p=0.1$; b) $Re_p=6$.  
Colors and symbols as in previous figures.}}
\end{figure}

\subsection{Stokesian Rheology: homogenisation approach }
\label{sec:Stokesian Rheology: homogenisation approach }
In this section, we briefly explain  the theoretical prediction  for the rheology of an inertia-less suspension of rigid spherical particles in generalized Newtonian fluids. This theoretical approach is based on
homogenisation theory and was first developed by \cite{Chateau08} assuming isotropic suspensions. Later, \cite{Ovarlez15},   \cite{Dagois-Bohy15} and \cite{Hormozi17} extended this theory to  estimate the rheology of anisotropic dilute and dense suspensions.

In order to predict the rheology of the suspensions we need to know the value of the local shear rate $\dot{\gamma}_{local}(x,y,z)$ 
for a bulk shear rate $\dot{\gamma}$. In homogenisation
theory, it is assumed that  the bulk rheology is determined by the mean value of the local shear rate, i.e., $\bar{\dot{\gamma}}_{local}$. Following  \cite{Chateau08}, we assume that viscous dissipation 
is responsible for the whole energy losses in the suspension \citep{Chateau08,DeGiuli15}. This gives the following estimate for the mean local shear rate 
 \begin{equation}
\bar{\dot{\gamma}}_{local}(\Phi) = \sqrt{\langle \dot{\gamma}^2_{local}(x,y,z) \rangle } = \dot{\gamma}  \sqrt{ \frac{G(\Phi)}{1-\Phi}}, 
\label{eq:Lshear}  
\end{equation}
 \begin{equation}
G(\Phi) = [ 1 + B \frac{\Phi}{1-\Phi/\Phi_{max}} ]^2,
\label{eq:Eilers}  
\end{equation}
where $< .>$ denotes the average of a quantity over the whole domain and 
$G(\Phi)$ is the dimensionless relative shear viscosity. The latter, a sole function of the particle volume fraction, increases monotonically with $\Phi$ and diverges when jamming occurs at the maximum packing fraction,
$\Phi_{max}$.  For non-dilute noncolloidal  suspensions various empirical fits to the experimental data have been proposed \cite[][]{Maron56,Krieger59,Quemada77,Mendoza09}. In this work we use Eilers fit \cite[][]{Stickel05}, 
where $B$ = 1.25-1.5 and the maximum packing fraction $\Phi_{max}$ = 0.58-0.64 \cite[][]{Zarraga00,Singh03,Kulkarn08,Shewan15}. The precise choice of the values for those parameters depend on particle shape, size, concentration 
and the shear rate \cite[][]{Konijn14}. The values used here are  $B$ = 1.5 and $\Phi_{max}$ = 0.61.

The apparent viscosity of the suspending fluid can be estimated via a linearization of the fluid behavior at each imposed shear rate, similar to equations (\ref{eq:CMV})  and  (\ref{eq:PLV})  for power law and Carreau suspending
fluids.  We assume  the same relative viscosity as in Newtonian suspensions can be adopted when the suspending fluid is non-Newtonian. Therefore, the shear stress of the suspension can be written as
 \begin{equation}
\tau_{ij}=G(\Phi)\mu_0\hat\mu\dot{\gamma}_{ij},
\label{eq:sus_stress}  
\end{equation}
where $\hat\mu$ is the dimensionless apparent viscosity seen by the particles; this depends on the local shear rate that we estimate  by its mean value $\bar{\dot{\gamma}}_{local}(\Phi)$.  $\hat\mu$ is of the following form 
for the Carreau and power law suspending fluids considered in this work
\begin{equation}
\hat{\mu}(\bar{\dot{\gamma}}_{local}(\Phi))= \frac{{\mu}_\infty}{{\mu}_0} + [ 1 - \frac{{\mu}_\infty}{{\mu}_0}][1+(\lambda ^2 \bar{\dot{\gamma}}_{local}^2(\Phi))]^{(n-1)/2} ,
\label{eq:CMV_local}  
\end{equation}
 \begin{equation}
\hat{\mu} (\bar{\dot{\gamma}}_{local}(\Phi))= \hat{m} \bar{\dot{\gamma}}_{local}^{n-1}(\Phi). 
\label{eq:PLV_local} 
\end{equation}
We can define the following apparent viscosity seen by the particles 
 \begin{equation}
\mu_{f}(\bar{\dot{\gamma}}_{local})= \mu_0\hat{\mu}(\bar{\dot{\gamma}}_{local}(\Phi)),\label{eq:local_mu}  
\end{equation}
and consequently a local particle Reynolds number as follows
\begin{equation}\label{eq:Re_plocal}
Re_{p,local} = \frac{{\rho}_f a^2 \dot\gamma }{ \mu_{f}(\bar{\dot{\gamma}}_{local})}.
\end{equation}
Finally, we may write the overall expression for the bulk effective viscosity of the suspension as follows
 \begin{equation}
\mu_{eff}(\Phi,\dot{\gamma}) =G(\Phi) \mu_0\hat{\mu}(\bar{\dot{\gamma}}_{local}(\Phi)).
\label{eq:GMueff}  
\end{equation}
We substitute for  the local shear rate from (\ref{eq:Lshear}) into (\ref{eq:CMV_local}) and (\ref{eq:PLV_local}), considering the definition of the suspension effective viscosity (\ref{eq:GMueff}) to obtain  the following dimensionless 
rheological formulation of the effective viscosity for shear thinning Carreau-law model) and shear thickening (power-law) fluids  
\begin{equation}
\hat{\mu}_{eff}(\Phi,\dot{\gamma}) = G(\Phi) (\frac{{\mu}_\infty}{{\mu}_0} + [ 1 - \frac{{\mu}_\infty}{{\mu}_0}][1+ (\lambda \dot{\gamma} )^2 (\frac{G(\Phi)}{1-\Phi})]^{(n-1)/2}),
\label{eq:Mueff_Thin}  
\end{equation} 
\begin{equation}
\hat{\mu}_{eff}(\Phi,\dot{\gamma}) = G(\Phi) \hat{m} (\dot{\gamma})^{n-1} ( \frac{G(\Phi)}{1-\Phi})^{\frac{n-1}{2}}.
\label{eq:Mueff_Thick}  
\end{equation}

It is noteworthy to mention that the above framework is developed for Stokesian suspensions. Also, $G(\Phi)$ includes all the information about the microstructure of the suspensions which might not be independent of the shear rate for 
the case of generalized Newtonian suspending fluids. We therefore investigate these assumptions by means of numerical simulations.  One of our  goals is to shed light on the recent experimental results in the Stokesian regime by \cite{Madraki17}
that show discrepancies with the homogenisation predictions.  Moreover, we will show how inertia affects the rheology, $G(\Phi)$, and provide a stress closure for non colloidal suspensions in the presence of inertia.

\begin{figure}
\centering{
\hspace{-0.3cm}
\includegraphics[width=0.5\linewidth]{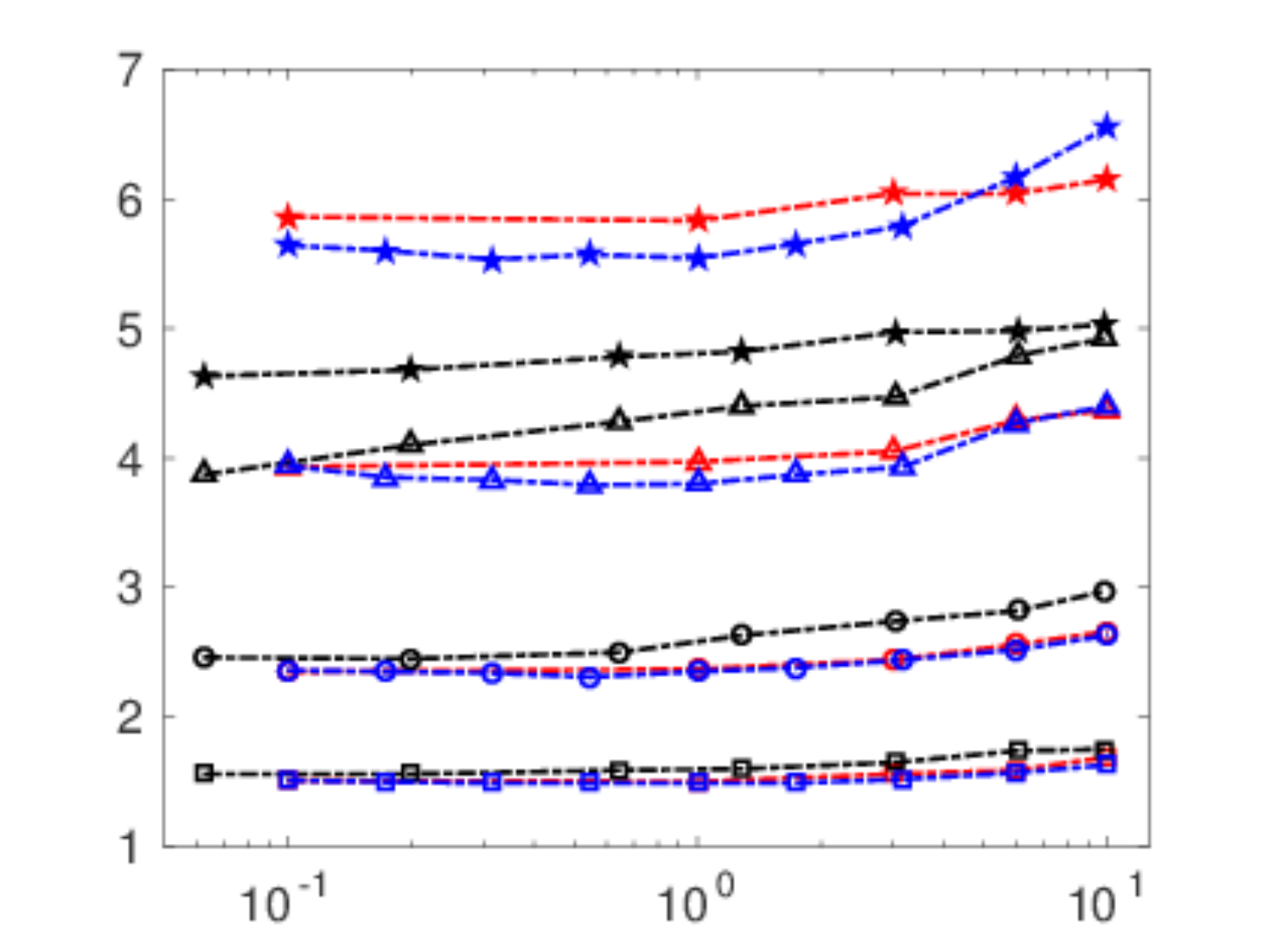}
\put(-170,140){{\large $(a) $}}
\put(-195,45){{\rotatebox{90}{$\bar{\dot{\gamma}}^2_{local}/\dot{\gamma}^2$}}}
\put(-100,-5){{$Re_p$}}
\includegraphics[width=0.5\linewidth]{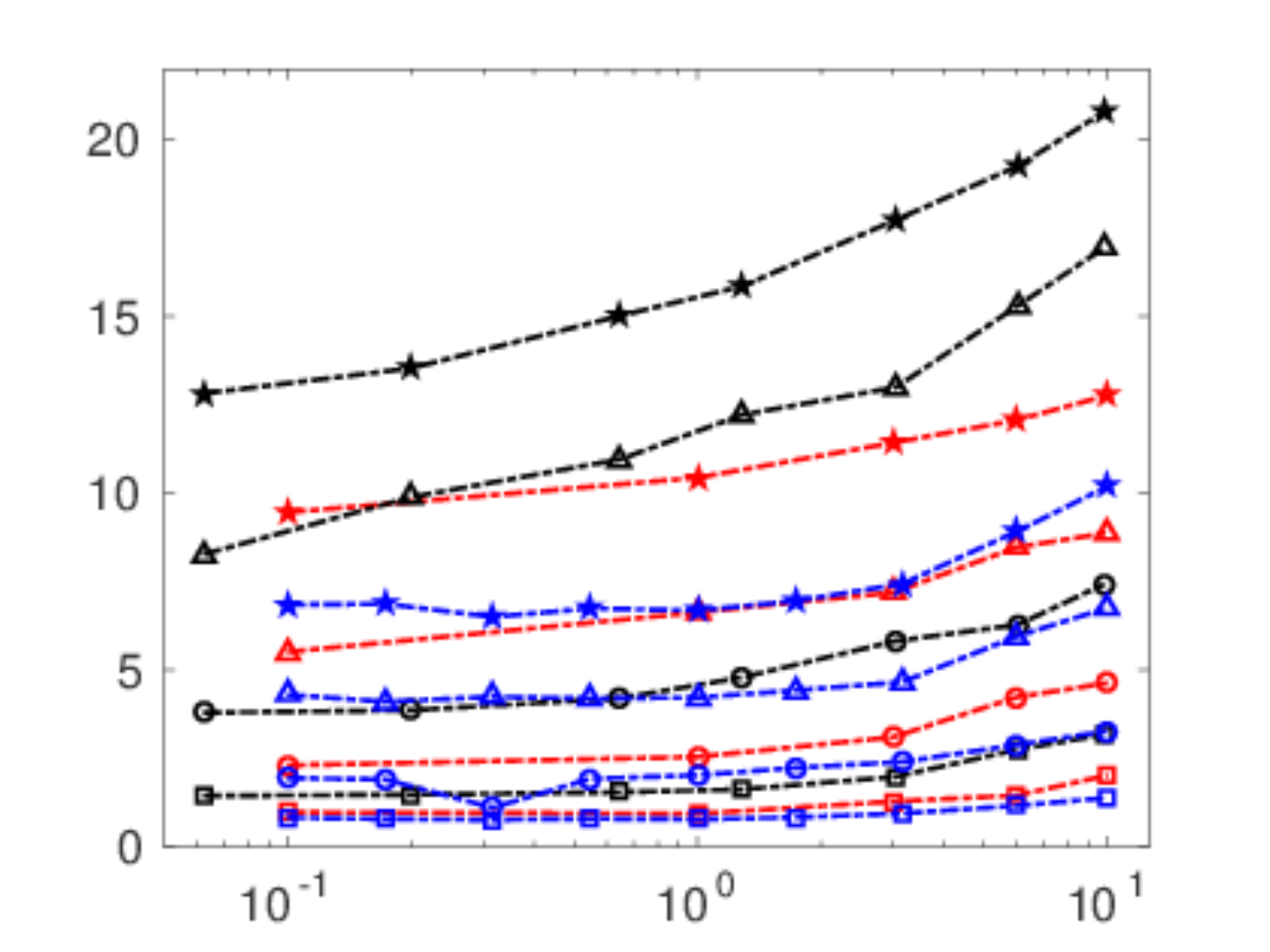}
\put(-170,140){{\large $(b) $}}
\put(-195,60){{\rotatebox{90}{$SD/\dot{\gamma}^2$}}}
\put(-100,-5){{$Re_p$}}
\caption{ Simulation results of a) {{$\bar{\dot{\gamma}}^2_{local}/\dot{\gamma}^2$}}; b) normalized standard deviation $SD/\dot{\gamma}^2$,   vs $Re_p$.  
Colors and symbols as in previous figures.}}
\label{fig:Local_shear_Newt_2}  
\end{figure}

\subsection{Local shear rate distribution} 
\label{sec:PDF}

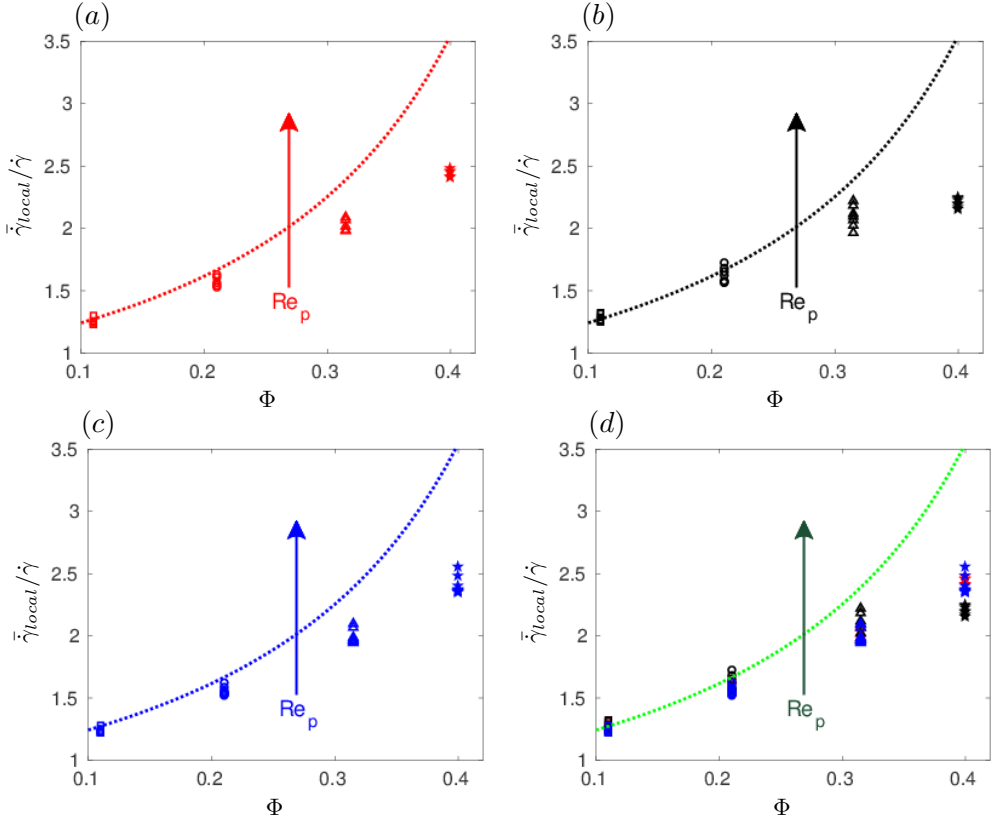
\begin{figure}
\centering{
\hspace{-0.3cm}
\includegraphics[width=0.5\linewidth]{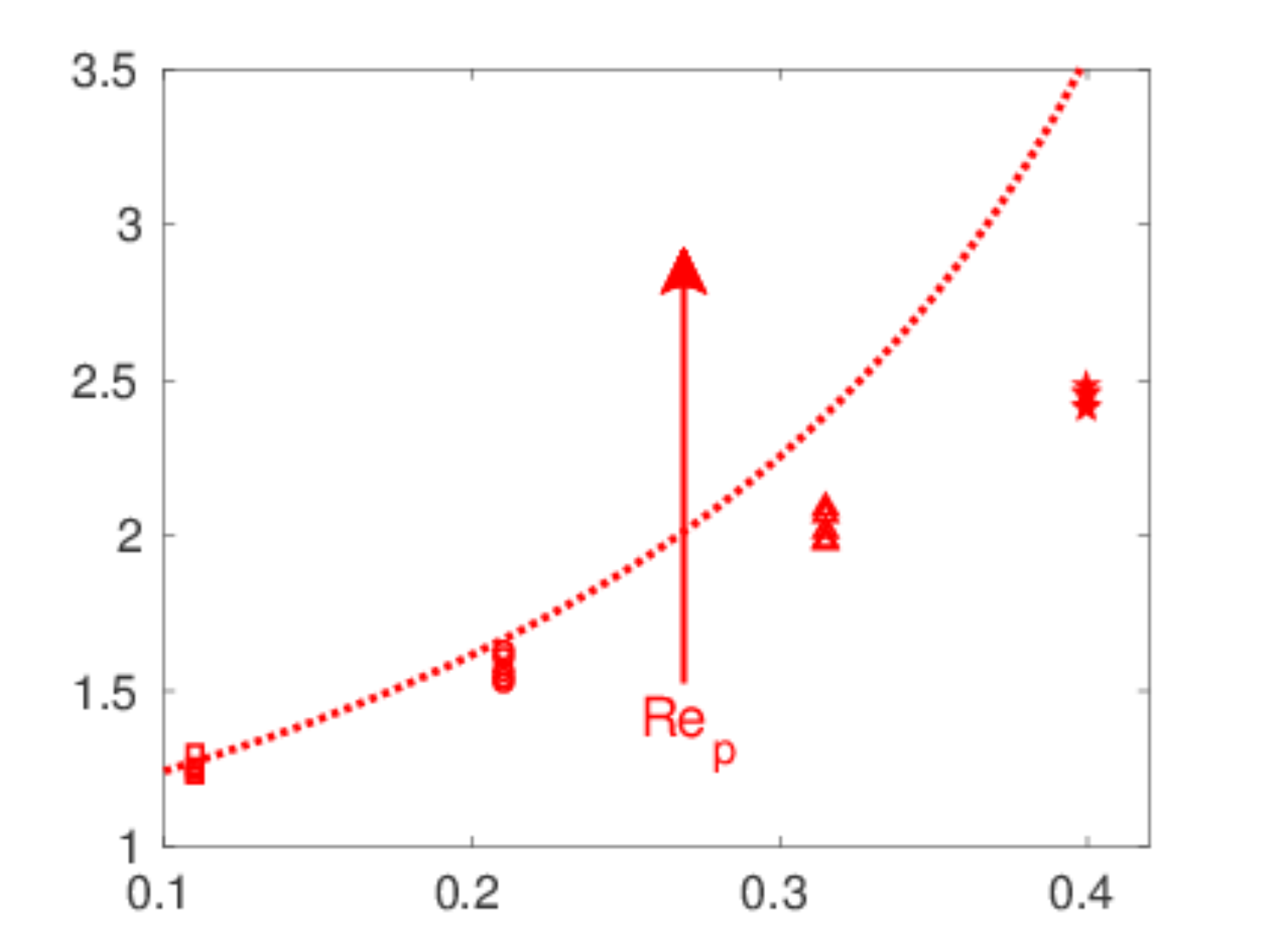}
\put(-170,140){{\large $(a) $}}
\put(-195,60){{\rotatebox{90}{$\bar{\dot{\gamma}}_{local}/\dot{\gamma}$}}}
\put(-100,-5){{$\Phi$}}
\includegraphics[width=0.5\linewidth]{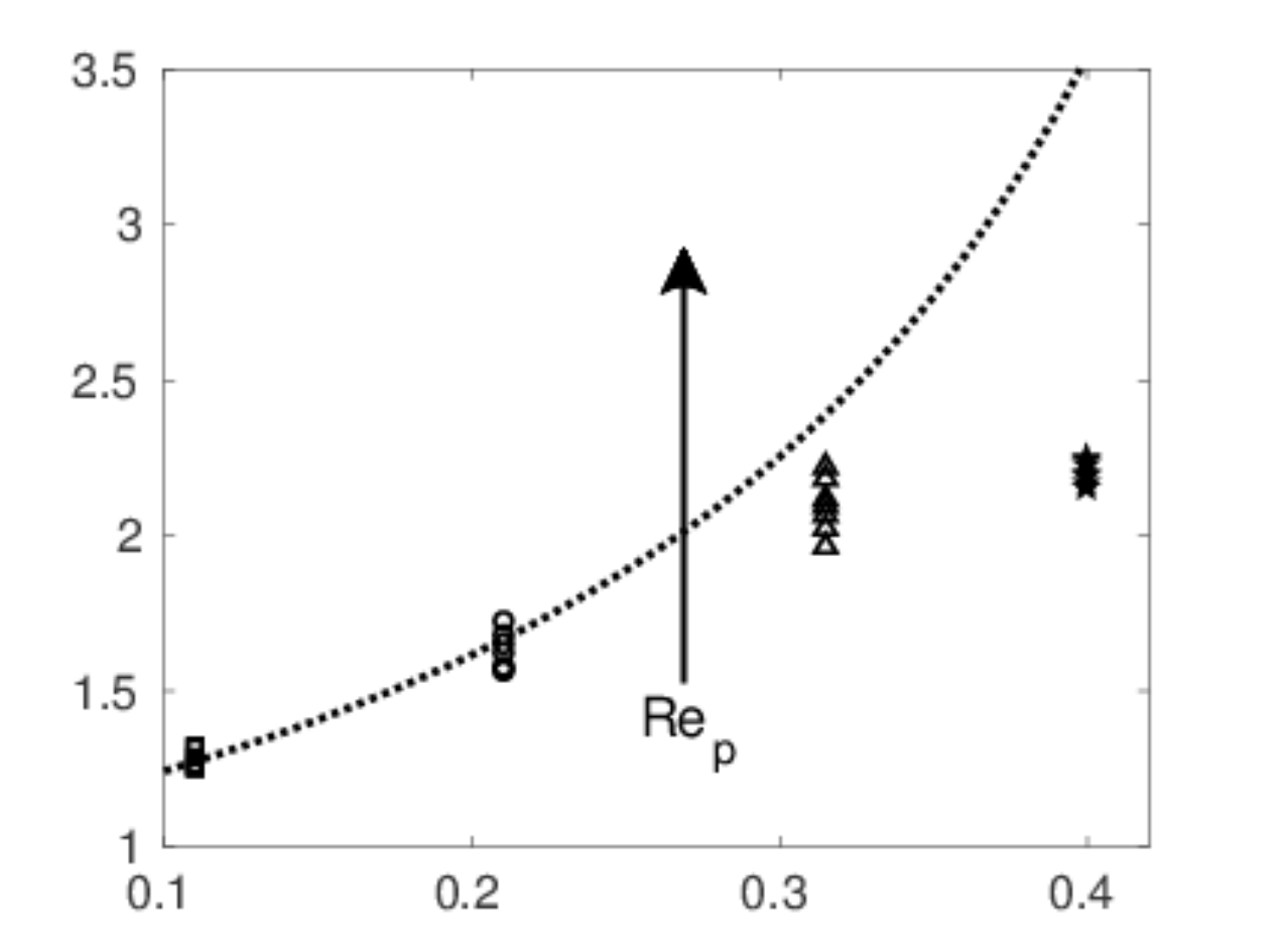}
\put(-170,140){{\large $(b) $}}
\put(-195,60){{\rotatebox{90}{$\bar{\dot{\gamma}}_{local}/\dot{\gamma}$}}}
\put(-100,-5){{$\Phi$}}
\\
\includegraphics[width=0.5\linewidth]{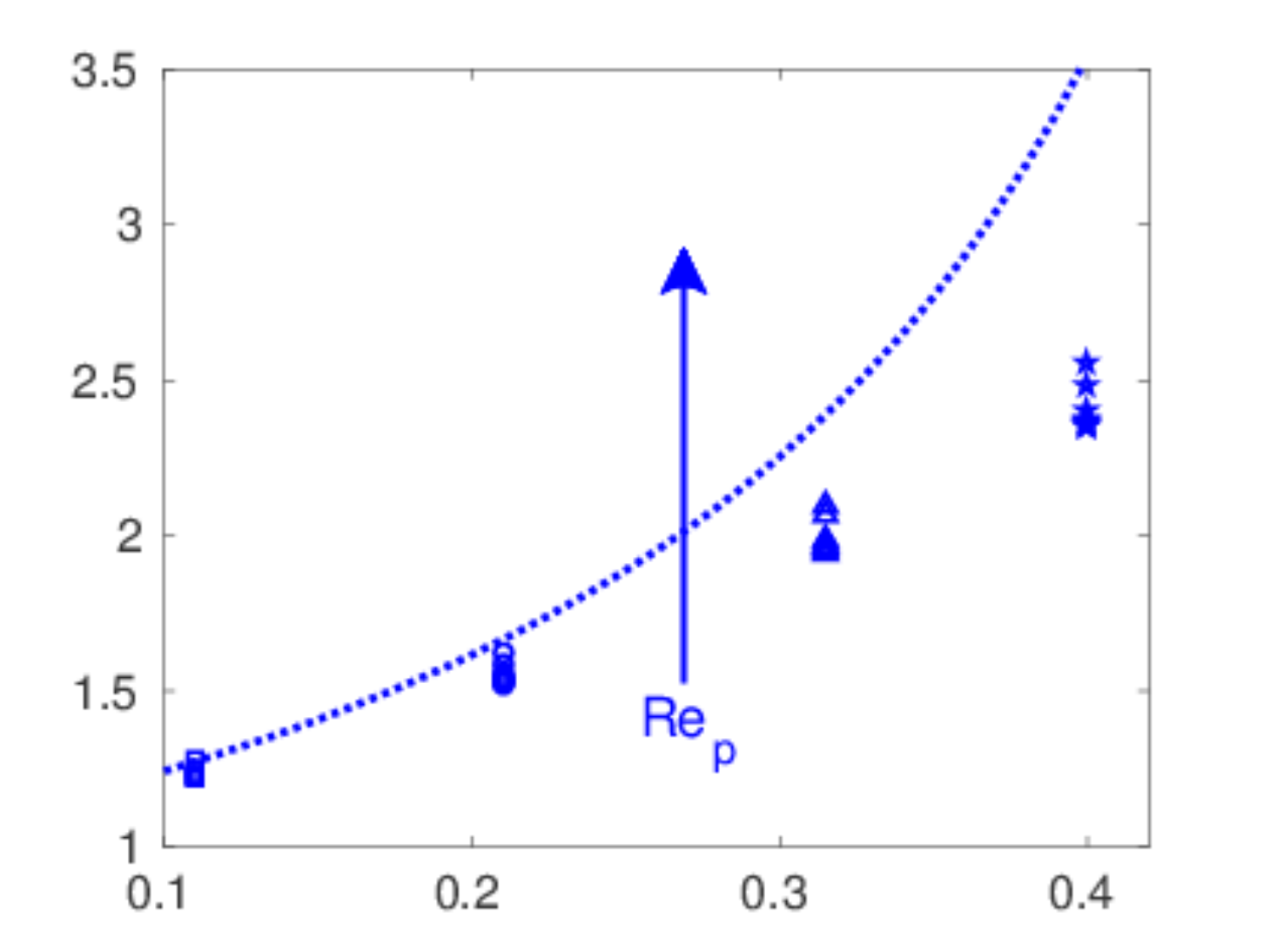}
\put(-170,140){{\large $(c) $}}
\put(-195,60){{\rotatebox{90}{$\bar{\dot{\gamma}}_{local}/\dot{\gamma}$}}}
\put(-100,-5){{$\Phi$}}
\includegraphics[width=0.5\linewidth]{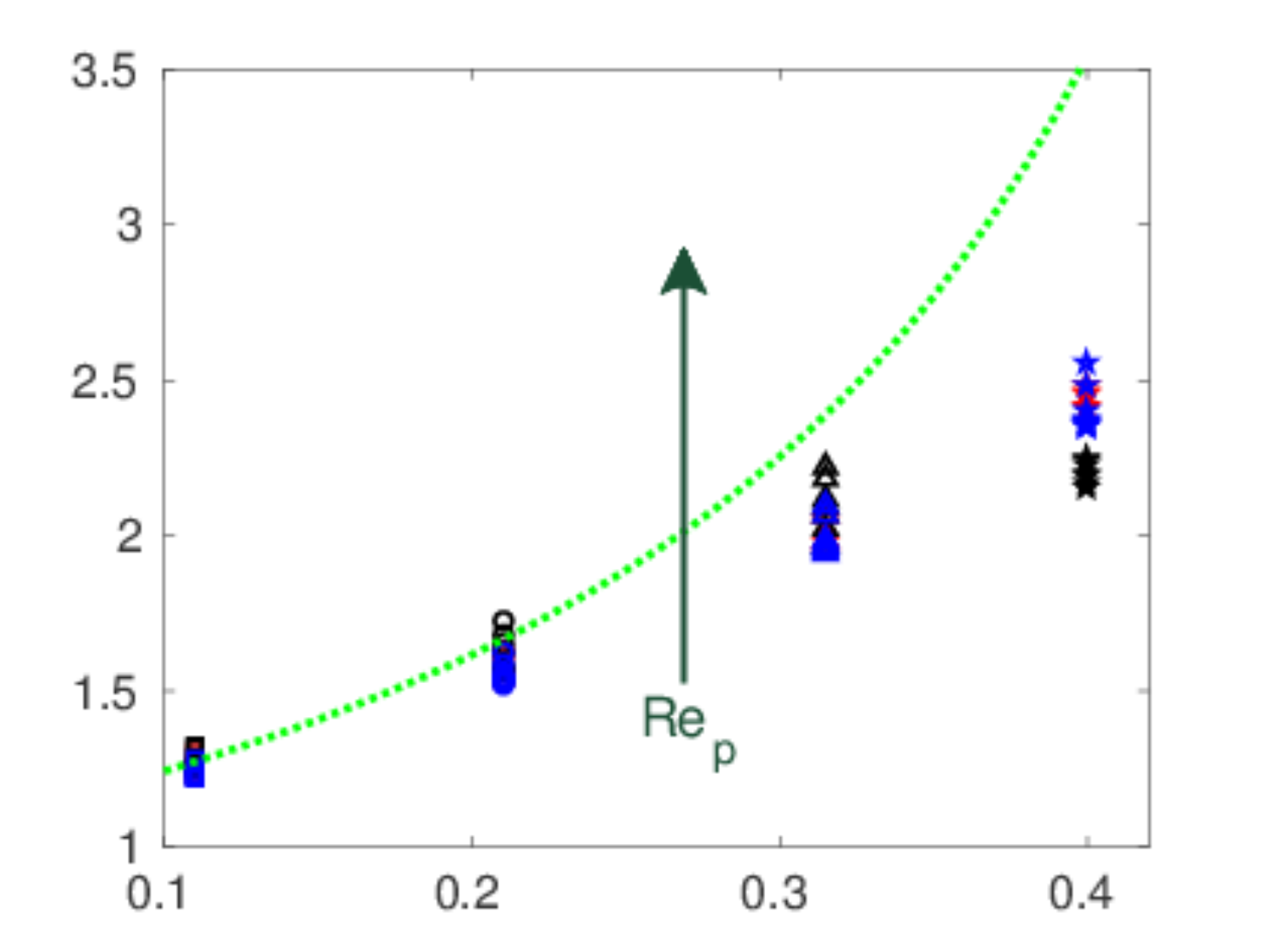}
\put(-170,140){{\large $(d) $}}
\put(-195,60){{\rotatebox{90}{$\bar{\dot{\gamma}}_{local}/\dot{\gamma}$}}}
\put(-100,-5){{$\Phi$}}
\\
\caption{\label{fig:Normalized_local_shear_all} 
Profiles of the average local shear rate (symbols for simulations, dashed lines for homogenisation theory), versus the particle volume fraction  $\Phi$ for $0.1\leq Re_p\leq10$. 
Colors and symbols as in previous figures.}}
\end{figure}

We  consider  the local shear rate distribution, $\dot{\gamma}_{local}(x,y,z)$, for the flow cases under investigation, see Table \ref{tab:allcases}, and compute the different statistical moments.  Fig. \ref{fig:Local_shear_Newt_1} depicts the Probability Density Function (PDF) of $\langle \dot{\gamma}^2_{local}(x,y,z) \rangle/\dot{\gamma}^2$  for four  particle volume fractions $\Phi =$ [0.11, 0.21, 0.315,  0.4] and two particle  Reynolds
numbers $Re_p =$ [0.1, 6 ] in the cases of Newtonian, shear thinning and shear thickening suspending fluids. The local shear rate is computed using the central finite difference scheme only at the grid points outside the particles when the 
neighboring points are also located in the fluid region. As shown in the figure, the Newtonian and shear thickening suspending fluids qualitatively have similar distributions of local shear rate, as expected from the profiles of the fluid phase
velocity shown in Fig.~\ref{fig:Velocity_wallnor_thin}a and  Fig.~\ref{fig:Velocity_wallnor_thin}b.

The normalized  mean square local shear rate $\bar{\dot{\gamma}}^2_{local}/\dot{\gamma}^2$ and its associated normalized standard deviation (SD) $SD/\dot{\gamma}^2$  are reported  in Fig. \ref{fig:Local_shear_Newt_2} for all of the simulations, where
$\bar{\dot{\gamma}}^2_{local} = \langle \dot{\gamma}^2_{local}(x,y,z) \rangle $. 
For all types of suspending fluids the mean local shear rate increases significantly with the solid volume fraction, and only marginally with $Re_p$. Moreover, the increment of the normalized standard deviation with both $\Phi$ and $Re_p$ is more than that of  the mean local shear rate, in other words the spectrum of the local shear rate distribution changes more than what the
variations of the mean values alone may suggest.
Such a distribution of the mean local shear rate, quantified by its standard deviation $SD$,  suggests that jamming and DST of a shear thickening  suspending fluid laden with noncolloidal particles cannot be understood only via the mean value of 
the local shear rate, but extreme, possibly rare, large values of the 
local shear rate should also be considered. 
This is in agreement with the recent observations by \cite{Madraki17}.  
It is also noteworthy to mention that the variation of the normalized standard deviation for the case of a Newtonian suspending fluid is smaller than that of a shear thinning  
and larger than that of a shear thickening suspending fluid. This difference is mainly due to the inertial effects (i.e., local particle Reynolds number, see Table \ref{tab:allcases}) and it is not related to the particle layering observed in Fig. \ref{fig:phi_wallnorm_thin}, as this is similar for all suspending fluids.

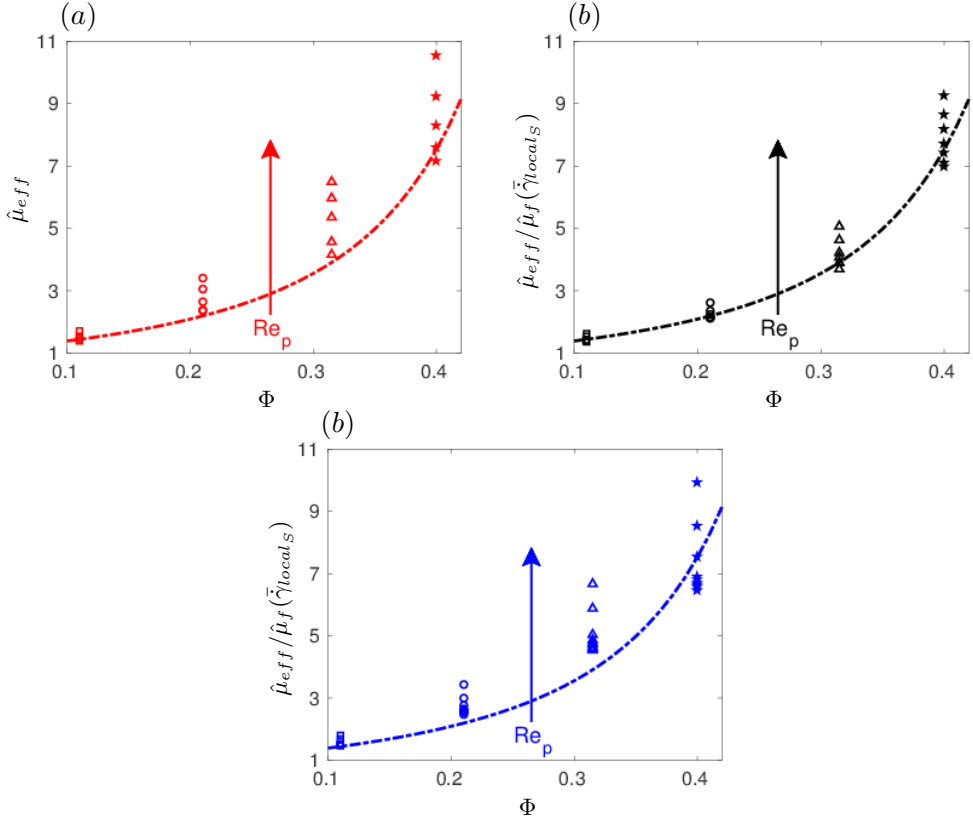
\begin{figure}
\centering{
\hspace{-0.3cm}
\includegraphics[width=0.5\linewidth]{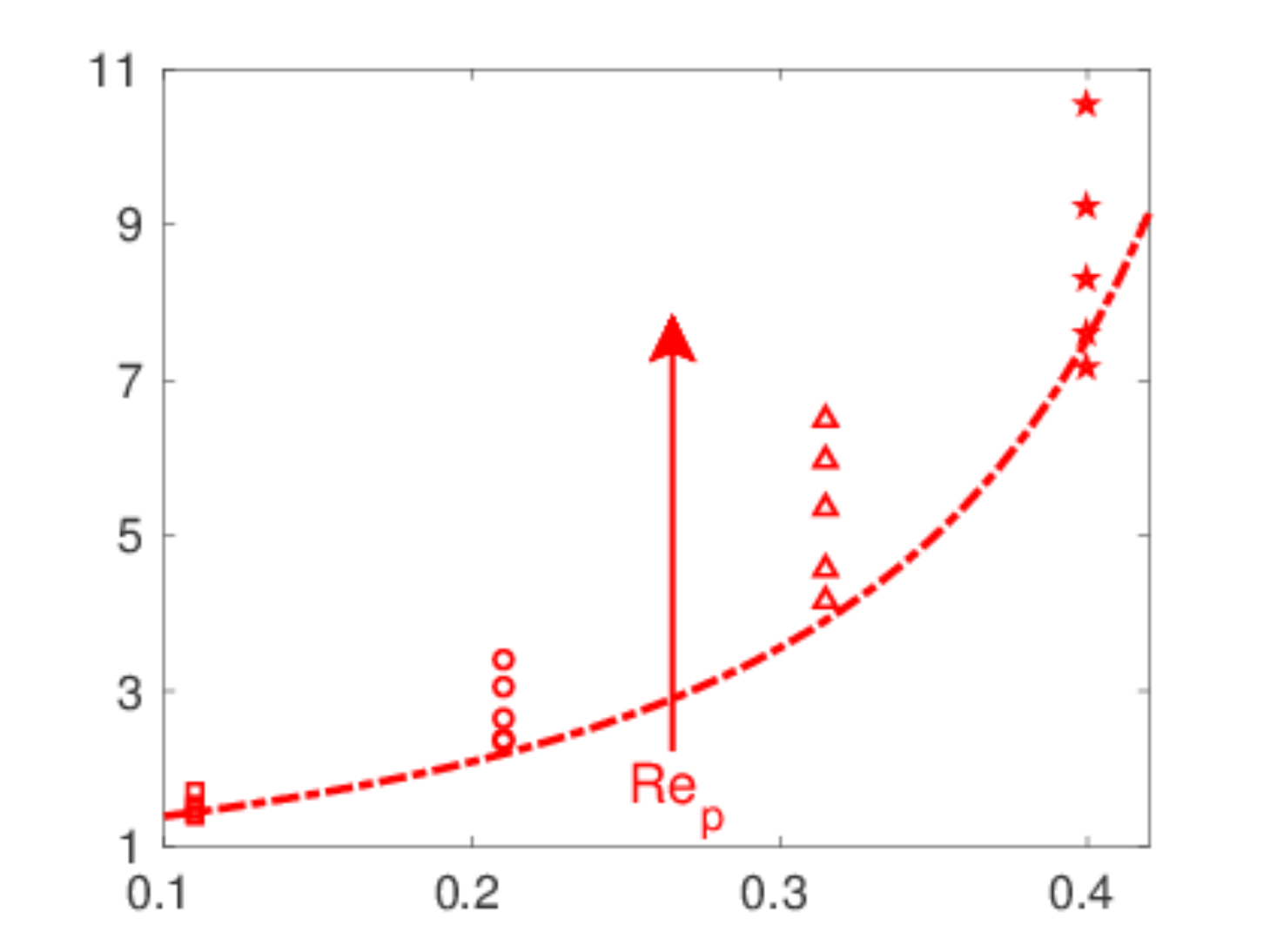}
\put(-170,140){{\large $(a)$}}
\put(-190,65){{\rotatebox{90}{$\hat{\mu}_{eff}$}}}
\put(-95,-5){{$\Phi$}}
\includegraphics[width=0.5\linewidth]{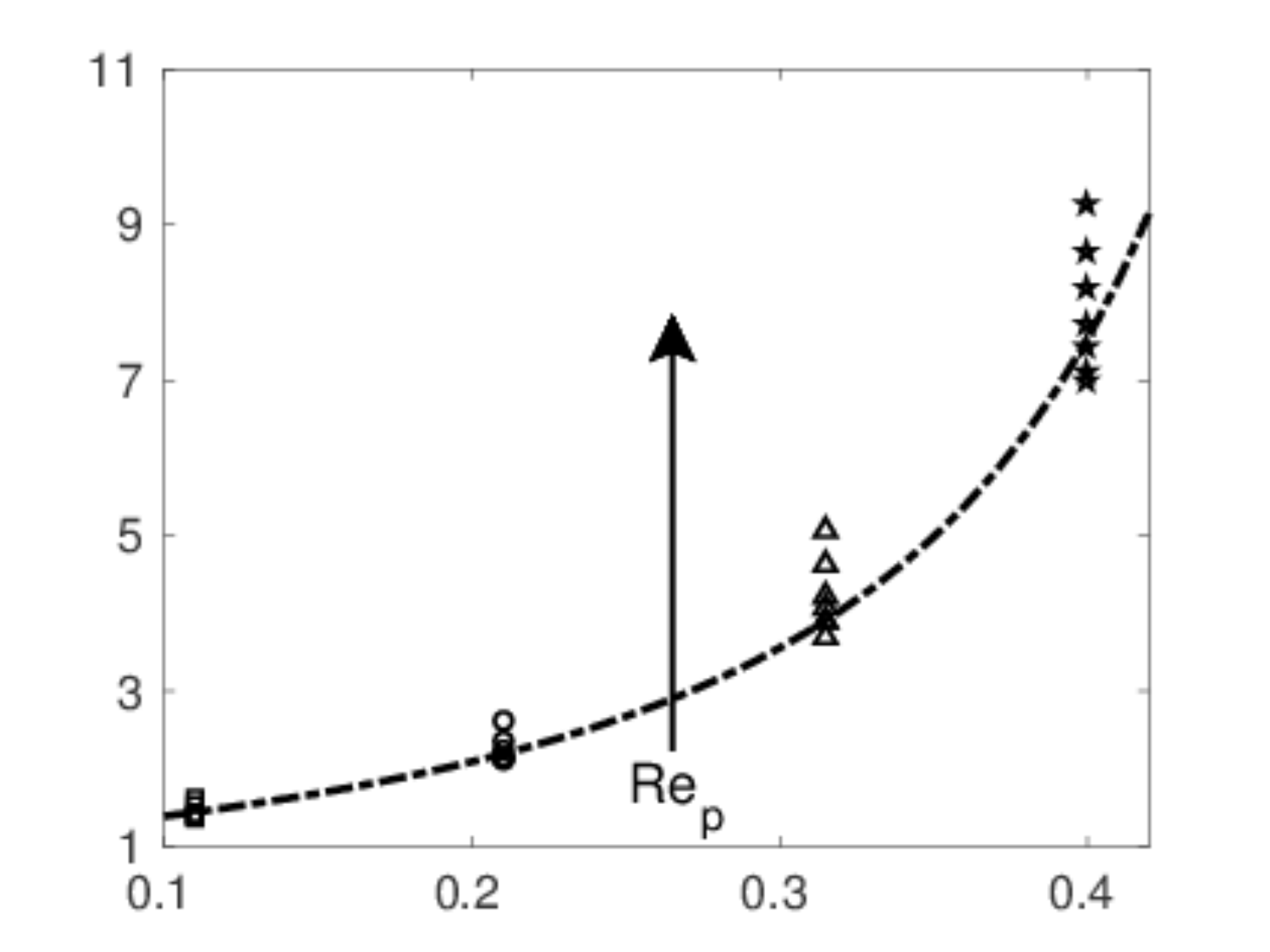}
\put(-170,140){{\large $(b)$}}
\put(-190,40){{\rotatebox{90}{$\hat{\mu}_{eff} / \hat{\mu} _f(\bar{\dot{\gamma}}_{local_S})$}}}
\put(-95,-5){{$\Phi$}}
\\
\includegraphics[width=0.5\linewidth]{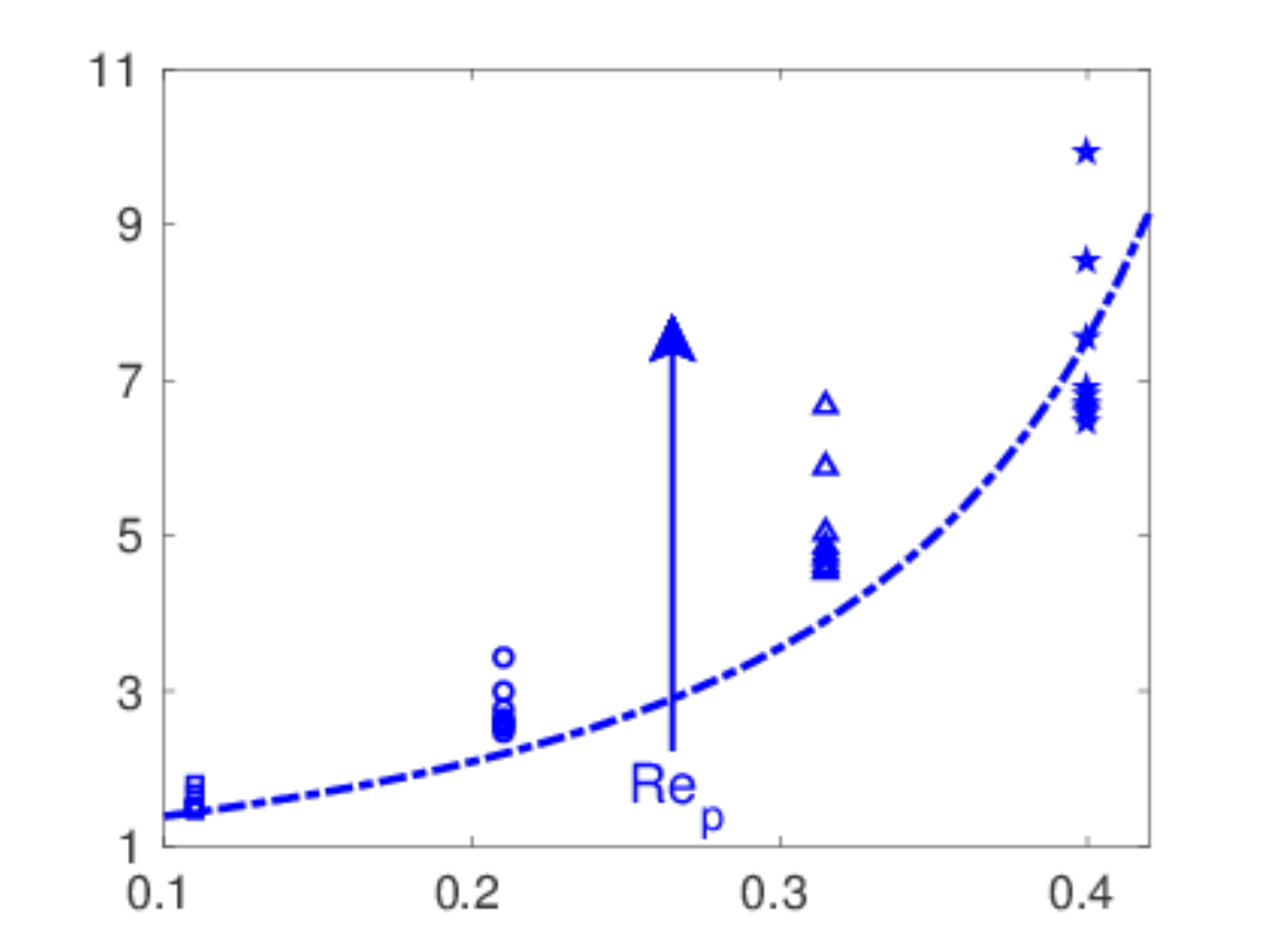}
\put(-170,140){{\large $(b)$}}
\put(-190,40){{\rotatebox{90}{$\hat{\mu}_{eff} / \hat{\mu} _f(\bar{\dot{\gamma}}_{local_S})$}}}
\put(-95,-5){{$\Phi$}}
\caption{\label{fig:Normalized mu_Re_all} 
The normalized effective viscosity  (simulation (symbols),  homogenisation theory (dashed line)), versus the particle volume fraction  $\Phi$ for $0.1\leq Re_p\leq10$. 
Colors and symbols as in previous figures.}}
\end{figure}

We show the normalized mean local shear rate for all the simulations in Fig. \ref{fig:Normalized_local_shear_all} (a-d). The dashed lines show the prediction of the homogenisation theory given by equation (\ref{eq:Lshear}).
It can be seen that the homogenisation theory gives a good estimate for the mean local shear rate when $\Phi\le0.3$. However, the theory does not provide accurate predictions of the mean local shear rate as we approach the dense regime, as also observed experimentally in the work of \cite{Dagois-Bohy15}.
In particular, it overestimates the values of the mean local shear rate when the solid volume fraction is larger than $0.3$.  This discrepancy between the mean local shear rate from our simulations and the  homogenisation theory is not due to the confinement 
effects, and consequently, particle layering. Indeed Fig. \ref{fig:phi_wallnorm_thin}a  shows a very slight variation of the fluid phase velocity across the gap even when strong particle layering exists, e.g., when  $\Phi=0.4$ for a shear thickening 
suspending fluid.  We can also see that, unlike the bulk solid volume fraction, inertia has a secondary effect on the  values of mean local shear rate. 
Fig. \ref{fig:Normalized_local_shear_all}d finally reports all the data in one graph to emphasize that the normalized mean local shear rate is almost independent of the type of suspending fluid. 

\subsection{Effective viscosity}
\label{sec:eff_visc}
The effective viscosity of the suspension is defined by the space and time-average wall shear stress divided by the bulk shear rate. Fig. \ref{fig:Normalized mu_Re_all} shows the non-dimensional effective viscosity for the suspensions considered here
(see Table \ref{tab:allcases}) versus the particle volume fraction $\Phi$ for   $0.1\leq Re_p\leq 10$. Note that the non-dimensional effective viscosity is calculated after the transient stage of the simulation and convergence tests have been performed
by comparing the statistical results with those obtained using half the number of samples. We also display the non-dimensional effective viscosity predicted by eq.~(\ref{eq:GMueff}) on the graphs of Fig. \ref{fig:Normalized mu_Re_all}. 
The results show that independent of the type of suspending fluid,  the  non-dimensional effective viscosity increases  with the solid volume fraction.  In the limit of small Reynolds number, $Re_p\sim0.1$ (when the inertia is negligible), 
the results of our simulations follow the prediction by the homogenization theory. However, as we increase the particle Reynolds number, we deviate from the homogenization theory, which  is developed for Stokesian suspensions. 
The increase of the the non-dimensional effective viscosity with $Re_p$ can be explained as an inertial shear thickening \cite[see e.g.,][]{Picano13} and will be discussed more  in \S\ref{sec:fric_view}.

\begin{figure}
\centering{
\hspace{-0.3cm}
\includegraphics[width=0.5\linewidth]{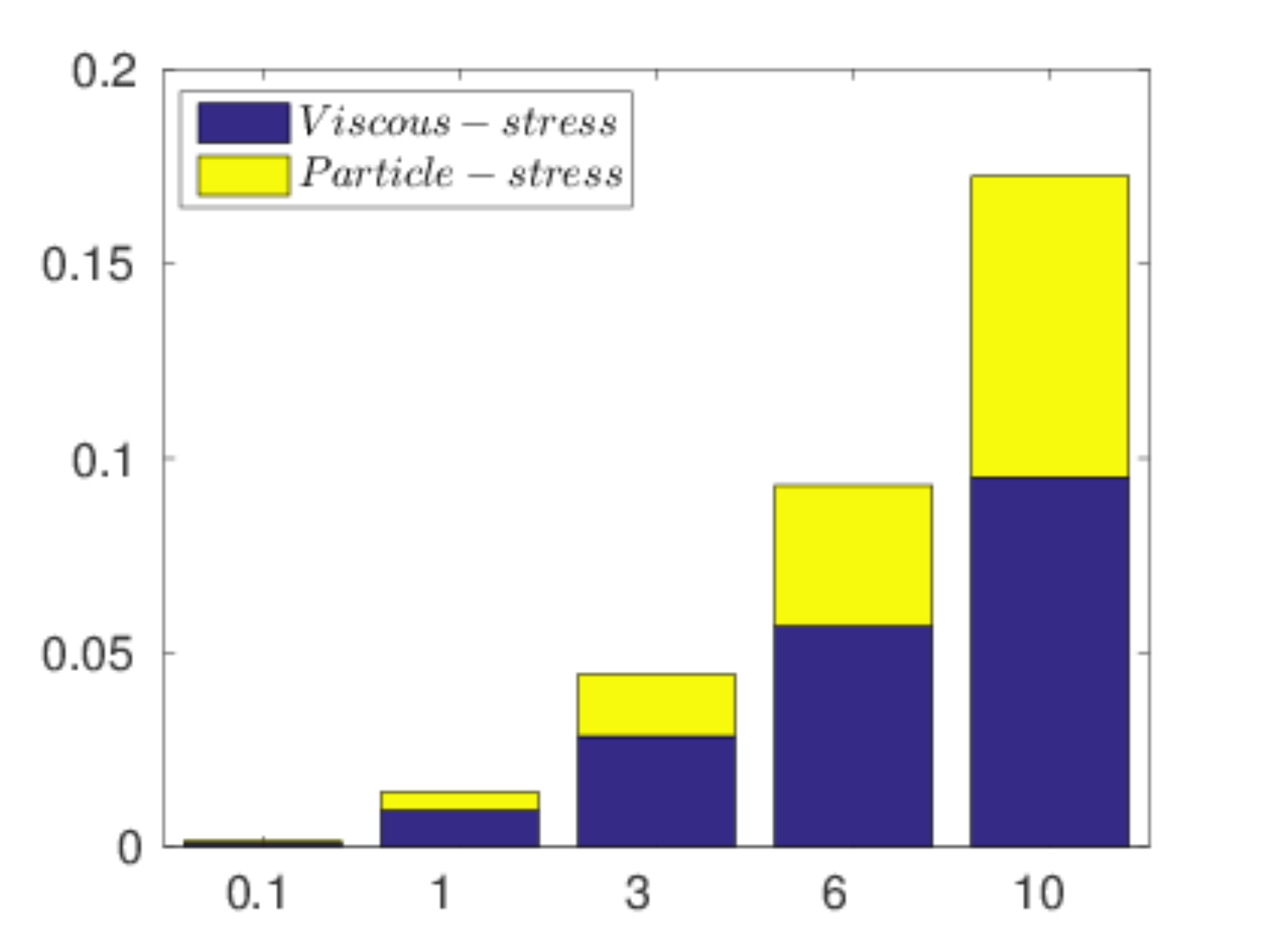}
\put(-170,140){{\large $(a) \Phi = 0.11 $}}
\put(-195,70){{\rotatebox{90}{$\tau_w$}}}
\put(-100,-5){{$Re_p$}}
\includegraphics[width=0.5\linewidth]{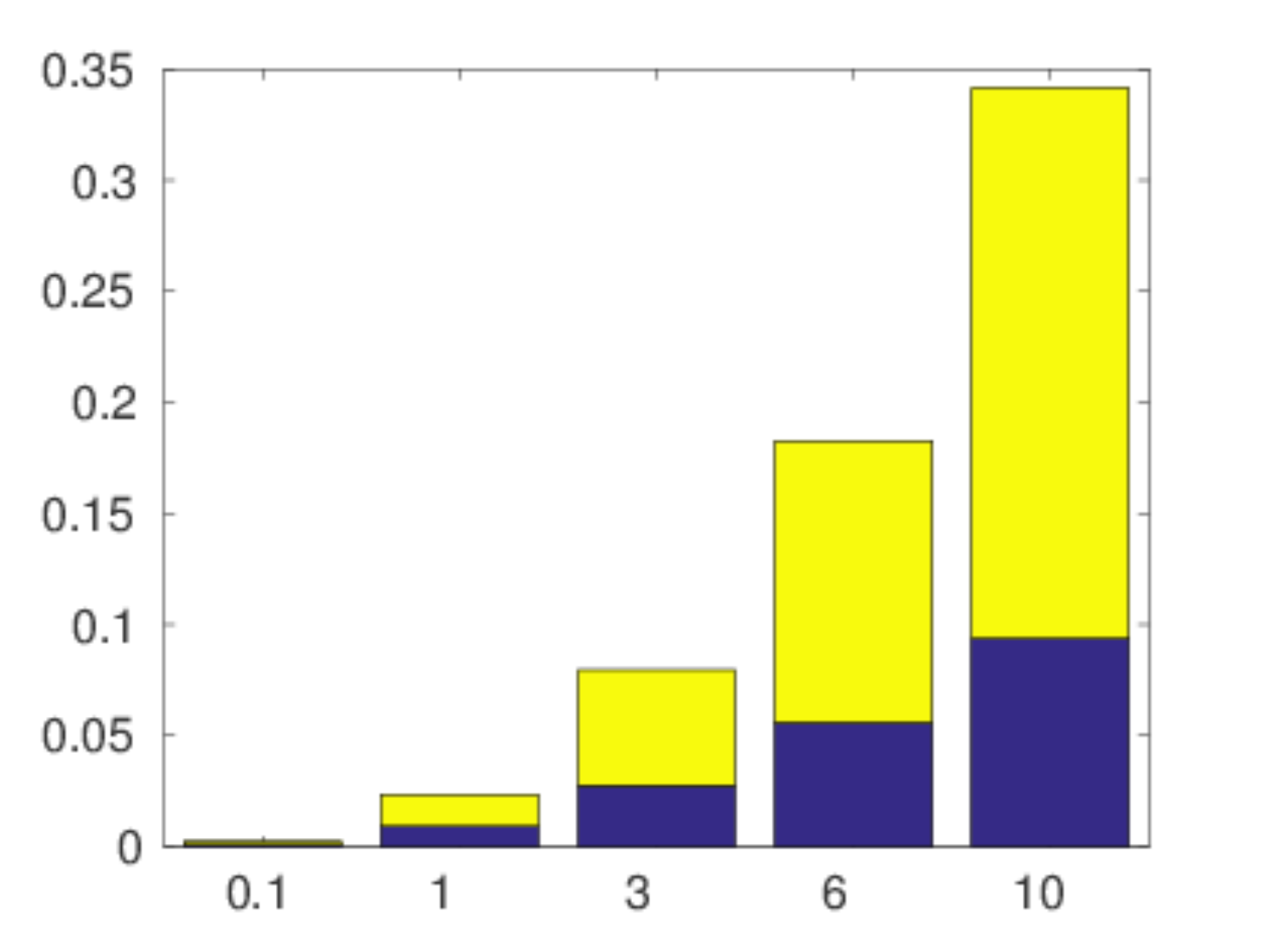}
\put(-170,140){{\large $(b) \Phi = 0.21 $}}
\put(-195,70){{\rotatebox{90}{$\tau_w$}}}
\put(-100,-5){{$Re_p$}}
\\
\includegraphics[width=0.5\linewidth]{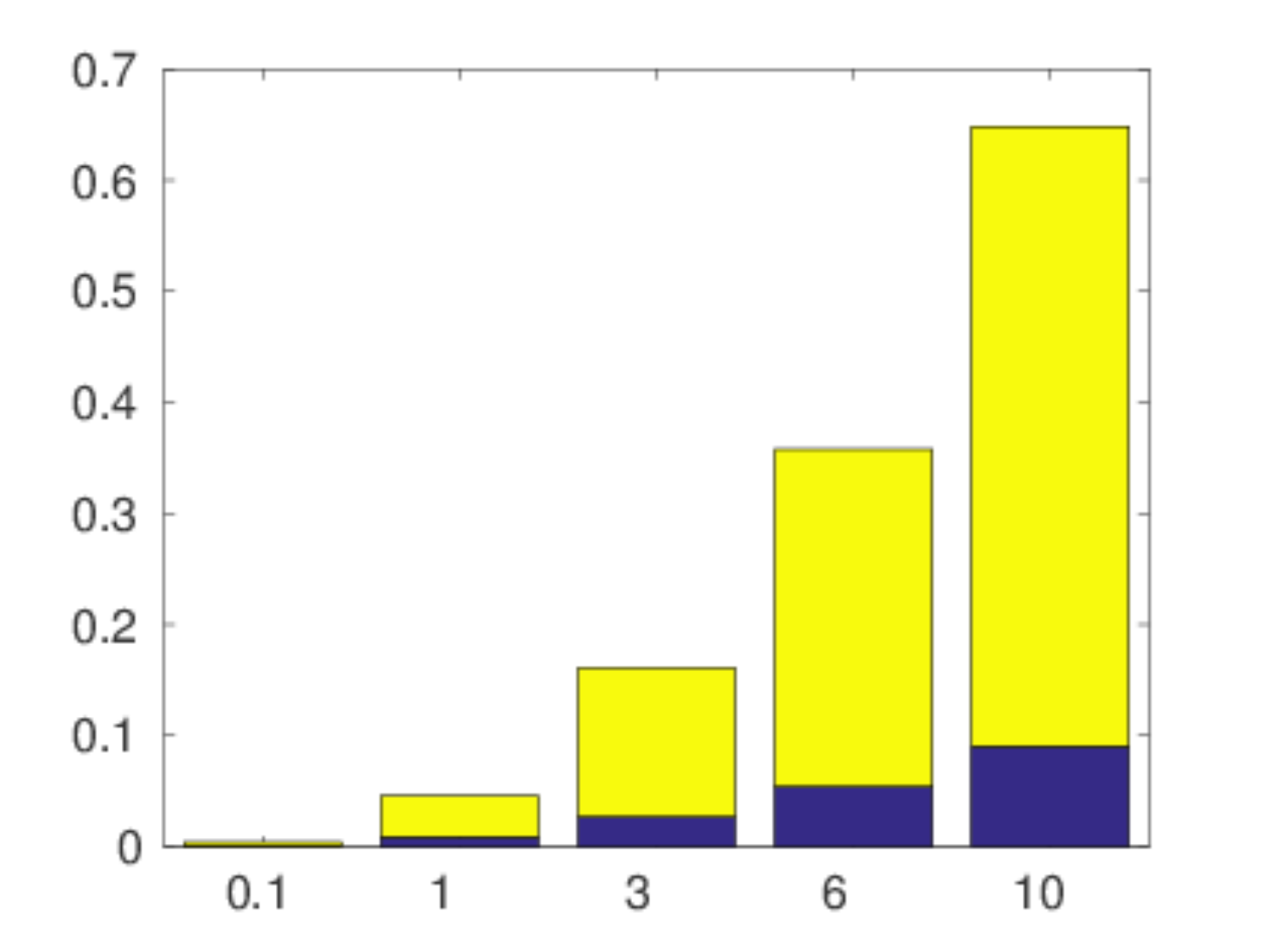}
\put(-170,140){{\large $(c) \Phi = 0.315$}}
\put(-195,70){{\rotatebox{90}{$\tau_w$}}}
\put(-100,-5){{$Re_p$}}
\includegraphics[width=0.5\linewidth]{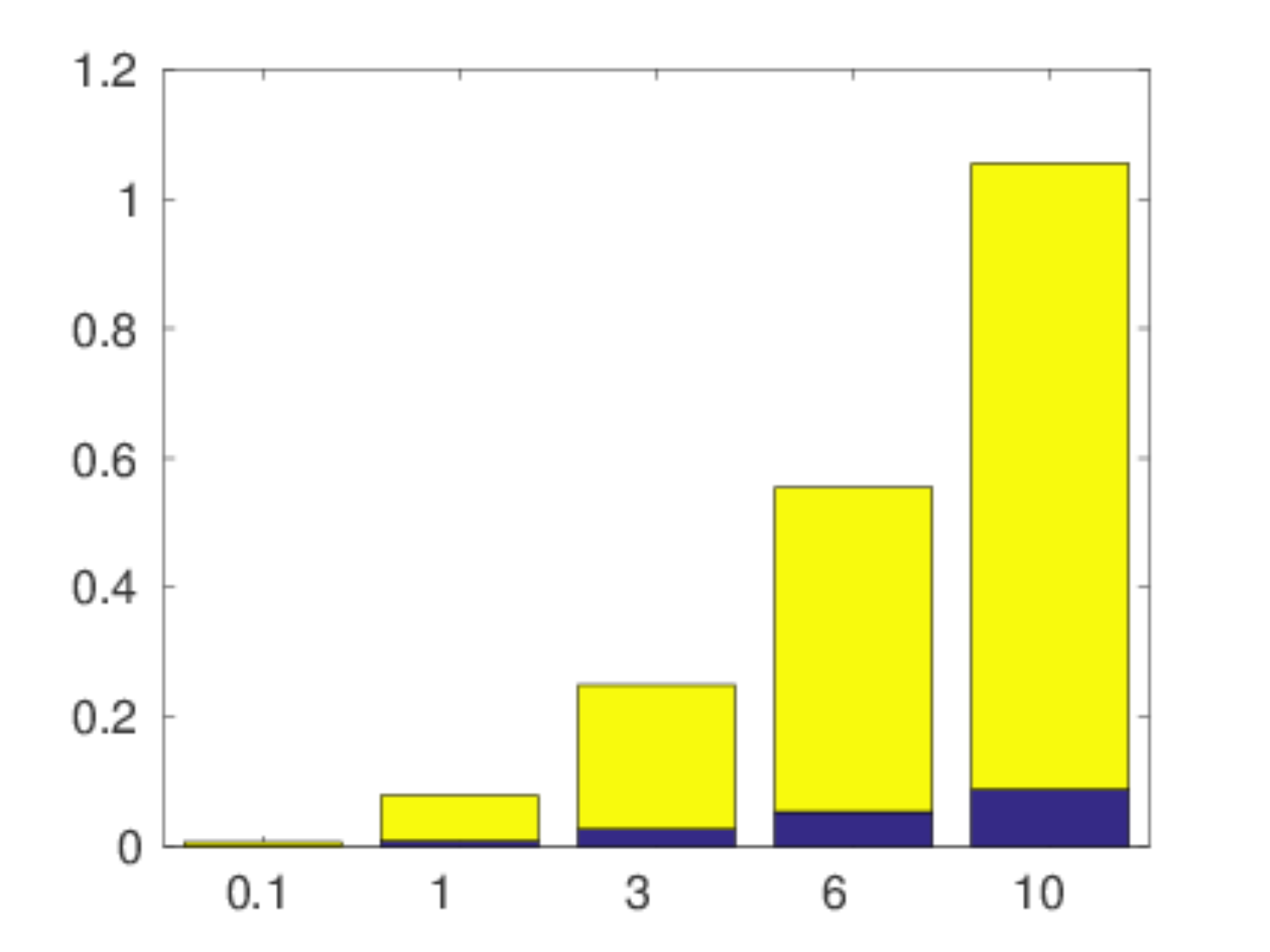}
\put(-170,140){{\large $(d) \Phi = 0.4$}}
\put(-195,70){{\rotatebox{90}{$\tau_w$}}}
\put(-100,-5){{$Re_p$}}
\caption{\label{fig:Tot_stress_Newt} 
Profiles of the wall stress budget $\tau_w$  versus particle Reynolds numbers $Re_p$ for various  volume fractions $\Phi$, for Newtonian cases.}}
\end{figure}

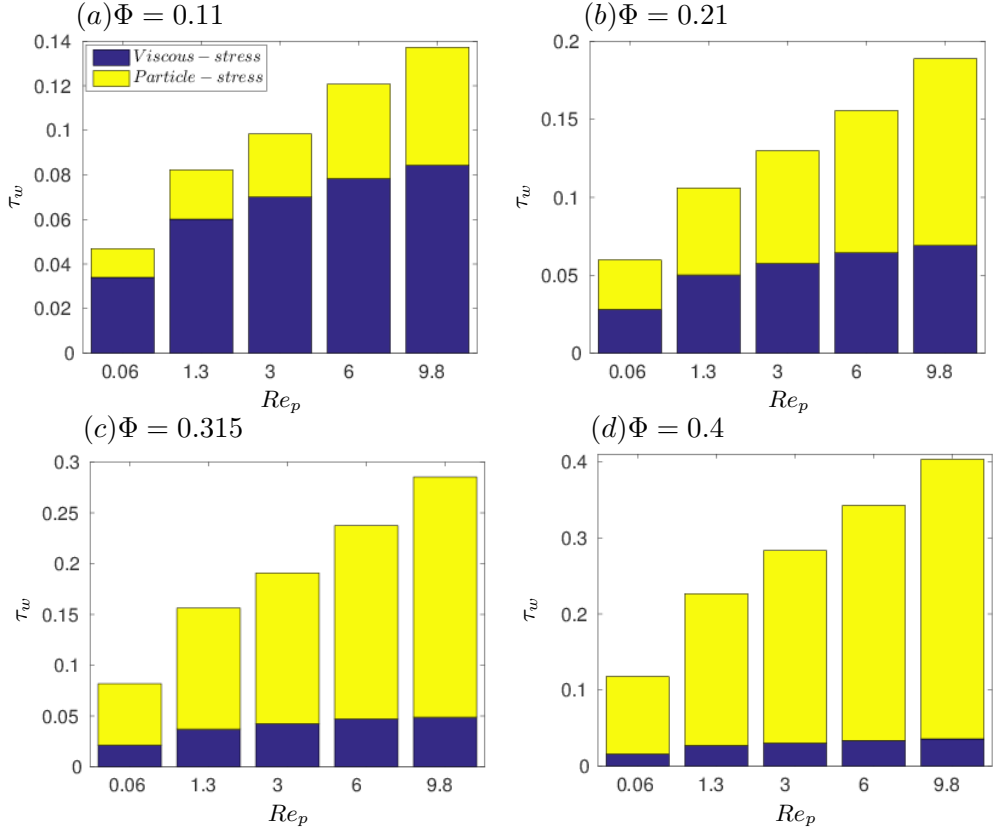
\begin{figure}
\centering{
\hspace{-0.3cm}
\includegraphics[width=0.5\linewidth]{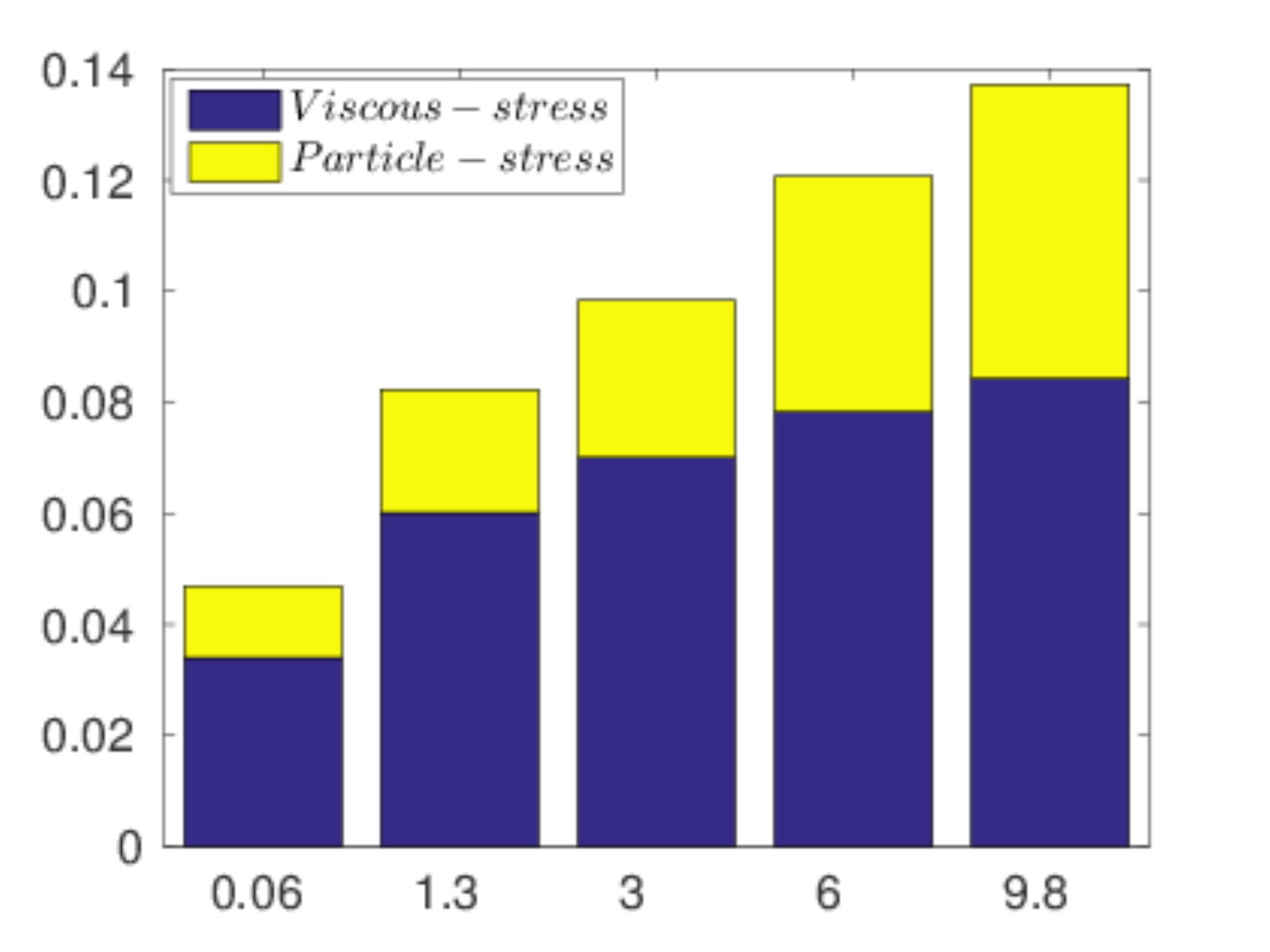}
\put(-170,140){{\large $(a) \Phi =0.11 $}}
\put(-195,70){{\rotatebox{90}{$\tau_w$}}}
\put(-100,-5){{$Re_p$}}
\includegraphics[width=0.5\linewidth]{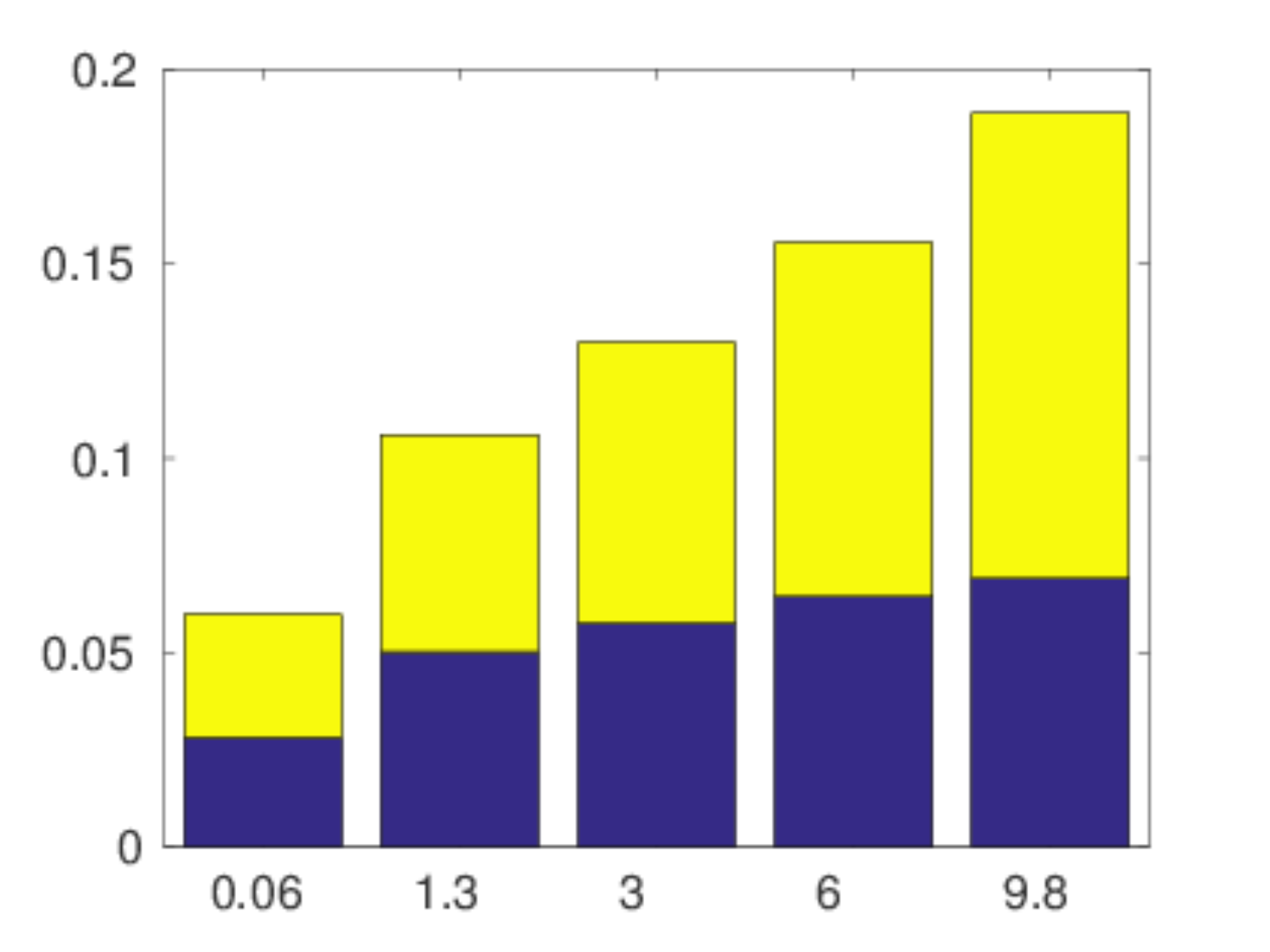}
\put(-170,140){{\large $(b)  \Phi = 0.21$}}
\put(-195,70){{\rotatebox{90}{$\tau_w$}}}
\put(-100,-5){{$Re_p$}}
\\
\includegraphics[width=0.5\linewidth]{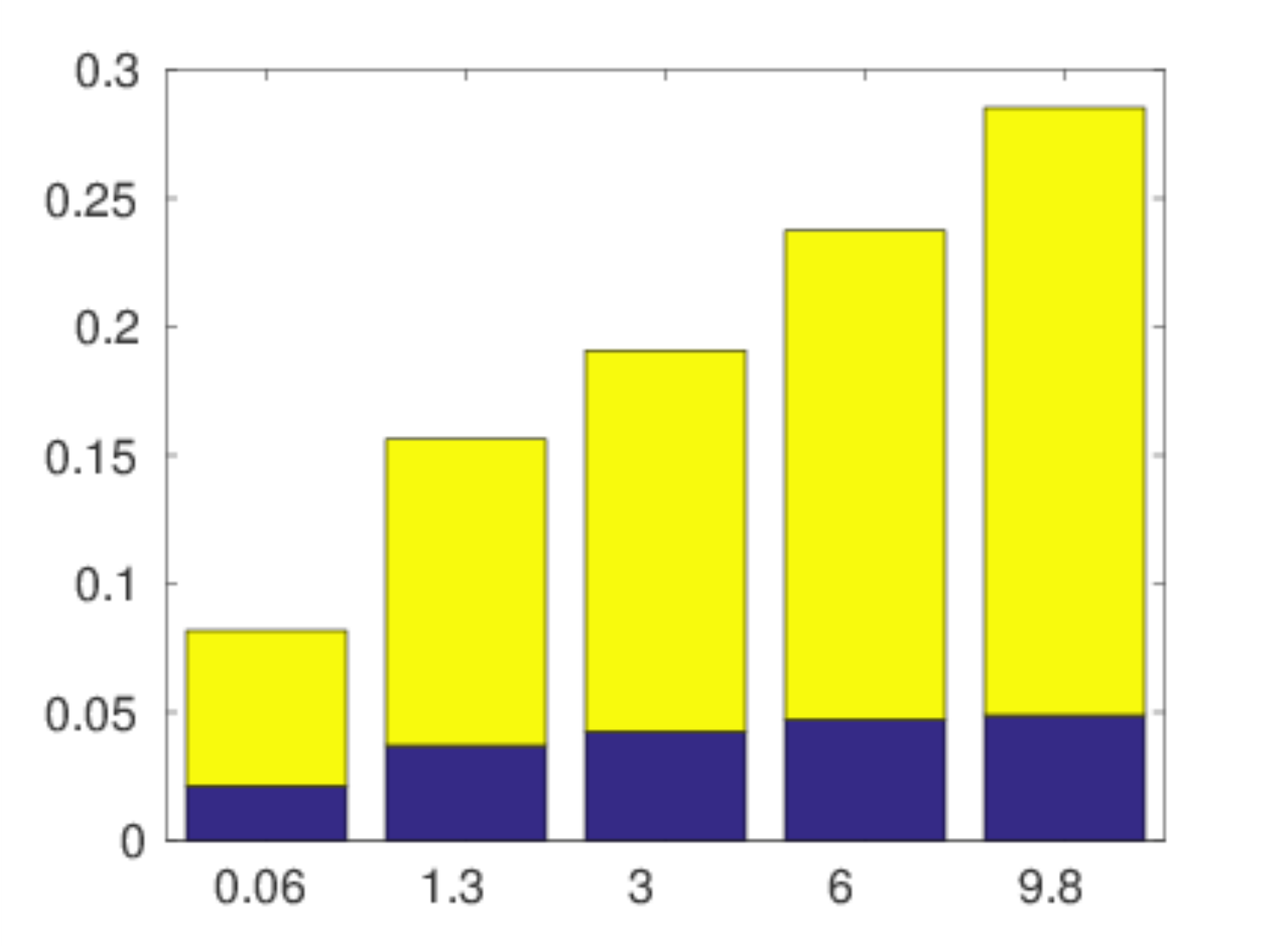}
\put(-170,140){{\large $(c) \Phi = 0.315 $}}
\put(-195,70){{\rotatebox{90}{$\tau_w$}}}
\put(-100,-5){{$Re_p$}}
\includegraphics[width=0.5\linewidth]{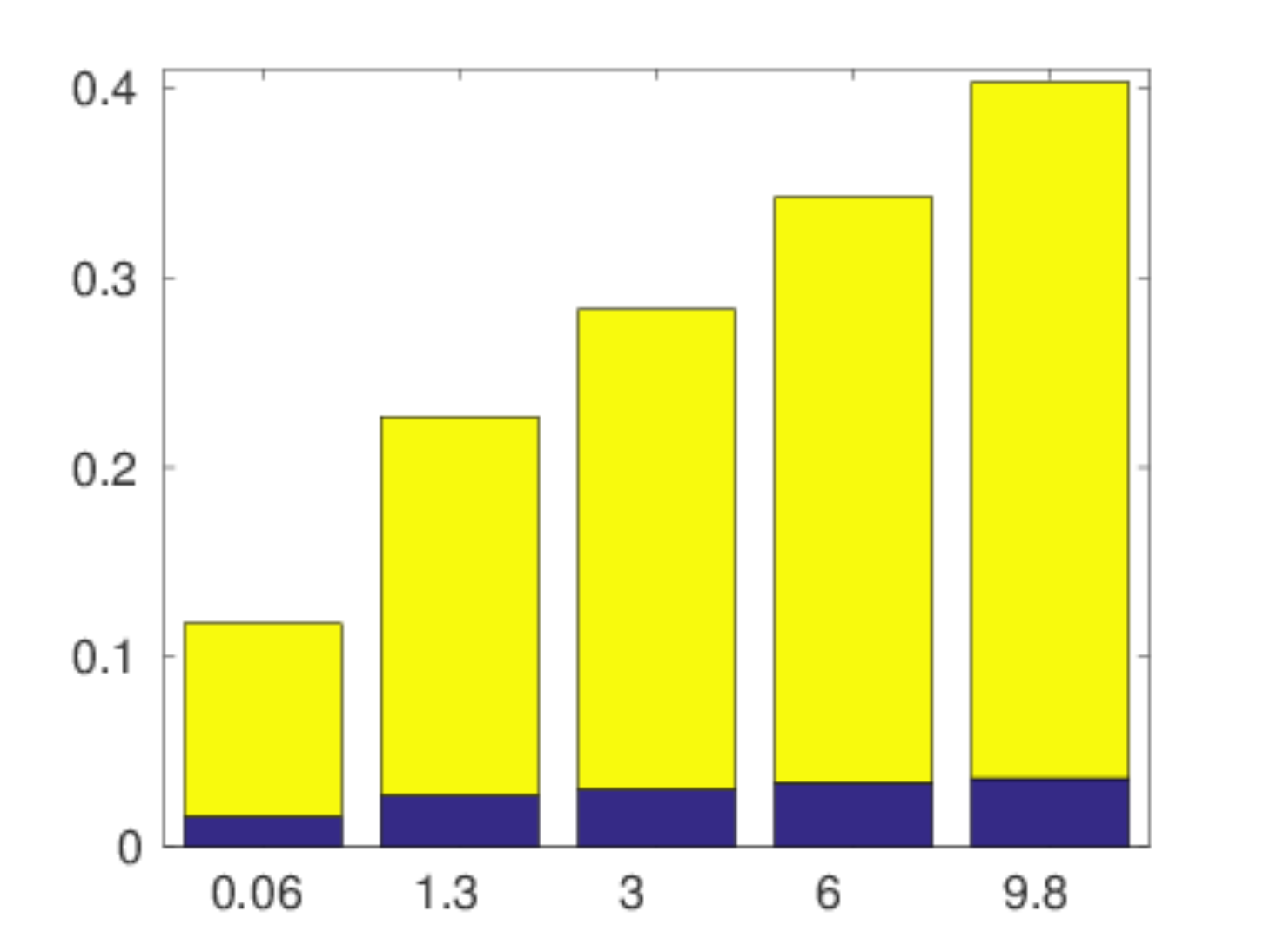}
\put(-170,140){{\large $(d)  \Phi = 0.4$}}
\put(-195,70){{\rotatebox{90}{$\tau_w$}}}
\put(-100,-5){{$Re_p$}}
\caption{\label{fig:Tot_stress_thin} 
Profiles of the wall stress budget $\tau_w$  versus particle Reynolds numbers $Re_p$ for various  volume fraction $\Phi$, for the shear thinning cases. }}
\end{figure}

\subsection{Role of inertia}
\label{sec:fric_view}
Here, we provide more detailed understanding of the role of inertia on the rheology of noncolloidal suspensions. To this end, we first calculate the total suspension stress ($\tau$}) as well as the contributions of viscous stress 
($\tau_v$), Reynolds stress  ($\tau_R$) and particle stress  ($\tau_p$) to the total stress. As detailed in \cite{Picano15} and \cite{Lashgari16}, we write 
\begin{equation}
\tau=\tau_v+\tau_R+\tau_p,
\label{eq:total_stress}  
\end{equation} 
where, $\tau_R$ includes both particle and fluid Reynolds stresses. Our calculation shows that for all the simulations performed in this study, the Reynolds stress $\tau_R$ is almost zero as the bulk Reynolds number ($Re\sim100 Re_p$) is in the range of laminar flows. 
The particle stress includes contributions from the hydrodynamic stresslet, particle acceleration, short-range lubrication correction and collision forces \citep{Lashgari14,Lashgari16}. 
It is noteworthy to mention that the total stress $\tau$ is essentially equal to the suspension stress at the wall $\tau_w$ as we are not in the dense regimes and particle-wall collisions are negligible.

\begin{figure}
\centering{
\hspace{-0.3cm}
\includegraphics[width=0.5\linewidth]{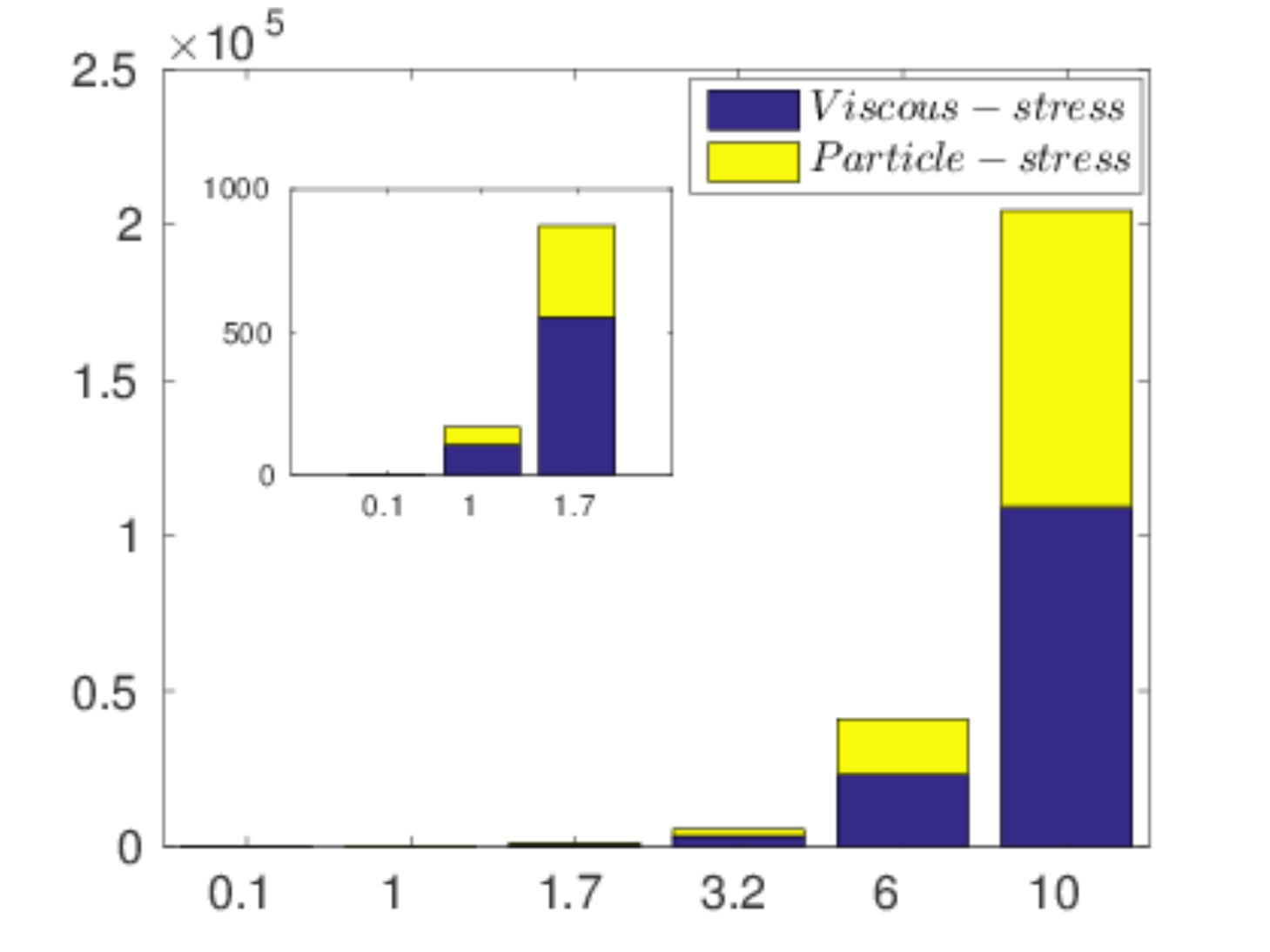}
\put(-170,150){{\large $(a) \Phi = 0.11$}}
\put(-195,70){{\rotatebox{90}{$\tau_w$}}}
\put(-100,-5){{$Re_p$}}
\includegraphics[width=0.5\linewidth]{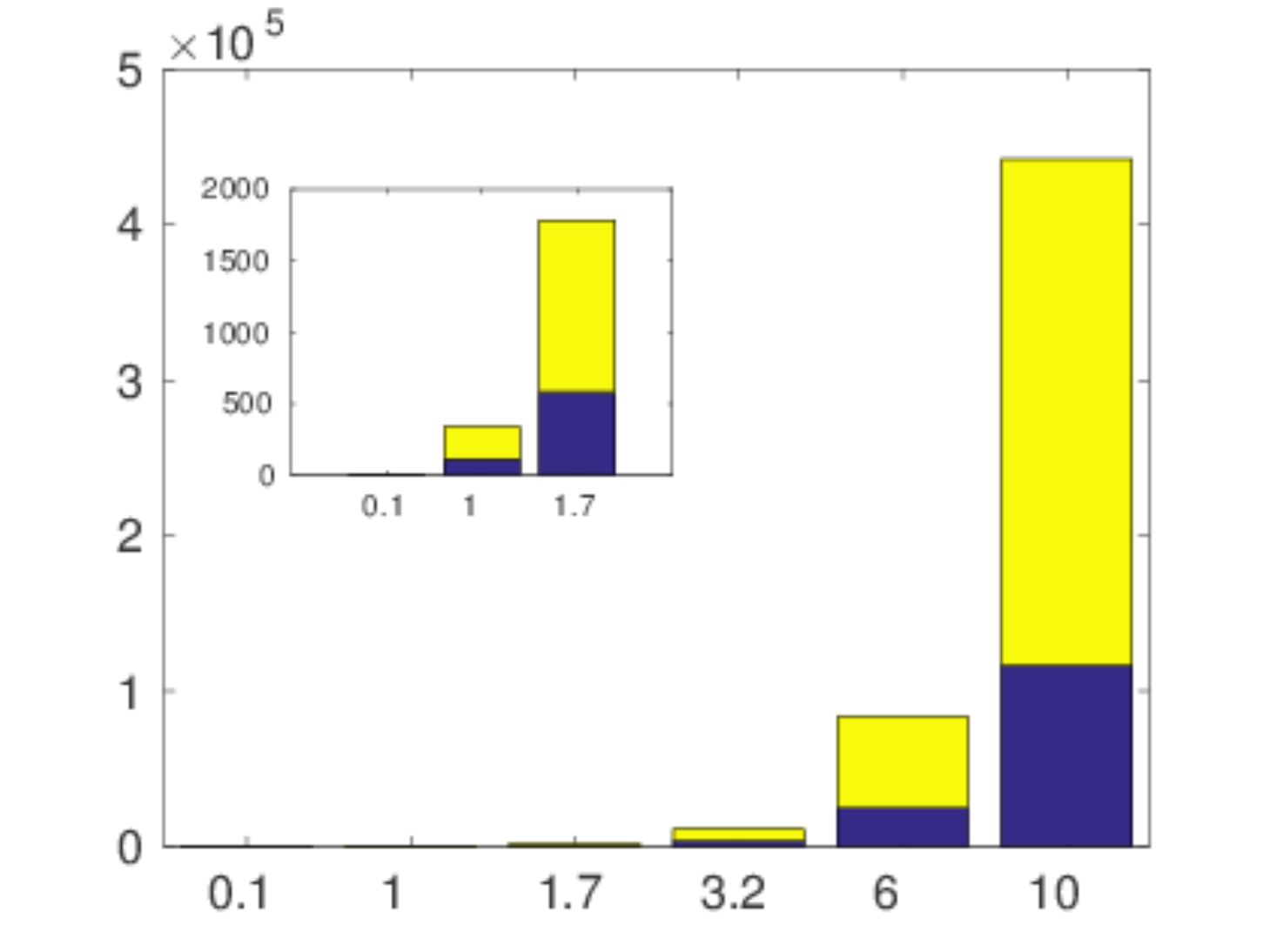}
\put(-170,150){{\large $(b) \Phi =0.21$}}
\put(-195,70){{\rotatebox{90}{$\tau_w$}}}
\put(-100,-5){{$Re_p$}}
\\
\includegraphics[width=0.5\linewidth]{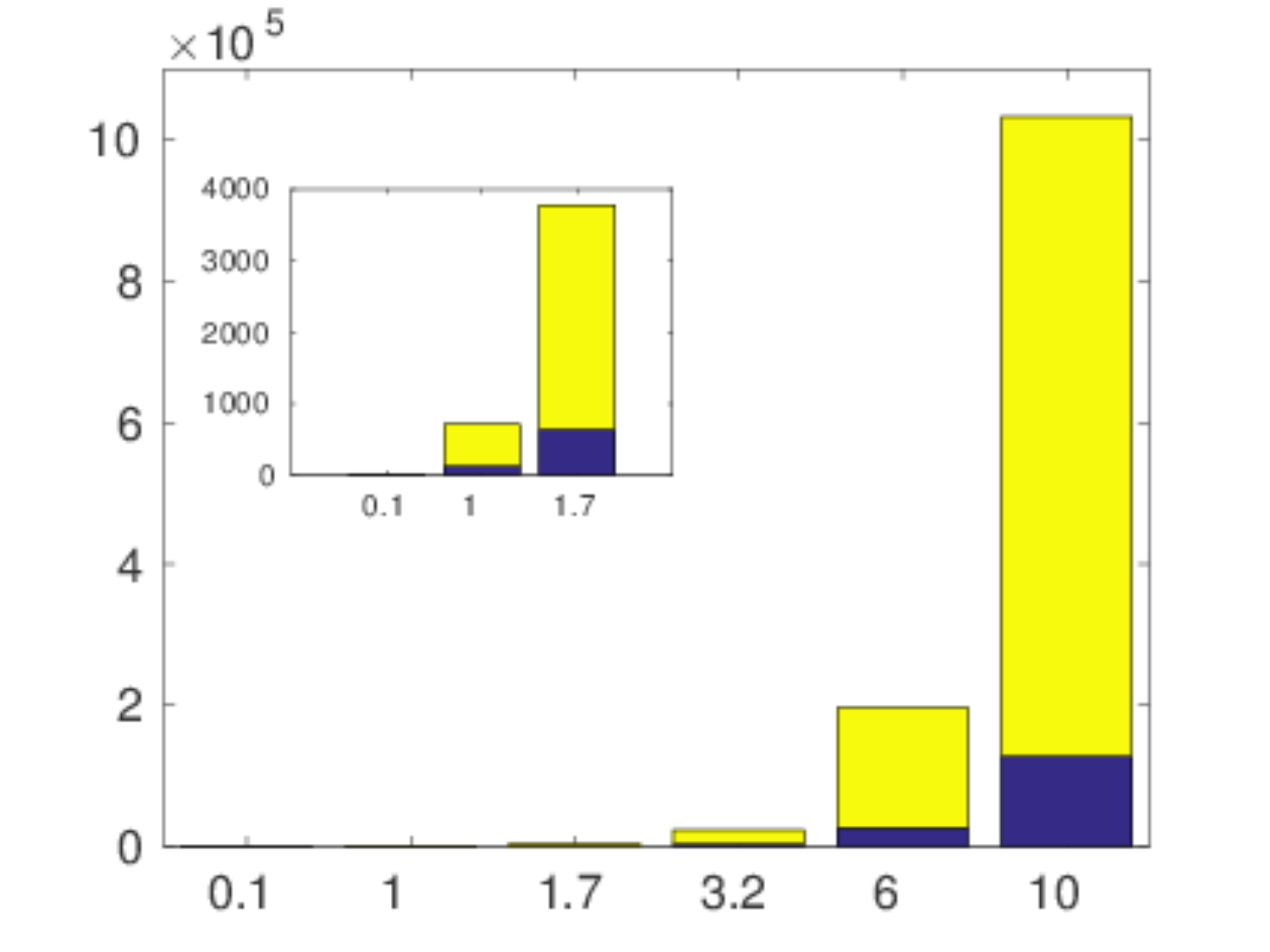}
\put(-170,150){{\large $(c) \Phi = 0.315$}}
\put(-195,70){{\rotatebox{90}{$\tau_w$}}}
\put(-100,-5){{$Re_p$}}
\includegraphics[width=0.5\linewidth]{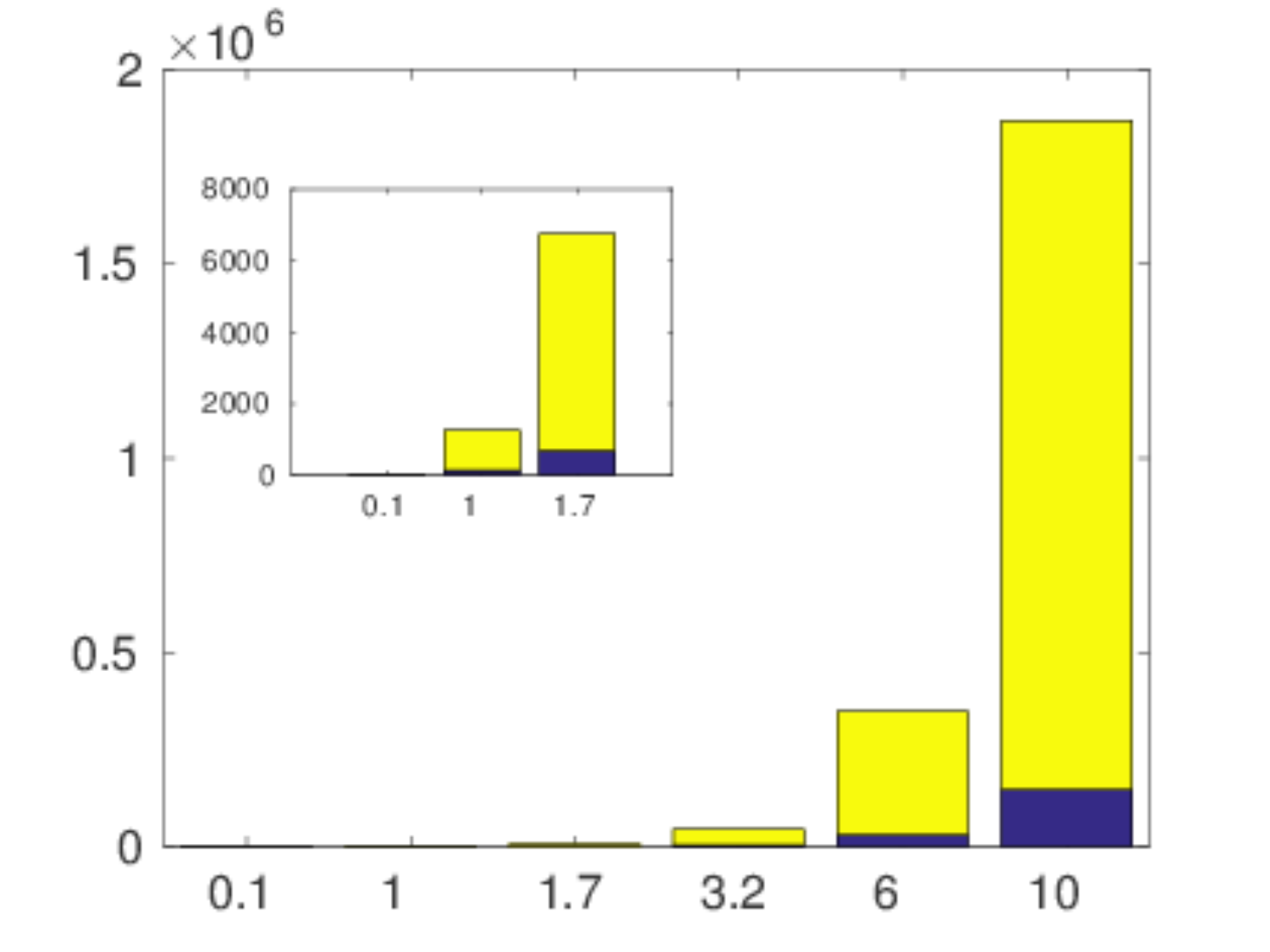}
\put(-170,150){{\large $(d)  \Phi = 0.4$}}
\put(-195,70){{\rotatebox{90}{$\tau_w$}}}
\put(-100,-5){{$Re_p$}}
\caption{\label{fig:Tot_stress_thick} 
Profiles of the wall stress budget $\tau_w$  versus particle Reynolds numbers $Re_p$ for various  volume fraction $\Phi$, for the shear thickening cases. }}
\end{figure}

\begin{figure}
\centering{
\hspace{-0.3cm}
\includegraphics[width=0.5\linewidth]{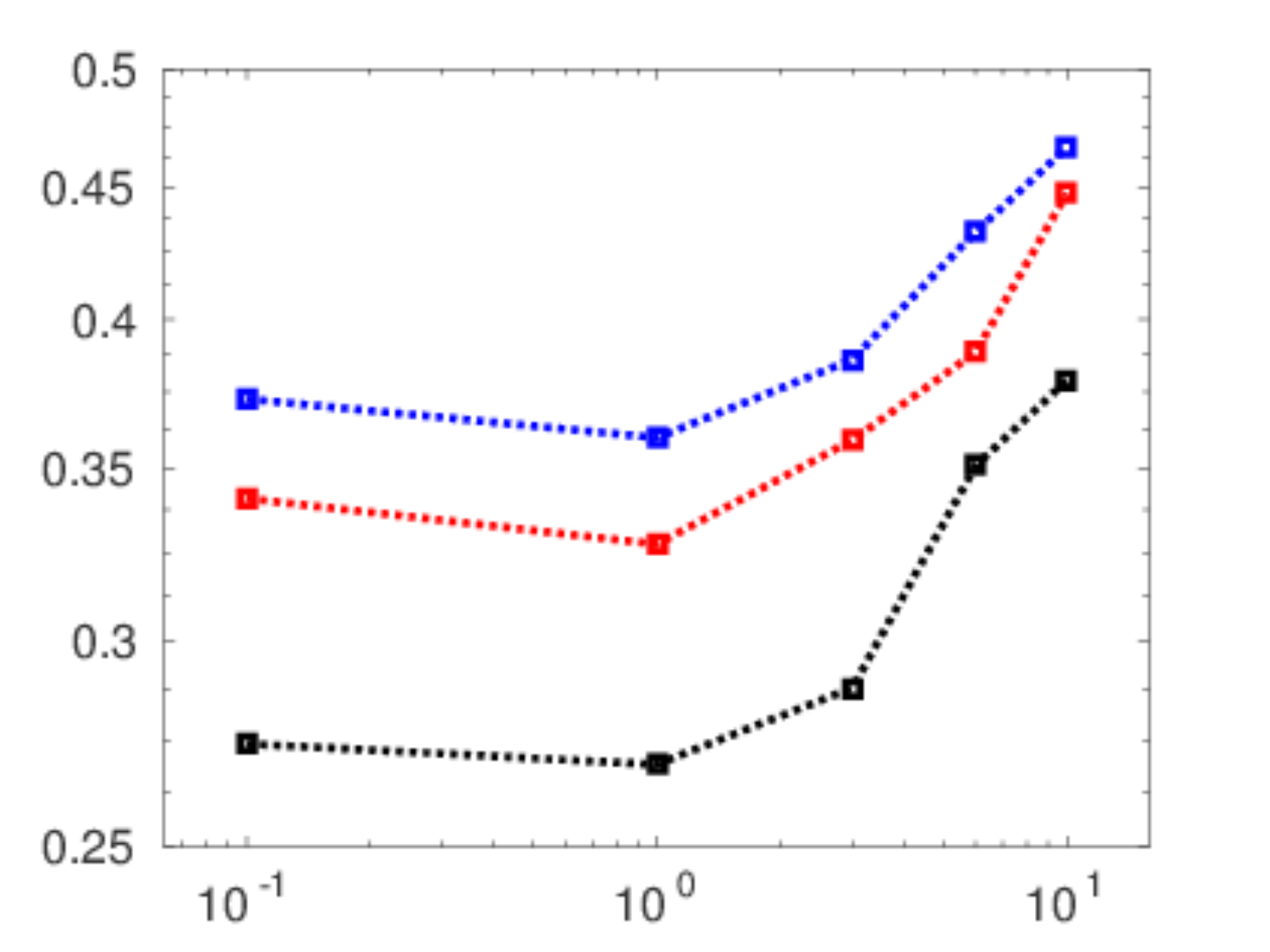}
\put(-170,140){{\large $(a) \Phi = 0.11$}}
\put(-200,65){{\rotatebox{90}{$\tau_p/\tau_w$}}}
\put(-100,-5){{$Re_p$}}
\includegraphics[width=0.5\linewidth]{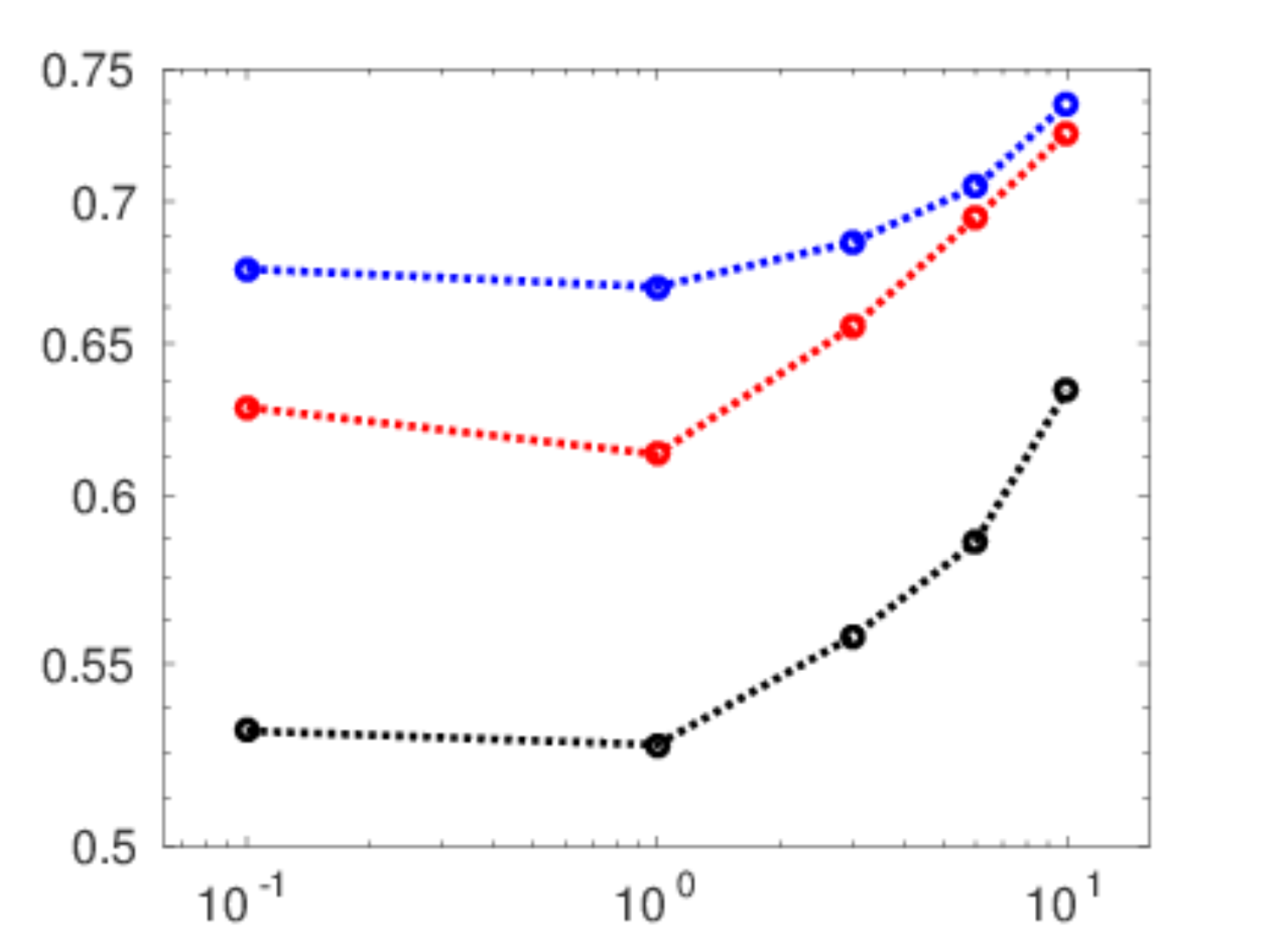}
\put(-170,140){{\large $(b) \Phi = 0.21$}}
\put(-200,65){{\rotatebox{90}{$\tau_p/\tau_w$}}}
\put(-100,-5){{$Re_p$}}
\\
\includegraphics[width=0.5\linewidth]{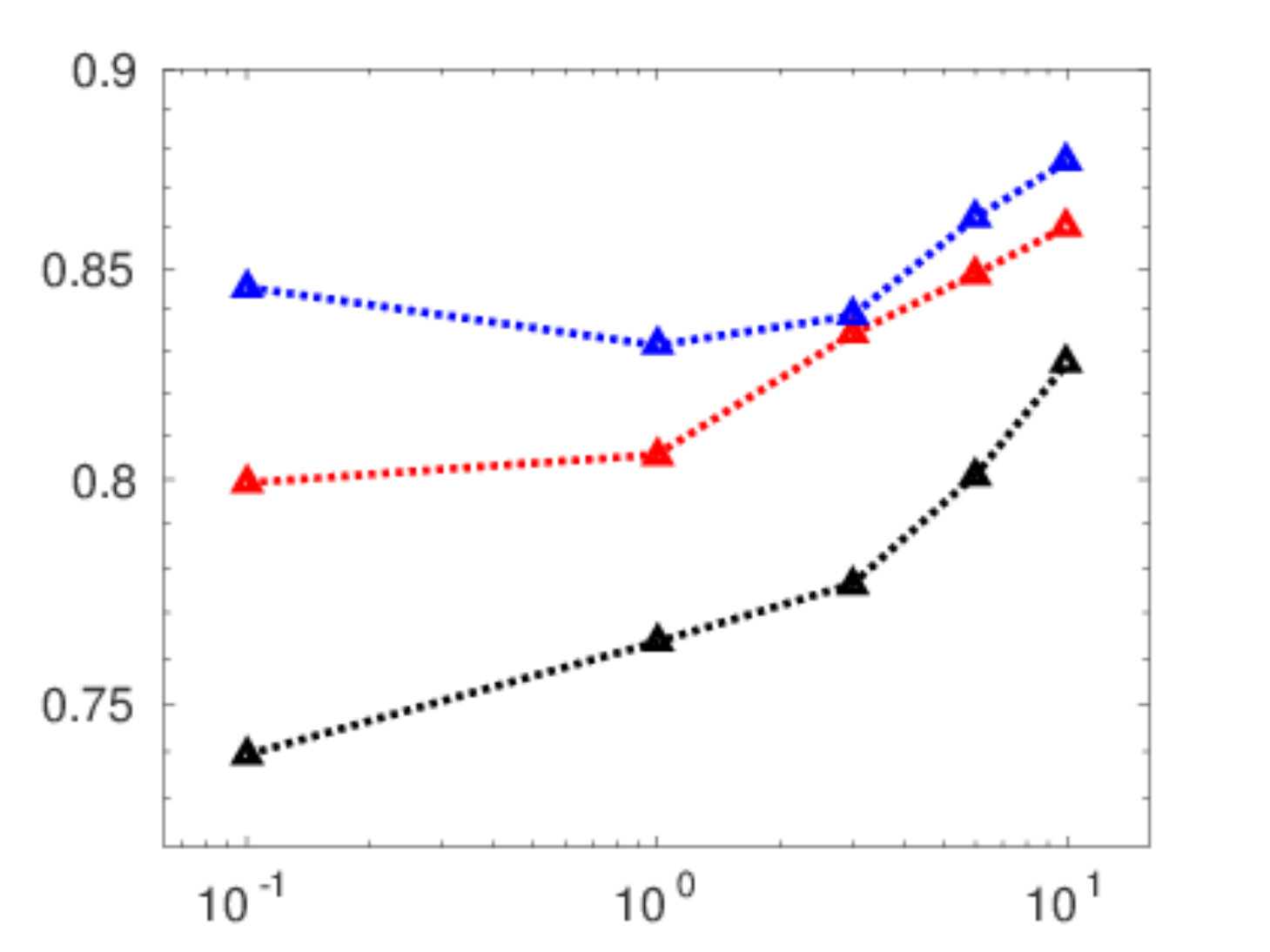}
\put(-170,140){{\large $(c) \Phi =0.315 $}}
\put(-200,65){{\rotatebox{90}{$\tau_p/\tau_w$}}}
\put(-100,-5){{$Re_p$}}
\includegraphics[width=0.5\linewidth]{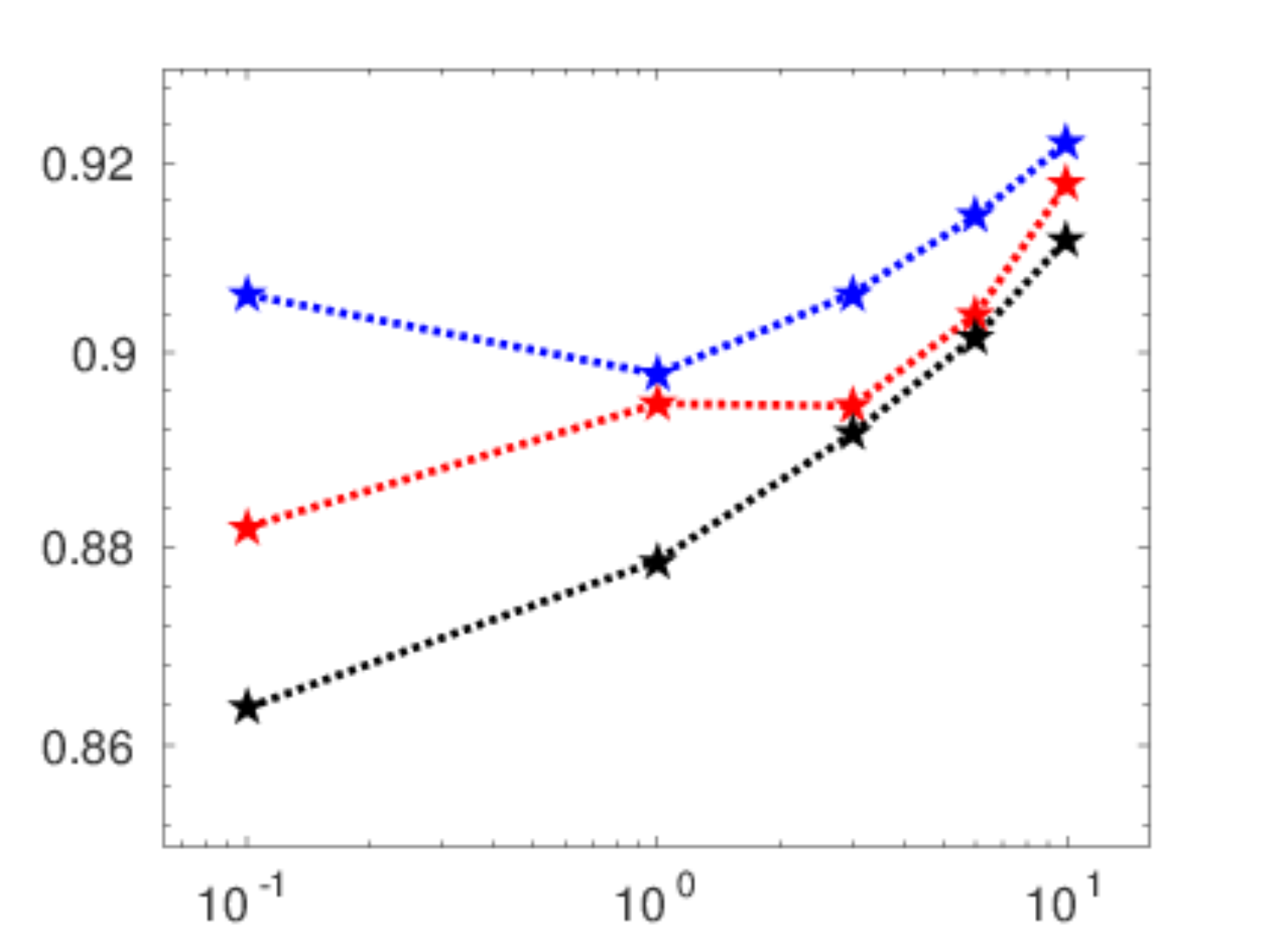}
\put(-170,140){{\large $(d) \Phi =0.4 $}}
\put(-200,65){{\rotatebox{90}{$\tau_p/\tau_w$}}}
\put(-100,-5){{$Re_p$}}
\caption{\label{fig:Part_stress} 
Profiles of the normalized particle stress $\tau_p/\tau_w$  versus particle Reynolds numbers $Re_p$ for different carrier fluids  and different values of volume fraction  $\Phi$. Colors and symbols as in previous figures.}}
\end{figure}

Figures \ref{fig:Tot_stress_Newt}-\ref{fig:Tot_stress_thick} show the stress budget for all of our simulations. The following trend appears immediately clear: independent of the type of suspending fluid, the particle stress mainly
contributes to the total stress for $\Phi\ge0.21$. This contribution magnifies as we increase the $Re_p$. The results are summarised in Fig. \ref{fig:Part_stress} a-d where we report $\tau_p/\tau_w$ versus $Re_p$ for
$\Phi=[0.11, 0.21, 0.315, 0.4]$ and all three types of suspending fluids.  It is evident that the particle stress contributes more to the total stress for the cases with shear thickening suspending fluids than for the cases
with Newtonian suspending fluids. This feature is reversed when we deal with shear thinning  fluids.  This may be due to the fact that the stresslet and particle-pair dynamics  changes in  non-Newtonian fluid, as suggested by the experiments of \cite{Firouznia17}. This would however deserve further investigations.

\begin{figure}
\centering{
\hspace{-0.3cm}
\includegraphics[width=0.5\linewidth]{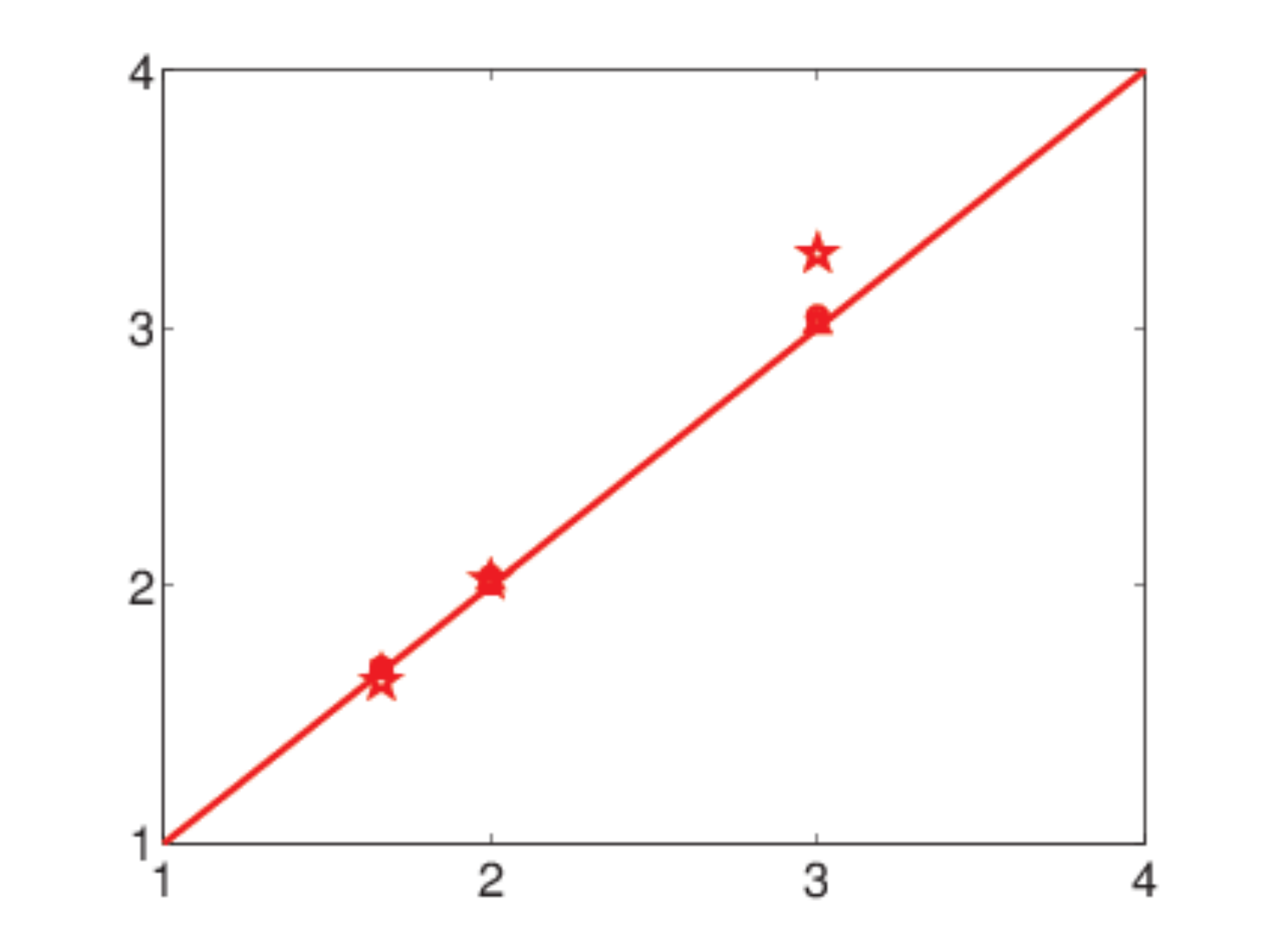}
\put(-170,140){{\large $(a) $}}
\put(-200,50){{\rotatebox{90}{$\tau_v(i+1)/\tau_v(i)$}}}
\put(-120,-5){{${\dot\gamma (i+1)/\dot\gamma (i)}$}}
\includegraphics[width=0.5\linewidth]{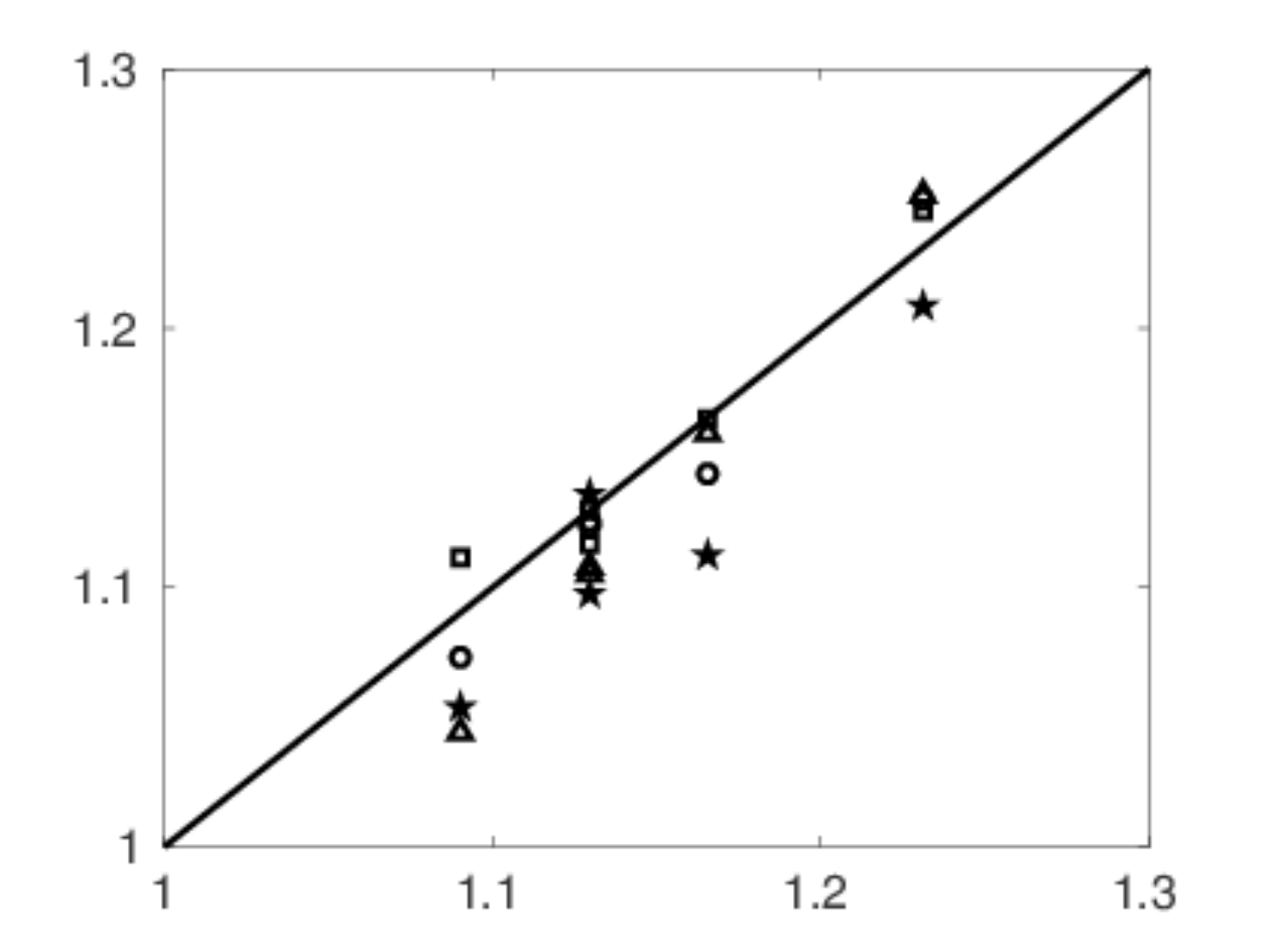}
\put(-170,140){{\large $(b) $}}
\put(-200,50){{\rotatebox{90}{$\tau_v(i+1)/\tau_v(i)$}}}
\put(-120,-5){{$[{\dot\gamma (i+1)} / {\dot\gamma (i)}]^{0.3}$}}
\\
\includegraphics[width=0.5\linewidth]{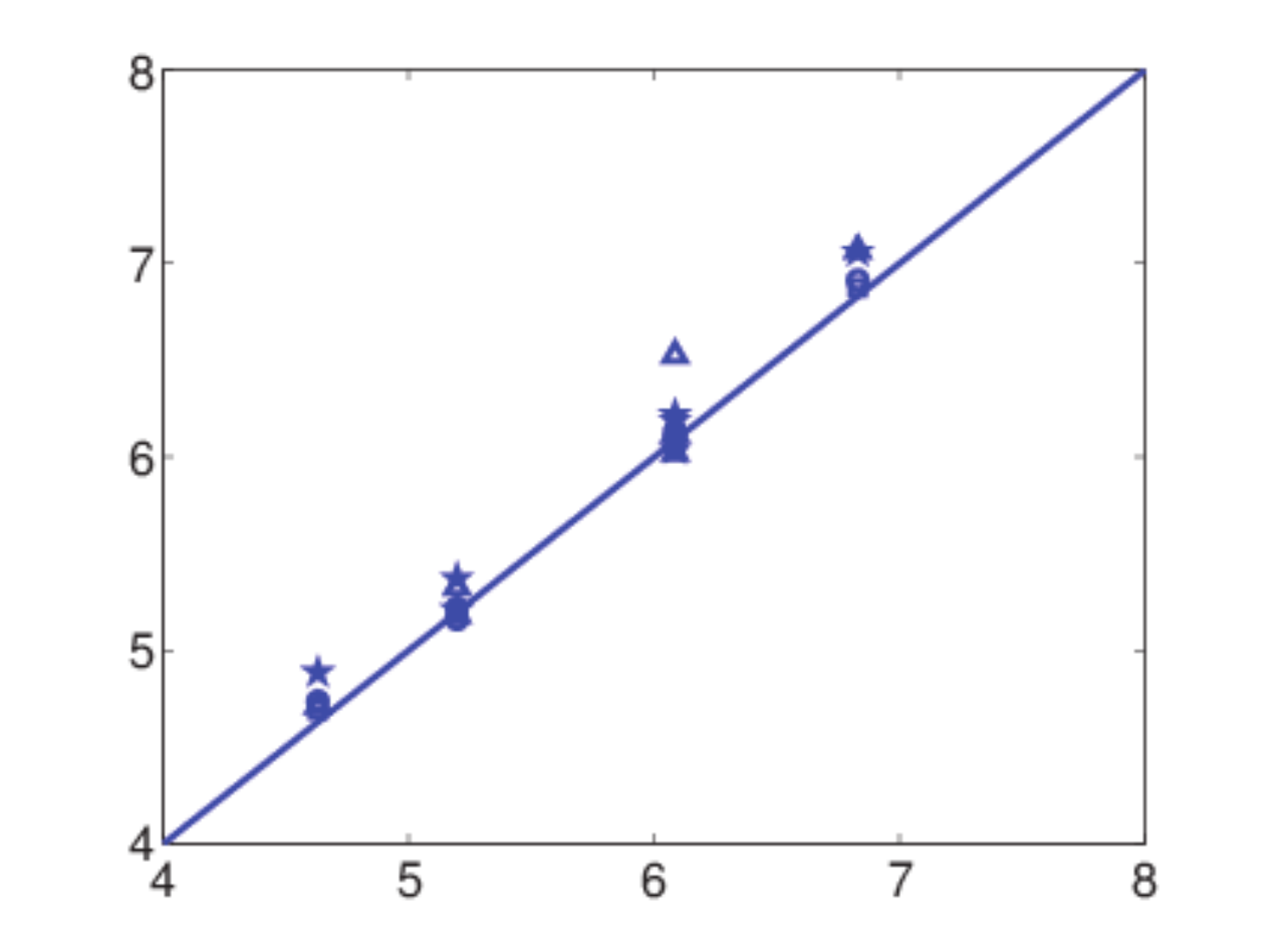}
\put(-170,140){{\large $(c) $}}
\put(-200,50){{\rotatebox{90}{$\tau_v(i+1)/\tau_v(i)$}}}
\put(-120,-5){{$[{\dot\gamma (i+1)} / {\dot\gamma (i)}]^{1.5}$}}
\\
\caption{\label{fig:tot_Vstress} 
{Scaling of viscous stresses for simulations with  (a) Newtonian, (b) shear thinning and (c) Shear thickening suspending fluids. 
Colors and symbols as in the previous figures.}}}
\end{figure}

\begin{figure}
\centering{
\hspace{-0.3cm}
\includegraphics[width=0.5\linewidth]{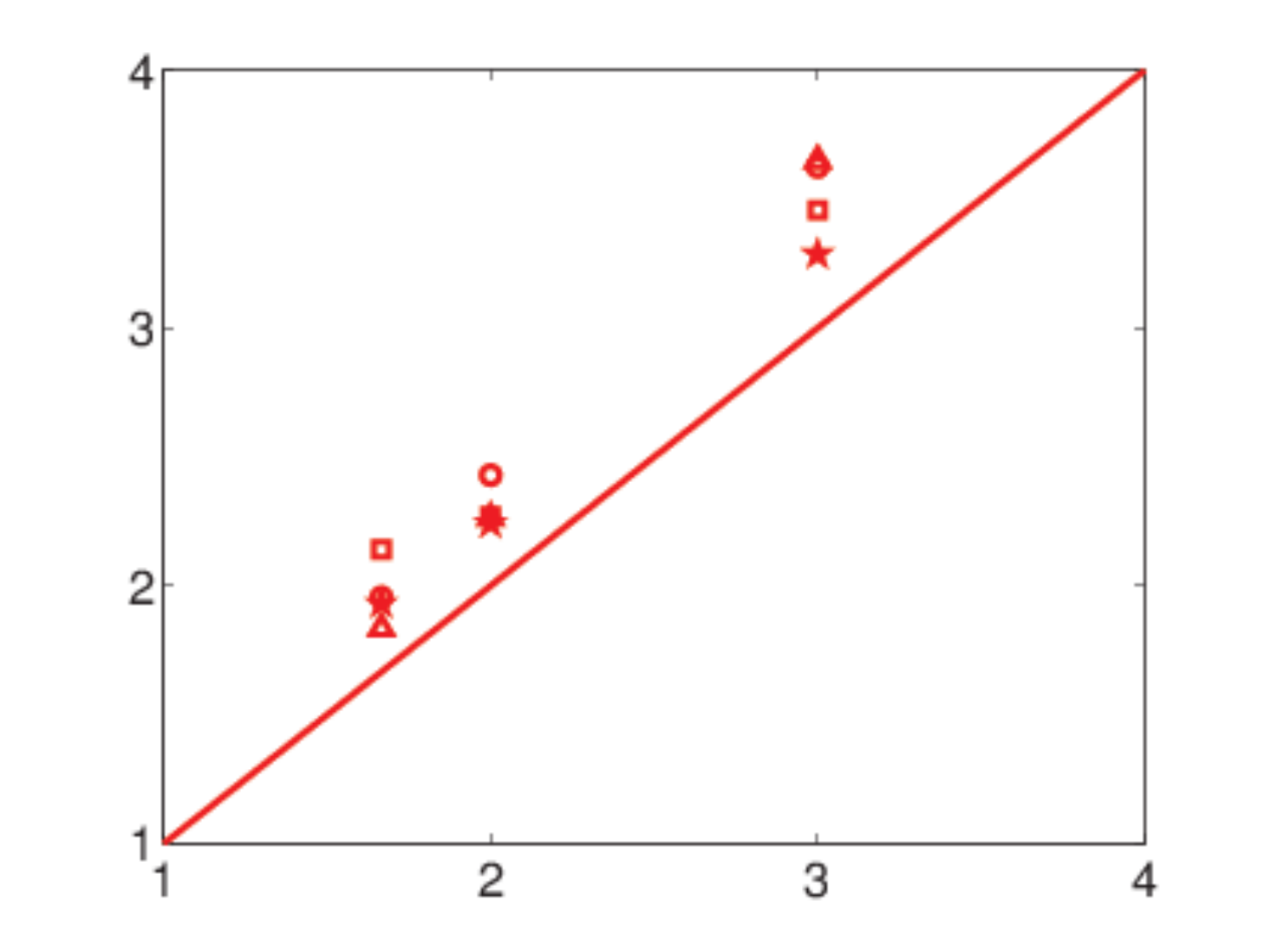}
\put(-170,140){{\large $(a) $}}
\put(-200,50){{\rotatebox{90}{$\tau_p(i+1)/\tau_p(i)$}}}
\put(-120,-5){{${\dot\gamma (i+1)/\dot\gamma (i)}$}}
\includegraphics[width=0.5\linewidth]{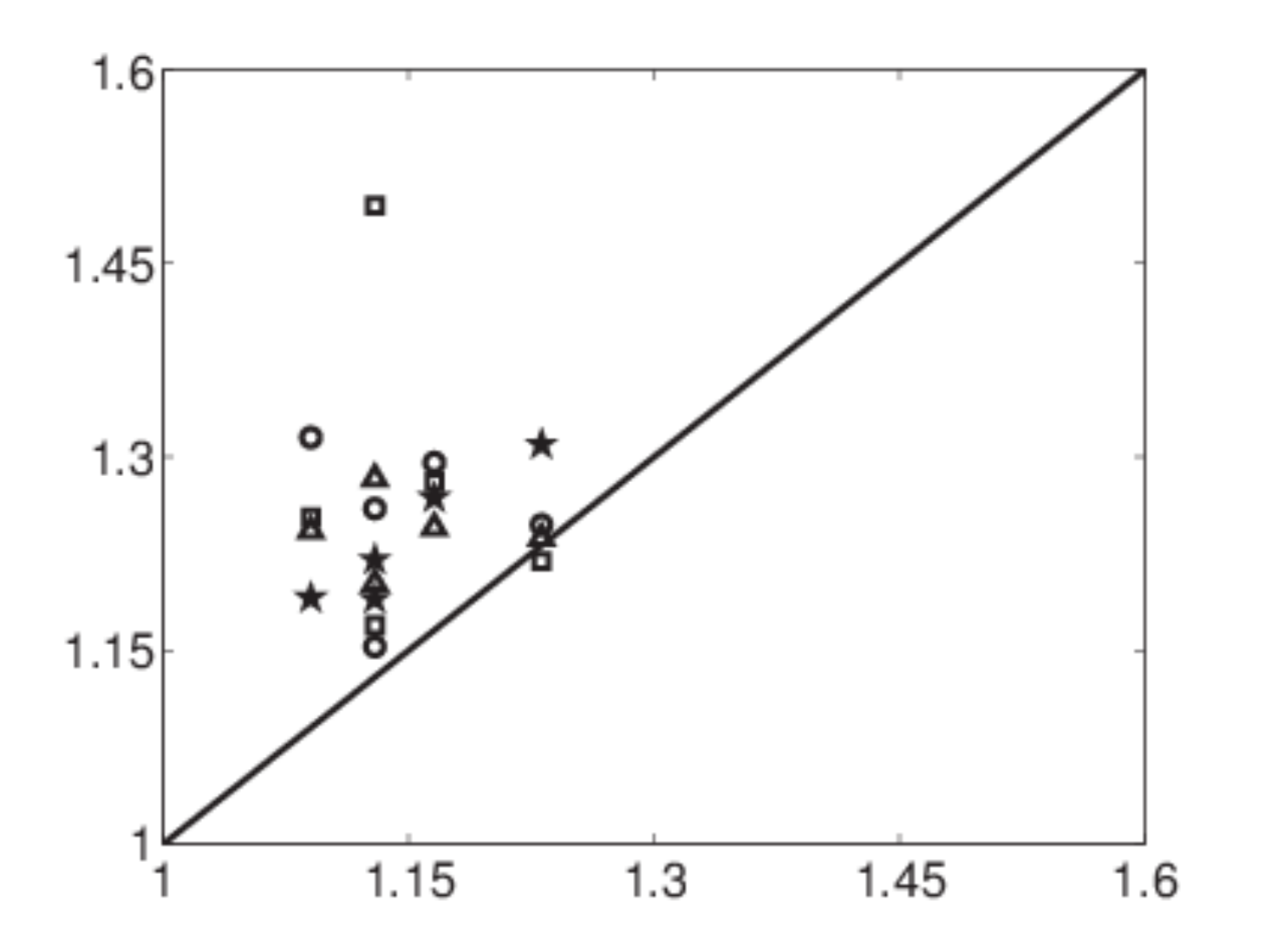}
\put(-170,140){{\large $(b) $}}
\put(-200,50){{\rotatebox{90}{$\tau_p(i+1)/\tau_p(i)$}}}
\put(-120,-5){{$[{\dot\gamma (i+1)} / {\dot\gamma (i)}]^{0.3}$}}
\\
\includegraphics[width=0.5\linewidth]{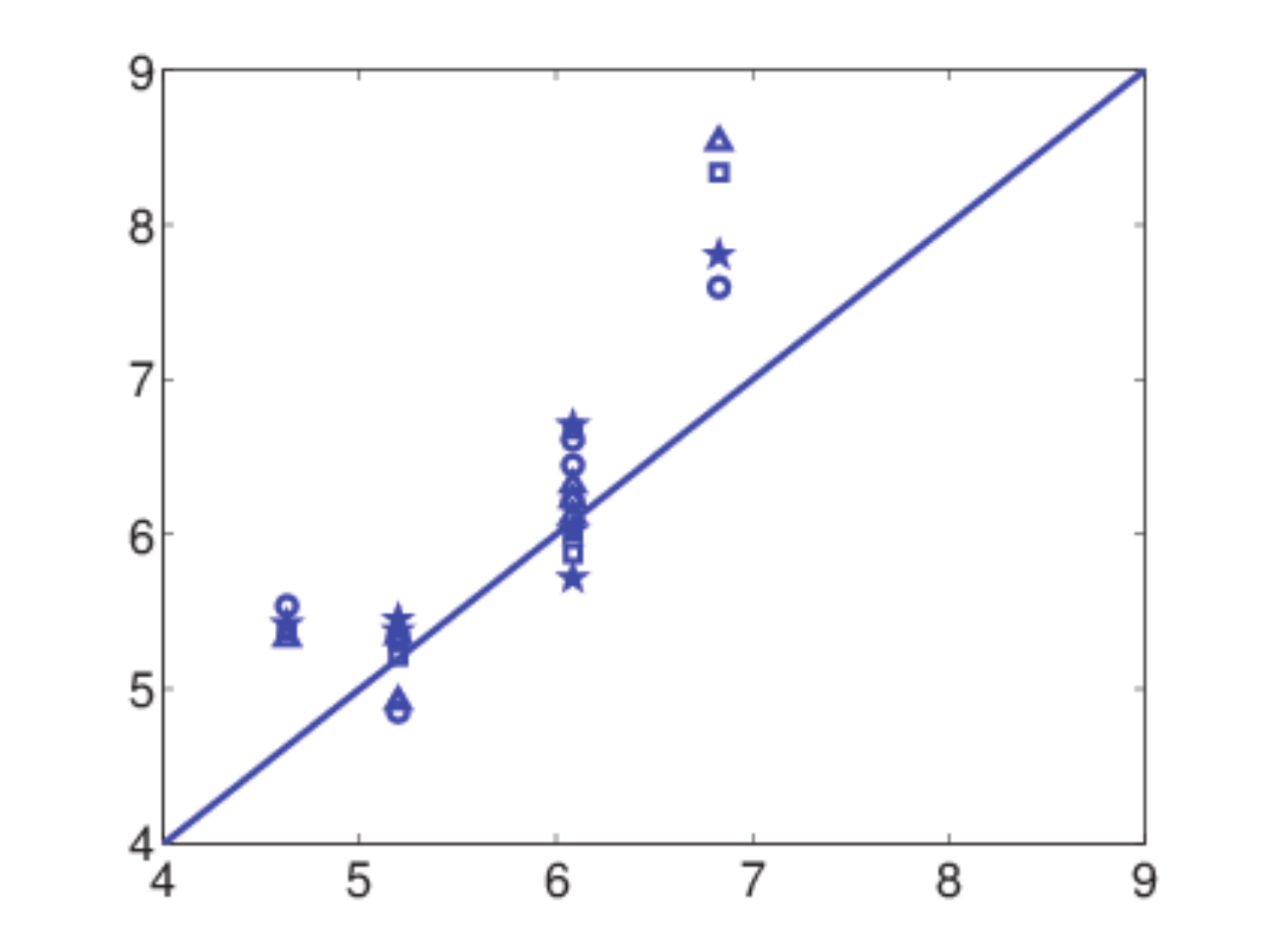}
\put(-170,140){{\large $(c) $}}
\put(-200,50){{\rotatebox{90}{$\tau_p(i+1)/\tau_p(i)$}}}
\put(-120,-5){{$[{\dot\gamma (i+1)} / {\dot\gamma (i)}]^{1.5}$}}
\includegraphics[width=0.5\linewidth]{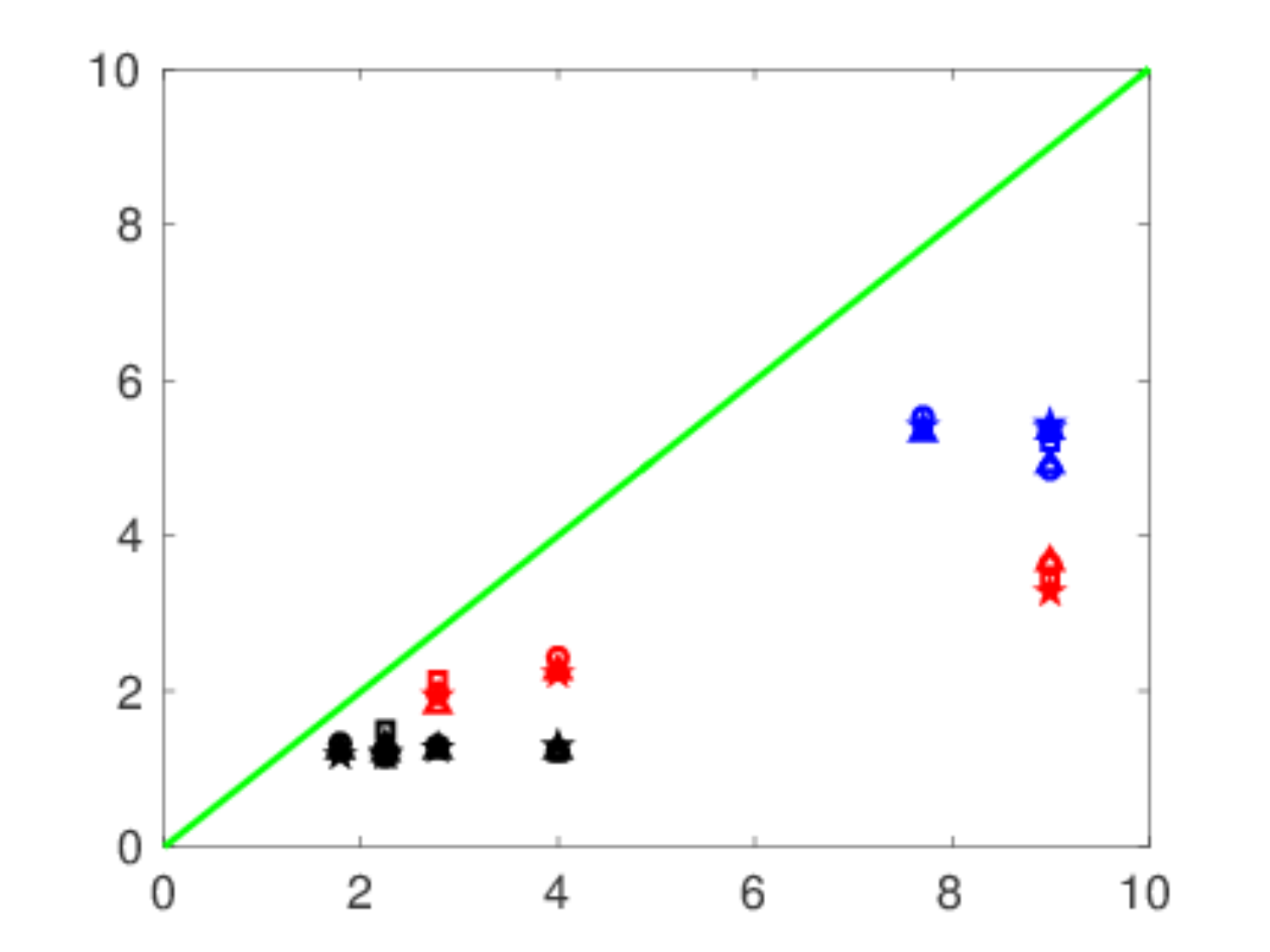}
\put(-170,140){{\large $(d) $}}
\put(-200,50){{\rotatebox{90}{$\tau_p(i+1)/\tau_p(i)$}}}
\put(-120,-5){{$[{\dot\gamma (i+1)} / {\dot\gamma (i)}]^{2}$}}
\\
\caption{\label{fig:tot_Pstress} 
{ Scaling of particle stresses for simulations with  (a) Newtonian, (b) shear thinning and (c) Shear thickening suspending fluids (d) all data. Colors and symbols as in the previous figures.}}}
\end{figure}

\begin{figure}
\centering{
\hspace{-0.3cm}
\includegraphics[width=0.9\linewidth]{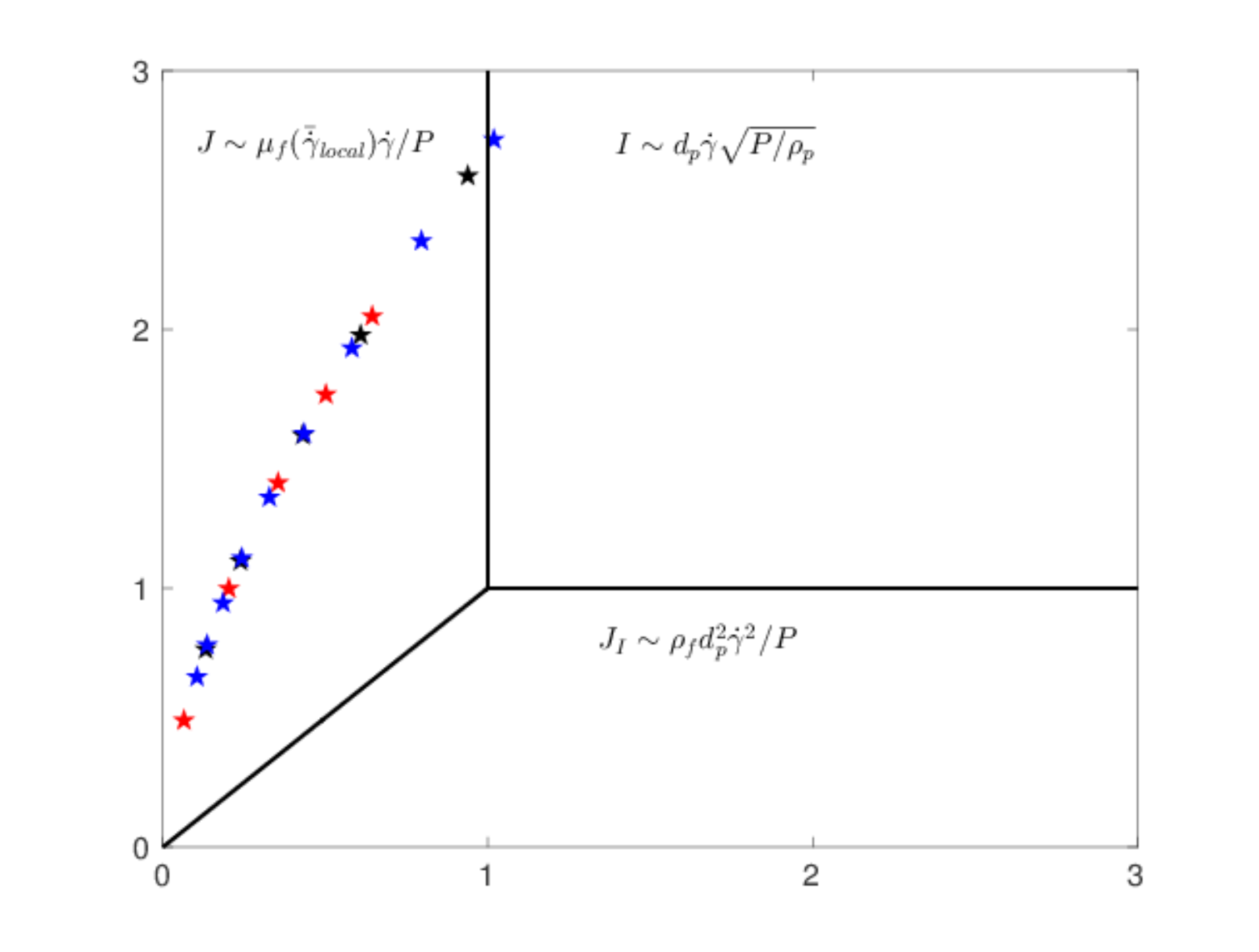}
\put(-330,70){{\rotatebox{90}{${t}_1/{t}_3 =(\rho_p /\rho_f)^{1/2} (4 \times Re_{p,local})^{5/16}$}}} 
\put(-240,0){{${t}_1/{t}_2 = \sqrt{(\rho_p /\rho_f)\times 4 \times Re_{p,local}/24}$}}
\caption{\label{fig:frictional_map} The phase diagram of three existing regimes for noncolloidal suspensions, styled after figure 7.13 in \cite{Andreotti13}.}}
\end{figure}

To shed light on the role of inertia, we provide scalings for the different components of the stress. First, we show that the viscous stress $\tau_v$ scales indeed viscously.  To this end, we  compute the ratio of  the values of the viscous stresses  in two successive simulations at constant 
solid volume fraction, cases $(i)$ and $(i+1)$, where we change the bulk shear rate from $\dot\gamma(i)$ to $\dot\gamma(i+1)$.  In the case of viscous scaling,  we expect $\tau_v(i+1)/\tau_v(i)=[ \dot\gamma(i+1)/\dot\gamma(i)]^n$, where $n=1$ for the Newtonian suspending 
fluid, $n=0.3$ for the shear thinning suspending fluid and $n=1.5$ for the shear thickening suspending fluid studied in this work. Fig. \ref{fig:tot_Vstress}a-c  show that our results follow the viscous scaling for all the cases. This also confirms the validity of our computational results. 

Second, we investigate the scalings for the particle stress. Unlike the viscous stresses, the scale of particle stresses is not known a priori due to non-zero $Re_p$ in our simulations.  We expect a scaling of  the order $\rho a^2 \dot\gamma^2$ when inertia is the
main mechanism to transport momentum in a suspension. 
We therefore  divide the values of the particle stresses  in two successive simulations  at constant solid volume fraction $(i)$ and $(i+1)$, differing for the bulk shear rate $\dot\gamma(i)$ and $\dot\gamma(i+1)$. Fig. \ref{fig:tot_Pstress}a-c show that for all
types of suspending fluids $\tau_p(i+1)/\tau_p(i)$ is slightly larger than $ [\dot\gamma(i+1)/\dot\gamma(i)]^n$ indicating that the viscous scaling is still a reasonably good approximation and the inertial contribution to the transport of momentum is a secondary effect.
To further confirm this, we report in Fig.  \ref{fig:tot_Pstress}d $\tau_p(i+1)/\tau_p(i)$ versus $ [\dot\gamma(i+1)/\dot\gamma(i)]^2$; the data points all fall below an inertial scaling, although $Re_p$ is finite. 
From the analysis in the figure, the results in \cite{Picano13} and the stress budget reported above, we therefore conclude that inertial effects alter the suspension microstructure, in particular the particle relative motion clearly breaking the fore-aft symmetry 
of Stokes flow, which induce shear thickening without altering the dominant viscous behavior. Different is the case presented in \cite{Lashgari14} where particles significant alter the turbulent flow of a suspension, triggering a transition from a Reynolds-stress 
inertial regime to a particle-dominated regime.

We therefore adopt a frictional view similar to that proposed in \cite{Cassar05,Andreotti13} to show that the main dimensionless number controlling the rheology of the suspensions studied here is associated with viscous stresses.  Assuming an assembly of
particles suspended in a viscous fluid that  is subject to steady shear under a confining particle pressure $P$, there exists only one dimensionless control parameter, i.e.,  the ratio of the microscopic timescale for the particles rearrangement to the macroscopic 
timescale of the flow ${\dot{\gamma}}^{-1}$. The microscopic timescale can be determined by balancing the forces applied on a particle, i.e., the imposed pressure $P$ and the drag force $F_d$ \citep{Cassar05,Andreotti13}. This implies the existence of the three
following microscopic timescales 
\begin{equation}
t_{1}\sim \frac{d_p}{\sqrt{P/\rho_p}}, ~~ t_{2}\sim \frac{d_p}{\sqrt{P/(\rho_f Cd_{St})}}, ~~ t_{3}\sim \frac{ d_p}{\sqrt{P/(\rho_f Cd_{Turb})}},~~ \mbox{where}~~ F_d= C_d d_p^2 \rho_f V^2.
\label{eq:eqintro}
\end{equation}
Here, $\rho_p$ and $d_p$ are the  particle density and  particle diameter, whereas the particle terminal velocity and its drag coefficient are denoted by $V$ and $C_d$. When the drag force is negligible and the rearrangement of particles is solely governed 
by the imposed pressure, the relevant microscopic timescale is $t_{1}$, and consequently, the controlling dimensionless parameter $I= t_1 \dot\gamma$. When  the imposed pressure balances the viscous drag force the microscopic timescale $t_2=\frac{d_p}{\sqrt{P/(\rho_f Cd_{St})}}$, 
leading to the controlling dimensionless parameter $J=t_2 \dot\gamma$.  Finally, the third scenario,  when  the imposed pressure balances the turbulent drag force on the particle, is associated to the microscopic timescale $t_3$  and the controlling dimensionless parameter is $J_I=t_3 \dot\gamma$. 

To calculate the values of the microscopic time scales $t_2$ and $t_3$ we need a closure for $C_d$. Following \cite{Hormozi17}, we adopt the   drag closure as follows
\begin{equation}
C_d(Re_{p,local}) = \left\{
                      \begin{array}{lcl}
                        \displaystyle{\frac{24}{4 \times Re_{p,local}}} & ~~~~ & 4 \times Re_{p,l} < 1.4 \\
                        \displaystyle{\frac{24/1.4^{0.375}}{(4 \times Re_{p,local})^{0.625}}} &  & 1.4 \leq 4 \times  Re_{p,l} \leq 500 \end{array}
\right. \label{CD2}
\end{equation}
 Note that the multiplier $4$ appears in front of  $Re_{p,local}$  since our local particle Reynolds number is based on the particle radius, not the particle diameter. The above closure for the drag coefficient allows us to calculate the ratio of the microscopic time scales:
 ${t}_1/{t}_2 = \sqrt{(\rho_p /\rho_f)\times 4 \times Re_{p,local}/24}$ and ${t}_1/{t}_3 =(\rho_p /\rho_f)^{1/2} (4 \times Re_{p,local})^{5/16}$.  Fig. \ref{fig:frictional_map} shows the phase diagram for the three  regimes of noncolloidal suspensions in the plane 
 $({t}_1/{t}_2, {t}_1/{t}_3)$. This map is styled after figure 7.13 in \cite{Andreotti13} \cite[see][for a detailed explanation]{Hormozi17}.

Three regimes can be identified on the frictional map in Fig.~\ref{fig:frictional_map}.  The first regime is associated with suspensions in which ${t}_1 \gtrsim {t}_2$ and ${t}_1 \gtrsim {t}_3$, where the dominant microscopic time scale is ${t}_1$ and the  primary
controlling dimensionless number is  $I \sim d_p \dot\gamma\sqrt{P/\rho_p}$. In this regime the main mechanism responsible for the momentum transport is that associated with particles contacts and collisions, i.e., in this regime particle inertia dominates.
The second regime, when ${t}_2 \gtrsim {t}_1$ and ${t}_2 \gtrsim {t}_3$, has ${t}_2$ as the dominant microscopic time scale and $J \sim \mu_{f}(\bar{\dot{\gamma}}_{local})\dot\gamma/P$ as the primary controlling dimensionless number. In this regime, 
the viscous stresses are the main mechanism for momentum transport, the viscous drag regime. The third regime, when ${t}_3$ is the dominant microscopic time scale, with relevant the dimensionless number $J_I \sim \rho_f d_p^2 \dot\gamma^2/P$, 
is dominated by the  Reynolds stresses of the fluid phase, i.e., the turbulent drag at  the particle scale  (inertial drag regime).

Note that this frictional framework  is relevant for dense suspensions in which the particle phase pressure is significant. Moreover, the boundaries between the three regimes discussed above and shown   in Fig.  \ref{fig:frictional_map} are unknown a
priori and finding the exact locations would require an extensive experimental or computational study. Here, we wish to determine the regime that our simulations for suspensions with $\Phi=0.4$ belong to.  It is reasonable to assume that a $40\%$ solid 
volume fraction is dense enough that the particle phase pressure $P$ is significant, and consequently, the frictional view just introduced can be used to characterize the suspension regimes. Fig.~\ref{fig:frictional_map}  shows that all of the dense 
simulations considered here (i.e.,  $\Phi=0.4$) fall into the regime where the viscous stresses play the main role in the momentum transport, and hence, the viscous number $J$ is the controlling dimensionless parameter. Therefore,  viscous forces are the main 
responsible for the dissipation of the power or energy given to the bulk of these suspensions, although $Re_p$ is non zero.

The above discussion rises the question of the role of inertia for the suspensions under investigation here and its influence on the effective viscosity. As shown in \S\ref{sec:eff_visc}, increasing  $Re_p$ results in an enhancement of the suspension
effective viscosity.  This is attributed to excluded volume effects in the work by  \cite{Picano13}: the effective volume fraction of the suspension is higher than the nominal one because of the ``shadow" region (a region with statistically vanishing 
relative particle flux) around the particles due to inertia in the relative particle motion. Therefore, finite inertia affects the suspension microstructure by increasing the effective $\Phi$, which, in turn, enhances the viscous dissipation.
This is why the enhancement of the effective viscosity with inertia is secondary with respect to increases of the nominal solid volume fraction  and the suspension viscosity can be scaled back to the  Eilers fit, obtained for Stokesian suspensions \cite[see][]{Picano13}.

Our study shows that the boundary between the inertial $I$-dominated and the viscous $J$-dominated regimes as estimated by scaling arguments with no prefactor is a good approximation of the real one at least for the suspensions examined here, see Fig.~\ref{fig:frictional_map}.
Despite non zero values of  $Re_p$ in our simulations, we are not in the regime where the particle inertia dominates, and therefore, closures for the suspension stresses should differ from those given in the recent studies by \cite{Trulsson12}. In fact the inertia affects the microstructure
and can be included into the suspension stress closure as follows 
 \begin{equation}
\tau=G(\Phi+\delta\Phi(\Phi,Re_{p,local})) \mu_0\hat{\mu}(\bar{\dot{\gamma}}_{local}(\Phi)) \dot\gamma,
\label{eq:sus_stress}  
\end{equation}
\begin{equation}
\delta\Phi=0.9 \times  Re_{p,local} \Phi^2(1-\Phi/\Phi_{max})^3.
\label{eq:delta_Phi}  
\end{equation}
Here, $\delta\Phi$ is the increase in the nominal volume fraction due to the excluded volume effects mentioned above. Equation (\ref{eq:delta_Phi}) provides an estimation of this increase of effective volume fraction by fitting the simulation results of \cite{Picano13} in which $\delta\Phi$ is obtained by computing the volume of the so-called ``shadow" regions. Note that the form of equation (\ref{eq:delta_Phi})  is chosen to satisfy the following conditions. First, at small $\Phi$, we expect $\delta\Phi$ to be proportional to $\Phi^2$, since at least two-particle interactions are needed to give rise to the excluded volume effect. Second,  the excluded volume must vanish as  $\Phi \rightarrow \Phi_{max}$. In summary, the closure (\ref{eq:sus_stress}) provides the suspension shear stress for Newtonian and generalized Newtonian suspending fluids where the microstructure modifications induced by inertia are coded into the $G$ function. We note again here that this closure is valid for suspension in which viscous effects are the  still main mechanism for momentum transport, while inertial effects are 
mainly responsible for changing the microstructure. Note that  the local Reynolds number seen by the particles, $Re_{p,local}$, is the appropriate dimensionless number  to model the effects of the microstructure. 

\section{Conclusion and remarks} 

We study suspensions of neutrally buoyant spheres in both Newtonian and  inelastic non-Newtonian fluids, using  interface-resolved direct numerical simulations. The Carreau and power-law models are employed to describe the rheological behaviour of shear thinning and shear thickening carrier fluids, where the fluid viscosity varies instantaneously with the local flow shear rate, $\dot{\gamma}_{local}(x,y,z)$. The simulations are performed in a plane-Couette configuration. The rheological parameters of the suspending fluids and the particle phase properties are kept constant while we change the volume fraction of the solid phase, $0 \le \Phi \le 40\%$, and the applied shear rate. This allows us to study inertial suspensions with $0 < Re_p \leq 10$. We focus on the bulk properties of the suspension as well as the local behavior of the particles and the carrier fluid. The main findings of our work are summarized as follows.

The local profile of solid volume fractions show particle layerings for nominal volume fractions $\Phi \ge 30\%$, with the peak of the layers located close to the wall. The distribution of particles across the Couette cell is mainly controlled by geometry and confinement effects, with a weaker dependency on the type of suspending fluid and inertial effects. However, the latter has strong influence on the velocity profiles of particle and fluid phase. In the cases with non-Newtonian suspending fluids, viscosity at the particle scale depends on the local shear rate. This results in a larger and smaller local particle Reynolds number as the shear thinning and thickening degree increase respectively. Inertia also induces a slip velocity between solid and fluid phases with a maximum value close to the wall and zero at the gap centre. This suggests that a suspension flow should be formulated  with  a two-phase  continuum  framework close to the walls, where inertial effects are first apparent, although a mixture 
continuum framework can still provide reasonable predictions of the flow away from the walls, i.e., around the central region. Note that this is also observed for a turbulent suspensions and has been successfully used in \cite{Costa16} to predict turbulent drag in channel flows.

We present the probability density function of the local shear rate, revealing the existence of a wide spectrum of local shear rate; this depends primarily on $\Phi$ and secondarily on $Re_p$, which implies the deficiency of theoretical approaches based on mean field values in explaining the mechanics of suspensions. Mean field  theories \citep[e.g.,][]{Chateau08}  should be refined including higher moments, including possibly minima and maxima, of the local shear rate distribution.
This requires further investigation and improvement of the available model frameworks.

We demonstrate that the non-dimensional relative viscosity of the  suspensions with non-Newtonian carrier fluids can be well predicted by the homogenization theory of \cite{Chateau08} in the limit of $Re_p \rightarrow 0$, and more  accurately for lower $\Phi$.
However, adding inertia to the system alters the microstructure and results in a deviation of the relative viscosity of the suspensions from the Stokesian prediction, while the main dissipation mechanism is still viscous. In fact, we show that for the parameter range explored here both particle stresses and fluid stresses are clustered about viscous scalings. We therefore adopt the frictional view of \cite{Cassar05,Andreotti13} to show that the main dimensionless number controlling the mechanics of suspensions is the so-called viscous number, $J \sim \mu_{f}(\bar{\dot{\gamma}}_{local})\dot\gamma/P$, confirming that viscous stresses are responsible for the momentum transport even when the particle Reynolds number is finite. However, due to inertial effects, the microstructure becomes anisotropic and so-called shadow regions form around particles \cite[these are regions with zero probability of finding another particle, see][]{Picano13}. 
This enhances the  effective particle volume fraction, and consequently, the viscous dissipation and relative viscosity. 
We have estimated the volume of these shadow regions from our simulations and  included this microstructural effect into a functional form for the relative viscosity. In this way, we provide a prediction for the added excluded volume due to inertia and a closure for the suspension stress in the case of both Newtonian and generalized Newtonian suspending fluids valid for $O(Re_p)\sim 10 $. Note also that once the effective volume fraction is considered, Eilers fit is able to predict the suspension viscosity, confirming that we are still in the viscous regime.

Concerning the frictional framework of  \cite{Cassar05} and \cite{Andreotti13}, 
we note that this view  provides scaling for the suspension stresses but no information about the microstructural effects. Therefore, refinements of existing rheological laws would need to include microstructural effects; these local details of the suspension flows can be obtained by  well-resolved 
experimental or computational data  as done here. Our study shows that the so-called inertial shear thickening mode introduced by  \cite{Picano13} belongs to the viscous class of suspensions, yet the inertial microstructure needs to be taken into account to accurately predict the suspension rheology. 

This study raises a series of research questions requiring further investigations. First, we show that as we change the type of suspending fluid the contribution of particle stress to the total suspension stress changes. In our system, this is most likely mainly due to the change in the stresslet.
A systematic study may therefore be devoted to investigate the  dependency of the stresslet on the non-Newtonian behaviour of the suspending fluid and its influence on the bulk rheology. 
Second, the simulations presented here fall into the viscous-dominated suspension regime, although we show how microstructural effects due to inertia are not at all negligible. One may therefore further explore the frictional map by performing simulations/experiments for a wider range of $\Phi$ and $Re_p$, one goal being to determine the boundaries between 
the viscous and inertial dominated transport mechanism. Such an analysis would also suggest how to include microstructural effects in order to refine rheological laws. 
Third, we consider here the simplest shear flow, with homogeneous bulk shear rate. It is not obvious 
how inhomogeneous bulk shear rates changes the results and how non-local rheology comes into play \cite[see e.g.,][]{ Pouliquen09, Bouzid13, Henann14, Kamrin15}.  Finally, theoretical approaches based on mean field theory should be improved to include the consequences of the wide spectrum of the local shear rate. To this end, it is essential to study complex features of suspensions with non-Newtonian suspending fluids.

\section*{Acknowledgments}
Parts of this research were supported  by NSF (Grant No. CBET-1554044-CAREER) and NSF-ERC  (Grant No. CBET-1554044 Supplementary CAREER) via the research awards (S.H.). L.B. acknowledges financial support by the European Research Council Grant no. ERC-2013-CoG- 616186, TRITOS. The authors acknowledge computer time provided by SNIC (Swedish National Infrastructure for Computing).

\bibliographystyle{jfm}

\end{document}